\documentclass[aip,amsmath,amssymb,preprint]{revtex4-2}
\usepackage[utf8]{inputenc}
\usepackage{graphicx}
\usepackage{bm}
\usepackage{makecell}
\usepackage{threeparttable}
\usepackage{float}
\usepackage{longtable}
\usepackage{multirow}
\usepackage{booktabs}
\usepackage{mathptmx}
\usepackage{etoolbox}
\usepackage[version=4]{mhchem}
\usepackage{mathrsfs}

\makeatletter
\def\@email#1#2{%
 \endgroup
 \patchcmd{\titleblock@produce}
  {\frontmatter@RRAPformat}
  {\frontmatter@RRAPformat{\produce@RRAP{*#1\href{mailto:#2}{#2}}}\frontmatter@RRAPformat}
  {}{}
}%
\makeatother

\begin{document}

\title{A new framework for frequency-dependent polarizable force fields}

\date{\today}

\author{YingXing Cheng}
\author{Toon Verstraelen}\email{toon.verstraelen@ugent.be}
\affiliation{Center for Molecular Modeling (CMM), Ghent University - Technologiepark-Zwijnaarde 46, B-9052 Gent, Belgium}

\begin{abstract}
A frequency-dependent extension of the polarizable force field ``Atom-Condensed Kohn-Sham density functional theory approximated to the second-order'' (ACKS2) [J. Chem. Phys. 141, 194114 (2014)] is proposed, referred to as ACKS2$\omega$. 
The method enables theoretical predictions of dynamical response properties of finite systems after a partitioning of the frequency-dependent molecular response function.
Parameters in this model are computed simply as expectation values of an electronic wavefunction, and the hardness matrix is entirely reused from ACKS2 as an adiabatic approximation is used.
A numerical validation shows that accurate models can already be obtained with atomic monopoles and dipoles.
Absorption spectra of 42 organic and inorganic molecular monomers are evaluated using ACKS2$\omega$, and our results agree well with the time-dependent DFT calculations.
Also for the calculation of $C_6$ dispersion coefficients, ACKS2$\omega$ closely reproduces its TDDFT reference.
When parameters for ACKS2$\omega$ are derived from a PBE/aug-cc-pVDZ ground state, it reproduces experimental values for 903 organic and inorganic intermolecular pairs with an MAPE of 3.84\%.
Our results confirm that ACKS2$\omega$ offers a solid connection between the quantum-mechanical description of frequency-dependent response and computationally efficient force-field models.
\end{abstract}

\maketitle

\section{Introduction}
Molecular simulation is a powerful tool to investigate and predict the properties and dynamics of a wide range of finite and condensed-phase systems.
Different levels of theory for such simulations can be roughly divided into two categories: quantum-mechanical (QM) electric structure methods and force-field (FF)  methods.
Kohn-Sham Density Functional Theory (DFT) offers a good trade-off between computational efficiency and accuracy, which has been successfully applied to study both ground and excited state properties of inorganic and organic molecular systems.
However, Kohn-Sham DFT (with density fitting) has an O($N^3$) computational complexity where $N$ is the number of basis functions, which impedes its application to large-scale systems, especially for molecular dynamics (MD) simulations.
On the other hand, FF methods, only using a set of functions of molecular geometry with several empirical parameters to determine molecular energy, have been developed and applied to large-scale MD simulations.
In general, FF methods have a  much lower computational cost than QM methods, at the expense of a reduced, yet acceptable, computational accuracy.
One significant problem in traditional FF methods is the lack of electronic polarization, i.e., the response of the electron distribution to the environment, which led the molecular modeling community to develop polarizable force fields (PFFs) \cite{Warshel1976,Kaminski2002,Ren2002,Jorgensen2007,Shi2013,Lamoureux2003,Yu2003,Lemkul2016,Mortier1985,Mortier1986,C7SC01181D,Cools-Ceuppens2022} to improve the predictive accuracy of long-range interactions by adding explicit polarization effects.

Classical PFF models can be classified into four broad categories: induced dipoles or multipoles,\cite{Warshel1976,Kaminski2002,Ren2002,Jorgensen2007,Shi2013} the Drude oscillator \cite{Lamoureux2003,Yu2003,Lemkul2016} (also known as ``core-shell''\cite{VanMaaren2001} or ``charge-on-spring''\cite{Kunz2009}), Electronegativity Equalization Method (EEM) \cite{Mortier1985,Mortier1986}, and the explicit electron model. \cite{C7SC01181D,Cools-Ceuppens2022}
It should be noted that several models similar to the EEM exist, such as fluctuating charges (FQ or FlucQ),\cite{Rick1994,Rick1996,Stern2001,Patel2004,Patel2004a} charge equilibration (QE, QEq,\cite{Rappe1991} or CHEQ\cite{Zhong2010}), and chemical potential equalization (CPE)\cite{York1996}. Note that EEM is often not considered to be a proper PFF: because it only uses variable atomic charges, it cannot describe out-of-plane polarization of planar molecules.\cite{York1996} 
One can represent these effects in EEM by including additional off-nuclear sites.\cite{Devereux2014,Unke2017,Devereux2020}
This leads to slightly more expensive models and additional model choices, such as the location of the extra sites and the associated parameters.
Here, we will treat EEM on the same footing as other PFFs because its mathematical structure is completely analogous. 

The EEM was first developed by Mortier \textit{et al.} from basic DFT equations according to Sanderson's principle of electronegativity equalization,\cite{Sanderson1951} i.e., electrons flow until all electronegativities are equalized during the formation of a molecule.
The ability of EEM (or one of its close relatives) to predict ground-state charge distributions has been validated extensively, e.g. for inorganic liquids,\cite{Rick1996} inorganic solids,\cite{VanDuin2003,Smirnov2003,Hallil2006} organic molecules,\cite{Bultinck2002,VanDuin2001,Bultinck2004,Patel2004} and biomolecular systems.\cite{Patel2004a,Yang2006,Verstraelen2012,Ionescu2013}
One appealing aspect of the EEM is that atomic monopolar fluctuations are included, and that one may systematically extend it with higher atomic multipoles, as in CPE.\cite{York1996} 
However, the EEM also has two fundamental issues in simulations of extended systems or charge transfer during chemical reactions as follows:\cite{chelli1999,nistor2006,warren2008,nistor2009,Verstraelen2013} (1) the EEM always predicts a cubic scaling of the dipole polarizability with system size, which is not correct for dielectric systems where a linear scaling exists in the macroscopic limit; (2) EEM obtains fractional molecular charges even when two molecules are well separated and gives a slowly decaying intermolecular charge transfer.
The main reason for both problems is that long-range charge flow is always allowed in EEM, which is only realistic in metallic systems.
Shortly after these shortcomings were first discussed in the literature, several \textit{ad hoc} improvements were proposed, for instance, by applying artificial constraints to molecular charges\cite{Patel2004,warren2008} or dipoles,\cite{Chelli2005} or introducing a distance-dependent function to penalize long-range charge transfer.\cite{Chen2007}
Later, more consistent models were proposed, such as the split-charge equilibration (SQE): by adding a bond-hardness term to EEM, SQE becomes capable of modeling dielectric materials. \cite{nistor2006,nistor2009}

Recently, Verstraelen \textit{et al.} developed a new polarizable force field (PFF), namely Atom-Condensed Kohn-Sham DFT approximated to second-order (ACKS2), \cite{Verstraelen2013} which has been validated and applied to molecular systems, and which addresses both fundamental issues of the EEM.
ACKS2 improves upon the EEM by adding an electronic kinetic energy term via the Legendre transform of the Kohn-Sham kinetic energy, enabling one to expand the kinetic energy to second-order in the atomic populations and atomic Kohn-Sham potentials.
It was shown that this new kinetic energy term is equivalent to the bond-hardness energy in the SQE.
Later, a generalization of the ACKS2 model for arbitrary variational wavefunctions was proposed.\cite{Verstraelen2014} The ACKS2 formalism was numerically validated by a direct computation of the ACKS2 parameters from a reference Kohn-Sham DFT calculation. 
With these parameters, the static linear response properties of molecules obtained with an underlying theory, e.g., Kohn-Sham DFT, can be reproduced precisely in the limit of a complete basis sets for  density and potential fluctuations. 

To use ACKS2 for approximate polarization energy calculations,
a tailored empirical parameterization was incorporated into ReaxFF, as an alternative to the EEM, which turned out to be essential in the development of eReaxFF. \cite{Islam2016}
Later, G{\"{u}}tlein \textit{et al.} developed a new PFF using Gaussian basis sets, based on the ACKS2 model, \cite{Gutlein2019} where generic parameters were proposed for elements C and H, such as widths of Gaussian basis functions.
This parametrization can predict DFT data, such as response properties and interaction energies, of a series of hydrocarbons.
Later, they proposed two ACKS2 variants, f-ACKS2 and scf-ACKS2, to handle condensed phases by dividing them into molecular fragments, which are coupled by Coulomb interactions.\cite{Gutlein2020}

This work extends ACKS2 to the time- and frequency-dependent domain to compute the dynamical response properties of finite molecules, e.g., frequency-dependent polarizabilities, $C_6$ dispersion coefficients, and molecular absorption spectra.
The resulting ACKS2$\omega$ model, through its use of atomic multipole expansions, allows one to investigate the contribution of charge-flow effects on dynamic response properties, which is a topic of ongoing research. 
One of the future applications envisioned for ACKS2$\omega$ is an efficient approximation of the non-local dispersion energy, i.e. for which interatomic charge fluctuations become important,\cite{Stone1989,Dobson2006,Hermann2017,Jackson2007,Jackson2016,Misquitta2010,Dobson2012,Misquitta2014} such as carbon nanomaterials,\cite{Dobson2006,Hermann2017} traditional semiconductors,\cite{Jackson2007,Jackson2016} and low-dimensional materials.\cite{Dobson2006,Misquitta2010,Dobson2012,Misquitta2014}
To our best knowledge, the time-dependent or frequency-dependent extension of ACKS2 has not been investigated yet.

The derivation of ACKS2$\omega$ relies on time-dependent DFT (TDDFT).
TDDFT has received much attention in recent decades because of its good trade-off between numerical efficiency and accuracy compared to wavefunction-based methods.
The standard formalism of TDDFT builds on the Runge-Gross theorem, \cite{Runge1984} stating that there exists a one-to-one correspondence between densities and potentials for any fixed initial many-body state.
TDDFT uses many DFT concepts, e.g. the whole interacting many-body system is replaced by a non-interacting Kohn-Sham system that shares the same density.
However, it also has unique requirements, such as the initial state dependence and time-dependent exchange-correlation functional.
Readers are referred to recent books \cite{ullrich2011time,marques2012fundamentals} and review articles \cite{Burke2005,Casida2012,Marques2004,Laurent2013} for further details about TDDFT.
Within the TDDFT regime, the linear response of the electron density can be calculated precisely, assuming one disposes of the exact exchange-correlation functional.
In practice, TDDFT is a useful and relatively simple scheme to compute molecular linear-response properties. \cite{Kjellgren2019,Osinga1997,Aiga1999,Castro2004,Hedegard2013}

To derive the ACKS2$\omega$ model, we use a quasi-energy formalism, also known as the Floquet theory, which is often used in time-dependent response model development \cite{Fromager2013,Saek2002,Hedegard2013,Tunell2003,Christiansen1998,Telnov1997,Salek2005} and can be seen as a special case of TDDFT. \cite{Runge1984,Deb1982} 
The benefits of the quasi-energy language have been discussed elsewhere.\cite{Salek2005} 
One of the advantages of quasi-energy DFT is that the initial state dependence of the Runge-Gross theorem is not necessary anymore, and instead a periodic boundary condition is applied to an external perturbation.
Moreover, the quasi-energy ansatz enables a unified framework for dynamic linear-response property calculations based on wavefunction and DFT methods, as it is generally valid for both variational and non-variational methods.\cite{Salek2005}
This allows dynamical response properties to be derived by the virtues of the energy derivative approach using the quasi-energy formalism.
However, the validity of Floquet theory has been a point of discussion elsewhere \cite{Maitra2002,Samal2006,Maitra2007,Kapoor2013} and principle requires the following approximations: (i) a finite basis is employed;\cite{Maitra2007,Kapoor2013} (ii) the driving external perturbation is weak and off-resonant.\cite{Langhoff1972,Maitra2007,Kapoor2013}
Point (i) ensures an adiabatic limit exists for the ``Floquet ground state'', on which the energy minimum principle is valid, and the latter determines the one-to-one density-potential mapping.
Point (ii) suggests that Floquet theory is only a proper approximate method to treat liner-response problems.
Nonetheless, for molecular force-field development, these two conditions are generally satisfied; thus, one can still use Floquet theory to develop frequency-dependent force-field methods.

The goal of this paper is to introduce ACKS2$\omega$ as a model with a structure reminiscent of conventional PFFs, yet with all parameters defined in terms of an underlying electronic structure theory.
This direct connection to electronic structure theory assures a solid foundation for more pragmatic parameterizations of frequency-dependent PFFs. \cite{mayer_electrostatic_2008, smalo_combined_2013, haghdani_complex_2014,Hermann2020, ambrosetti_long-range_2014, wildman_nonequilibrium_2019}
The main distinction with the work of Misquitta and Stone, e.g. ISA-Pol,\cite{misquitta_isa-pol_2018} is that ACKS2$\omega$ explicitly describes the transformation from non-interacting to interacting response functions through electron-electron interactions, whereas ISA-Pol directly models the interacting response relying on time-dependent Hartree-Fock or TDDFT results.
Potential applications of frequency-dependent PFFs include modeling of optical spectra of complex systems at a force-field cost,\cite{mayer_electrostatic_2008, smalo_combined_2013, haghdani_complex_2014} approximations of dispersion in force fields,\cite{misquitta_isa-pol_2018} computationally efficient non-local dispersion corrections including many-body effects for DFT calculations\cite{ambrosetti_long-range_2014,Hermann2020} or frequency-dependent polarizable embedding.\cite{wildman_nonequilibrium_2019}
In this paper, ACKS2$\omega$ parameters are always computed from a Kohn-Sham DFT calculation for any given molecular geometry.
This leads to relatively accurate and expensive parametrizations of test systems, primarily intended to validate the ACKS2$\omega$ equations.
Obviously, for large-scale simulations, more efficient parameterizations should be developed, not involving a prior Kohn-Sham DFT calculation, for which machine-learning approaches have shown promising results.\cite{schriber_cliff_2021}

A benchmark database including 42 inorganic and organic molecules from Ref.~\onlinecite{Tkatchenko2009}, referred to as TS42, is used for a numerical validation of the ACKS2$\omega$ model.
First, the absorption spectra of the 42 molecules are investigated using ACKS2$\omega$ in comparison to TDDFT calculations.
Then, the $C_6$ dispersion coefficients are compared against experimental data and Tkatchenko-Scheffler van der Waals method (called TS in what follows).
More specifically, the accuracy of the ACKS2$\omega$ model is tested by checking the $C_6$ coefficients of 903 molecular pairs constructed from the database.
Furthermore, we also compare the performance of ACKS2$\omega$ to the range-separated calculations of Toulouse \textit{et al.} on 27 homodimers from a subset of TS42, referred to hereafter as the TS27 database.\cite{Toulouse2013}

The remainder of this paper is organized as follows: the basic theory of the ACKS2$\omega$ model is introduced in Section \ref{sec:theory}, followed by computational details of the numerical validation in Section \ref{sec:cp_details}. Results and discussion are presented in Section \ref{sec:results}.
A summary is given in Section \ref{sec:summary}. Atomic units are used, unless noted otherwise.

\section{Theory}
\label{sec:theory}
This section derives the ACKS2$\omega$ formalism from TDDFT, using the quasi-energy formalism.
Hence, a brief review of the quasi-energy method is first presented in Section \ref{subsec:quasienergy}. 
The quasi-energy expressions have been employed in response theory because of their straightforward definition.\cite{Rice1991,Christiansen1998}
The theory of ACKS2$\omega$ within quasi-energy ansatz is presented in Section \ref{subsec:part_functional}, where a set of equations is derived in analogy with ACKS2.\cite{Verstraelen2014} 
A dynamic linear-response theory is then provided in the frequency domain in Section \ref{subsec:LR}.
As a result, a set of linear-response equations are generated when two finite basis sets are applied (Section \ref{subsec:basis_sets}). 
Finally, in Section \ref{subsec:KS}, we apply ACKS2$\omega$ in Kohn-Sham DFT with a semi-local exchange-correlation (xc) functional.

\subsection{Quasi-energy formalism}
\label{subsec:quasienergy}
The quasi-energy formalism or Floquet theory is employed in the original TDDFT derivation.\cite{Runge1984}
The definitions of time-dependent exchange-correlation potentials and kernels are much more accessible using the quasi-energy notation.
In this section, we use the notation of Salek \textit{et al.} \cite{Salek2005} and Christiansen \textit{et al.}\cite{Christiansen1998}
From the time-dependent Schr{\"{o}}dinger equation, we have
\begin{gather}
    \left ( \hat{H} - i\frac{\partial}{\partial t} \right ) | \overline{0}(t) \rangle = 0 \\
    \hat{H} = \hat{H_0} + \hat{V}_1(t)
\end{gather}
where $\hat{H}_0$ and $\hat{V}_1(t)$ are the time-independent zeroth-order Hamiltonian and time-dependent perturbation operator, respectively.
The exact solution of $\hat{H}_0$ is given by a ground-state calculation and perturbation theory can be developed through a generalization of the Rayleigh-Schr{\"o}dinger theory.\cite{Salek2005}
We consider the time-dependent perturbation theory \cite{Langhoff1972,Christiansen1998}
\begin{equation}
    |\overline{0}(t) \rangle = e^{-iF(t)}|\tilde{0}(t) \rangle
\end{equation}
where $F(t)$ is a purely time-dependent function. 
The vector $|\overline{0} \rangle$ represents the complete wavefunction whose phase-isolated part is denoted by $|\tilde{0}\rangle$, where time dependence of the vectors is understood.
The time-dependent Schr{\"{o}}dinger equation is then rewritten as 
\begin{equation}
    \left( \hat{H} - i\frac{\partial}{\partial t}\right) |\tilde{0} \rangle = L(t) |\tilde{0} \rangle 
\end{equation}
in which $L(t) = \dot{F}(t)$ is the time-dependent quasi-energy.
Langhoff \textit{et al.} showed that $|\tilde{0} \rangle$ can be normalized at all times.\cite{Langhoff1972}
Consequently, the time-dependent quasi-energy can be expressed as 
\begin{equation}
    L(t) = \langle \tilde{0} | \hat{H} - i\frac{\partial}{\partial t} | \tilde{0} \rangle
    \label{eq:tdse}
\end{equation}

In order to use Hellmann-Feynman theorem, in analogy to the static case, the external time-dependent perturbation is assumed to be periodic, i.e., $\hat{V}_1(t) = \hat{V}_1(t+\mathscr{T})$.\cite{Salek2005}
Here, we use the symbol $\mathscr{T}$, referred to as the time period, instead of $T$ used in Ref.~\onlinecite{Salek2005} to avoid ambiguity with kinetic energy functional $T$, which is well-known in DFT or TDDFT. 
Using curly brackets for time-averaging,
\begin{equation}
    \{f(t)\}_\mathscr{T} = \frac{1}{\mathscr{T}} \int_{-\mathscr{T}/2}^{+\mathscr{T}/2} f(t) dt,
\end{equation}
the quasi-energy is defined as 
\begin{equation}
    Q = \{L(t)\}_\mathscr{T}
\end{equation} 
Techniques for time-independent problems, such as  the variational principle\cite{Christiansen1998} as well as the Hellman-Feynman theorem, \cite{Salek2005} can be applied to the quasi-energy, to obtain a solution or a response to a perturbation for the time-dependent case.
The quasi-energy formalism is also generally applicable to wavefunction methods and DFT \cite{Telnov1997,Salek2005}, at least for weak and off-resonant perturbation and when using finite basis sets approximations.\cite{Maitra2007,Kapoor2013}

In the context of TDDFT, the quasi-energy can be written a functional of the time-dependent density, $\rho(\bm{r}, t)$: \cite{Aiga1999,Salek2005}
\begin{equation}
    Q[\rho] = \left\{ \langle \tilde{0}| \hat{H} - i \frac{\partial}{\partial t} |\tilde{0} \rangle \right\}_\mathscr{T} = T[\rho] + V[\rho] + J[\rho] + Q_\text{ncl}[\rho] - S[\rho]
\end{equation}
where $T[\rho]$ and $V[\rho]$ are the kinetic energy and the interaction with the external potential, respectively.
The non-classical quasi-energy $Q_\text{ncl}[\rho]$ includes the effects of exchange and correlation, whereas $J[\rho]$ and $S[\rho]$ are formally defined by 
\begin{align}
    J[\rho] &= \frac{1}{2} \left\{ \iint \frac{\rho(\bm{r}, t)\rho(\bm{r}', t)}{|\bm{r} - \bm{r}'|} d\bm{r} d\bm{r}' \right\}_\mathscr{T} \\
    S[\rho] &= \left\{ \langle \tilde{0} | i\frac{\partial}{\partial t} | \tilde{0} \rangle \right\}_\mathscr{T}
\end{align}
A more extensive discussion of the quasi-energy formalism in DFT can be found in Ref.~\onlinecite{Salek2005,Maitra2007,Kapoor2013}.

\subsection{Decomposition of the quasi-energy into explicit and implicit terms}
\label{subsec:part_functional}
The functional in DFT can be written as the sum of a universal functional and the interaction of the electrons with external potential. 
The universal functional is a general term for all systems, which can be divided into two parts suggested by the ACKS2 model: the explicit functional ($E^{\text{exp}}$), which is a known functional of the density, and the implicit functional ($E^{\text{imp}}$). 
Similarly, in quasi-energy TDDFT the corresponding explicit functional $Q^{\text{exp}}$ and implicit functional $Q^{\text{imp}}$ are given as
\begin{equation}
    Q_v[\rho] = Q^{\text {exp}}[\rho] + Q^{\text{imp}}[\rho] + \left\{ \int \rho(\bm{r},t)v(\bm{r}, t)d\bm{r} \right \}_\mathscr{T}
\end{equation}
The implicit functional uses an auxiliary $N$-fermion wavefunction through a constrained-search formulation
\begin{equation}
    Q^{\text{imp}}[\rho] = \min_{\Psi \rightarrow{\rho}} W[\Psi]
\end{equation}
With the method of Lagrange multipliers, it can also be written out explicitly 
\begin{gather}
    Q^{\text{imp}}[\rho] = \sup_{u} \left( Q^{\text{o}}[u, N] - \left \{\int \rho(\bm{r},t)u(\bm{r},t)dr \right \}_\mathscr{T} \right) \\
    Q^{\text{o}}[u, N] = \min_{\Psi} \left( W[\Psi] + \left \{ \int \rho[\Psi](\bm{r},t) u(\bm{r},t)dr \right \}_\mathscr{T} \right)
\end{gather}
where $u(\bm{r},t)$ is a function that specifies a Lagrange multiplier at every point in terms of both space and time, and $\rho[\Psi](\bm{r},t)$ is the electron density of the auxiliary wavefunction. 
All terms that are dependent on the auxiliary wavefunction are collected in $Q^{\text{o}}[u, N]$. 
This quasi-energy can be interpreted as the ground-state energy of $W[\Psi]$ in a specified auxiliary potential, $u(\bm{r},t)$.

For a given external potential, $v(\bm{r},t)$, the $N$-electron ground state is solved by minimizing the following Lagrangian with respect to $\rho(\bm{r}, t)$ and maximizing it with respect to $u(\bm{r},t)$ and $\mu(t)$:
\begin{equation}
    L_v[\rho, u, \mu] = Q_v[\rho, u] - \left\{\mu(t) \left( \int  \rho(\bm{r},t)d\bm{r} - N \right) \right\}_\mathscr{T}
\end{equation}
where the quasi-energy is now a functional of the density and auxiliary potential
\begin{equation}
    Q_v[\rho, u] = Q^{\text{exp}}[\rho] + Q^{\text{o}}[u, N] + \left \{ \int \rho(\bm{r},t)[v(\bm{r},t) - u(\bm{r},t)] d\bm{r} \right \}_\mathscr{T}
\end{equation}
Finally, the stationary point is defined by the following sets of Euler-Lagrange equations, \cite{Aiga1999}
\begin{align}
    \frac{\delta Q^{\text{exp}}[\rho]}{\delta \rho(\bm{r},t)} + v(\bm{r},t) - u(\bm{r},t) &= \mu(t) 
    \label{eq:euler_dens}
    \\
    \frac{\delta Q^{\text{o}}[u, N]}{\delta u(\bm{r},t)} - \rho(\bm{r},t) &= 0
    \label{eq:euler_pots}
    \\
    \int \rho(\bm{r},t) d\bm{r} &= N
    \label{eq:euler_mu}
\end{align}

\subsection{Linear response}
\label{subsec:LR}
In linear-response theory, it is more convenient to first apply a Fourier-transform to all time-dependent functions,\cite{ullrich2011time} which yields formally similar Euler-Lagrange equations:
\begin{align}
    \frac{\delta Q^{\text{exp}}[\rho]}{\delta \rho(\bm{r},\omega)} + v(\bm{r},\omega) - u(\bm{r},\omega) &= \mu(\omega) 
    \label{eq:euler_dens_omega}
    \\
    \frac{\delta Q^{\text{o}}[u, N]}{\delta u(\bm{r},\omega)} - \rho(\bm{r},\omega) &= 0
    \label{eq:euler_pots_omega}
    \\
    \int \rho(\bm{r},\omega) d\bm{r} &= N
    \label{eq:euler_mu_omega}
\end{align}
In analogy with the ACKS2 method, a static DFT ground state is taken as the reference to which fluctuations are considered, including the external potential $v_0(\bm{r})$, density $\rho_0(\bm{r})$,  auxiliary potential $u_0(\bm{r})$, and equalized chemical potential $\mu_0$. 
The main difference with the original ACKS2 method, is that we now consider frequency-dependent fluctuations:
\begin{align*}
    v(\bm{r},\omega) &= v_0(\bm{r}) + \Delta v(\bm{r},\omega) &
    \rho(\bm{r}, \omega) &= \rho_0(\bm{r}) + \Delta \rho(\bm{r},\omega) \\
    u(\bm{r},\omega) &= u_0(\bm{r}) + \Delta u(\bm{r},\omega) &
    \mu(\omega) &= \mu_0 + \Delta \mu(\omega)
\end{align*}
where $\Delta v(\bm{r},\omega)$ is a frequency-dependent perturbation and $\Delta \rho(\bm{r},\omega)$, $\Delta u(\bm{r},\omega)$, and $\Delta \mu(\omega)$ are corresponding frequency-dependent responses to the perturbation.
After substitution in the Euler-Lagrange equations, one obtains
\begin{align}
    \left. \frac{\delta Q^\text{exp}[\rho]}{\delta \rho(\bm{r},\omega)} \right|_{\rho=\rho_0 + \Delta \rho}
    - \left. \frac{\delta Q^\text{exp}[\rho]}{\delta \rho(\bm{r},\omega)} \right|_{\rho=\rho_0} 
    + \Delta v(\bm{r},\omega) - \Delta u(\bm{r},\omega) &= \Delta \mu(\omega)
    \\
    \left. \frac{\delta Q^\text{o}[u, N]}{\delta u(\bm{r},\omega)} \right|_{u=u_0 + \Delta u} 
    - \left. \frac{\delta Q^\text{o}[u, N]}{\delta u(\bm{r},\omega)} \right|_{u=u_0} 
    - \Delta \rho(\bm{r},\omega) &= 0 
    \\
    \int \Delta \rho(\bm{r},\omega) dr &= 0
    \label{eq:fixed_particles_cond}
\end{align}
In the limit of a small perturbation, the first two equations can be linearized
\begin{align}
    \int \left.\frac{\delta^2 Q^{\text{exp}}[\rho]}{\delta \rho(\bm{r},\omega) \delta \rho(\bm{r}', \omega)} \right |_{\rho=\rho_0} \Delta \rho(\bm{r}', \omega) d\bm{r}' + \Delta v(\bm{r},\omega) - \Delta u(\bm{r},\omega) &\approx \Delta \mu(\omega)
    \label{eq:2nd_order_exp} \\
    \int \left. \frac{\delta^2 Q^\text{o}[u, N]}{\delta u(\bm{r},\omega)\delta u(\bm{r}',\omega)} \right |_{u=u_0} \Delta u(\bm{r}', \omega) d\bm{r}' - \Delta \rho(\bm{r},\omega) &\approx 0
    \label{eq:2nd_order_imp}
\end{align}
The second order functional derivatives in Eqs.~\eqref{eq:2nd_order_exp} and \eqref{eq:2nd_order_imp} are the hardness kernel of the explicit functional and response kernel of implicit functional, referred to as $\eta^{\text {exp}}(\bm{r}, \bm{r}', \omega)$ and $\chi^{\text{imp}}(\bm{r}, \bm{r'}, \omega)$, respectively.

\subsection{Expansion in a finite basis}
\label{subsec:basis_sets}
In a practical simulation, one must expand the density and the auxiliary potential fluctuations in a finite basis,
\begin{gather}
    \Delta \rho(\bm{r},\omega) = \sum_m^M C_m(\omega) f_m(\bm{r}) 
    \label{eq:density_basis} \\
    \Delta u(\bm{r},\omega) = \sum_n^N U_n(\omega) g_n(\bm{r})
    \label{eq:pot_basis} 
\end{gather}
where $f_m$ ($g_n$) denote density (potential) basis functions, while $C_m(\omega)$ ($U_n(\omega)$) denote the expansion coefficients of the induced density (potential) changes $\Delta \rho$ ($\Delta u$).
It is worth mentioning that both basis sets are time-independent, and only the corresponding coefficients are dynamic, suggesting that the basis sets used in the ACKS2 model can also be employed herein.

Substitution of the basis-set expansion in Eq.~\eqref{eq:2nd_order_exp}, followed by a multiplication with $f_k(\bm{r})$ and integration over $\bm{r}$, leads to:
\begin{equation}
    \sum_m^M \eta^\text{exp}_{km}(\omega)C_m(\omega) + V_k(\omega) - \sum_n^N O_{kn} U_n(\omega) = \Delta \mu(\omega) D_k \quad \forall k\in\{1\ldots M\}
\end{equation}
with 
\begin{align}
    \eta^\text{exp}_{km}(\omega) &= \iint \eta^{\text{exp}}(\bm{r},\bm{r}',\omega) f_k(\bm{r}) f_m(\bm{r}') d\bm{r} d\bm{r}' \\
    O_{kn} &= \int f_k(\bm{r})g_n(\bm{r}) d\bm{r} 
    \label{eq:basis_overlap}\\
    V_k(\omega) &= \int f_k(\bm{r}) \Delta v(\bm{r},\omega) d\bm{r} 
    \label{eq:ext_vec}\\
    D_k &= \int f_k(\bm{r}) d\bm{r}.
    \label{eq:charge_vec}
\end{align}
Similar manipulations of Eqs.~\eqref{eq:2nd_order_imp} and \eqref{eq:fixed_particles_cond}, lead to algebraic equations as follows:
\begin{align}
    \sum_n^N \chi^{\text{imp}}_{kn}(\omega)U_n(\omega) - \sum_m^M O_{km} C_m(\omega) &= 0 \quad \forall k\in\{1\ldots N\} \\
    \sum_m^M D_m C_m(\omega) &= 0,
\end{align}
with 
\begin{equation}
    \chi^{\text{imp}}_{kn}(\omega) = \iint \chi^{\text{imp}}(\bm{r},\bm{r}',\omega) g_k(\bm{r})g_n(\bm{r}') d\bm{r} d\bm{r}'.
    \label{eq:chi0}
\end{equation}
The linear system can be written in block matrix notation as follows:
\begin{align}
    \begin{bmatrix}
        -\bm{\eta}^\text{exp}_{M,M} & \bm{O}_{M,N}                  & \bm{D}_{M,1} \\
        \bm{O}^T_{N,M}   & -\bm{\chi}^\text{imp}_{N,N} & \bm{0}_{N,1}  \\
        \bm{D}^T_{1,M}   & \bm{0}_{1,N}                   & 0  
    \end{bmatrix} 
    \begin{bmatrix}
        \bm{C}_{M,1}(\omega) \\
        \bm{U}_{N,1}(\omega) \\
        \Delta \mu(\omega) \\
    \end{bmatrix}
    =
    \begin{bmatrix}
        \bm{V}_{M,1}(\omega) \\
        \bm{0}_{N,1} \\
        0 
    \end{bmatrix}
    \label{eq:acks2w_P}
\end{align}
where a bold letter denotes a block matrix and the subscript gives its dimension.
For example, $\bm{\eta}^\text{exp}_{M,M}$ represents the $M \times M$ hardness submatrix, $\bm{O}_{M,N}$ is the $M \times N$ overlap submatrix, and the $\bm{0}_{N,1}$ ($\bm{0}_{1,N}$) is a column (row) vector with all elements equal 0.
Solving this block matrix equation gives the expansion coefficients for the induced density, i.e., $\bm{C}_{M,1}(\omega)$, which is also known as the response vector.
The shape subscript will be omitted below for the sake of visual clarity.

According to linear response theory, the induced density can  be expressed through an \textit{interacting} response function, $\chi$:
\begin{align}
    \Delta \rho(\bm{r}, \omega) = \int d\bm{r'} \chi(\bm{r}, \bm{r}', \omega) V(\bm{r}', \omega) 
\end{align}
where $V(\bm{r}, \omega)$ is the frequency-dependent external perturbation. ACKS2$\omega$ can be used to approximate the interacting response, for which two strategies can be followed. 
The first and most straightforward option is to invert the block matrix in Eq.~$\eqref{eq:acks2w_P}$ directly.
The top-left block of the inverse transforms the input $\bm{V}(\omega)$ into the output $\bm{C}(\omega)$, and is therefore the ACKS2$\omega$ approximation of the interacting response.

One may also derive a useful closed expression for the interacting response matrix.
This derivation starts by solving the second row of Eq.~\eqref{eq:acks2w_P} to obtain $\bm{U}(\omega)=(\bm{\chi}^\text{imp})^* \bm{O}^T\bm{C}(\omega)$.
A pseudo-inverse of the implicit response matrix is needed, because it contains at least one zero eigenvalue, due to the charge distribution not being sensitive to a constant shift in Kohn-Sham potential.
Next, one uses this solution to eliminate $\bm{U}(\omega)$ from the first row of Eq.~\eqref{eq:acks2w_P}, yielding
\begin{align}
    \bm{A} \bm{C}(\omega)
    &=\bm{V}(\omega)-\bm{D}\Delta \mu(\omega)
\end{align}
where we introduced a shorthand $\bm{A}=-\bm{\eta}+\bm{O}(\bm{\chi}^\text{imp})^* \bm{O}^T$ to facilitate the derivation.
This equation can be solved, assuming $\bm{A}$ is non-singular:
\begin{align}
    \bm{C}(\omega)
    &=
    \bm{A}^{-1}
    \Bigl(\bm{V}(\omega)-\bm{D}\Delta \mu(\omega)\Bigr)
\end{align}
When $\bm{A}$ is singular, the density response is not well-defined.
While we cannot exclude this possibility of $\bm{A}$ being singular, we never encountered this issue in our numerical results.
The Lagrange multiplier is found by substituting this result in the third row of Eq.~\eqref{eq:acks2w_P}:
\begin{align}
    \Delta \mu(\omega)
    &=
    \frac{
    \bm{D}^T 
    \bm{A}^{-1}
    \bm{V}(\omega)}
    {\bm{D}^T \bm{A}^{-1} \bm{D}}
\end{align}
The final expression for the interacting response is:
\begin{align}
    \bm{C}(\omega)
    &=
    \bm{\chi}\bm{V}(\omega)
    =
    \left(
    \bm{A}^{-1} - \frac{\bm{A}^{-1} \bm{D} \bm{D}^T \bm{A}^{-1}}{\bm{D}^T \bm{A}^{-1} \bm{D}}
    \right)
    \bm{V}(\omega)
    \label{eq:acsk2_interacting_response}
\end{align}
The second term in $\bm{\chi}$ (due to normalization) is responsible for a proper zero eigenvalue in the interacting response, which also guarantees the constraint of particle conservation, i.e., $\bm{D}^T \bm{\chi} = \bm{0}$.

Note that $\bm{C}(\omega)=\bm{\chi}\bm{V}(\omega)$ does not contain an overlap matrix, unlike $\bm{O}^T\bm{C}(\omega)=\bm{\chi}^\text{imp}\bm{U}(\omega)$, because $\bm{V}(\omega)$ is defined with respect to the density basis $f_m$, whereas $\bm{U}(\omega)$ is expanded in terms of $g_n$. Hence, $\bm{\chi}$ is the interacting response expanded in the density basis. If desired, one may also express it in the potential basis by assuming that perturbations in the external potential have the following form:
\begin{align}
    \Delta v(\bm{r},\omega) &= \sum_n^N V'_n(\omega) g_n(\bm{r})
\end{align}
The new expansion coefficients, $V'_n(\omega)$, can always be transformed to those from Eq.~\eqref{eq:ext_vec}:
\begin{align}
    V_k(\omega) &= \sum_n^N O_{kn} V'_n(\omega)
\end{align}
With that, one can left-multiply Eq.~\eqref{eq:acsk2_interacting_response} with $\bm{O}^T$ and substitute $\bm{V}(\omega)$ with $\bm{O}\bm{V}'(\omega)$:
\begin{align}
    \bm{O}^T\bm{C}(\omega)
    =
    \bm{O}^T\bm{\chi}\bm{O}\bm{V}'(\omega)
    =
    \bm{O}^T
    \left(
    \bm{A}^{-1} - \frac{\bm{A}^{-1} \bm{D} \bm{D}^T \bm{A}^{-1}}{\bm{D}^T \bm{A}^{-1} \bm{D}}
    \right)
    \bm{O}
    \bm{V}'(\omega)
    & =
    \bm{\chi}'\bm{V}'(\omega)
\end{align}
where $\bm{\chi}'$ is finally identified as the interacting response matrix in the potential basis.

\subsection{ACKS2$\omega$ matrix elements}
\label{subsec:KS}

So far, the precise forms of the implicit and explicit functionals were not specified, and one may use, in principle, any definition whose sum equals (or approximates) the total quasi-energy.
In this section, we show that for Kohn-Sham DFT, a convenient choice can be made, leading to straightforward equations for all coefficients in the ACKS2$\omega$ equations.

The Kohn-Sham DFT expression of the quasi-energy is 
\begin{equation}
    Q^\text{o}[\rho] = T_s[\rho] + V[\rho] + J[\rho] + Q_{xc}[\rho] - S_s[\rho]
\end{equation}
where, $T_s[\rho$] and $S_s[\rho]$ are defined using the determinant of the non-interacting system ($| \tilde{0}_s \rangle$):
\begin{gather}
    T_s[\rho] = \left\{ \langle \tilde{0}_s| \hat{T} | \tilde{0}_s \rangle \right\}_\mathscr{T} \\
    S_s[\rho] = \left\{ \langle \tilde{0}_s | i\frac{\partial}{\partial t} | \tilde{0}_s \rangle \right\}_\mathscr{T}
\end{gather}
The quasi-energy xc functional is thereby defined as
\begin{equation}
    Q_{xc}[\rho] = Q_{ncl}[\rho] + (T[\rho] - T_s[\rho]) + (S[\rho] - S_s[\rho])
\end{equation}
Within the quasi-energy formalism, we employ the adiabatic approximation by replacing $Q_{xc}[\rho]$ with a time-independent counterpart $\{E_{xc}[\rho]\}_\mathscr{T}$.

With a semi-local xc functional, the explicit part of the energy functional in Kohn-Sham DFT takes the following form:
\begin{equation}
    Q^{\text{exp}}[\rho] = 
    \left\{
        \frac{1}{2} \iint \frac{\rho(\bm{r},t)\rho(\bm{r}',t)}{|\bm{r}-\bm{r}'|} d\bm{r} d\bm{r}' 
    + E_{xc}[\rho]
    \right\}_\mathscr{T} 
\end{equation}
where the first term is the Hartree quasi-energy, $J[\rho]$, while the second is the xc functional. 
The auxiliary wavefunction $\Phi$ is a single Slater determinant of Kohn-Sham orbitals, and $W[\Phi]$ is the Kohn-Sham kinetic energy minus functional $S_s[\rho]$
\begin{equation}
    W[\Phi] = 
    \left\{ 
        \sum_{i \in \text{Occ.}} \int \phi_i^{*}(\bm{r}, t) 
        \left( 
            -\frac{1}{2} \nabla ^2 
        \right) 
        \phi_i(\bm{r}, t) d\bm{r} 
    \right\}_\mathscr{T} 
    - S_s[\rho]
\end{equation}
where, $\phi_i$ is the $i$-th occupied (Occ.) molecular spatial orbital.

The expressions for the explicit hardness matrix and implicit response matrix with the Lehmann representation \cite{ullrich2011time,marques2012fundamentals} take the following forms in frequency space, respectively:
\begin{align}
    \eta^{\text{exp}}_{km} = \iint 
    \left(
        \frac{1}{|\bm{r}-\bm{r}'|} 
        + \frac{\delta^2 E_{xc}[\rho]} {\delta \rho(\bm{r}) \delta\rho(\bm{r}')} 
    \right) 
    f_k(\bm{r})f_m(\bm{r}') d\bm{r} d\bm{r}'  
    \label{eq:eta_exp_para} 
\end{align}
\begin{align}
    \chi^{\text{imp}}_{kn}(\omega) = 
    \lim_{\eta \rightarrow 0^{+}} \sum_{\substack{ i\in \text{Occ.} \\ a \in \text{Vir.} }} (n_i - n_a)   
    &\left[
        \frac{
            \int \phi_i^*(\bm{r}) g_k(\bm{r}) \phi_a(\bm{r}) d\bm{r} 
            \int \phi_i(\bm{r}') g_n(\bm{r}') \phi_a^*(\bm{r}') d\bm{r}'
        }{
            \omega - (\epsilon_a - \epsilon_i) + i\eta
        } 
    \right. 
    \nonumber \\
    &-\left. 
        \frac{
            \int \phi_i(\bm{r}) g_n(\bm{r}) \phi_a^*(\bm{r})d\bm{r}
            \int \phi_i^*(\bm{r}') g_k(\bm{r}') \phi_a(\bm{r}') d\bm{r}'
        }{
            \omega + (\epsilon_a - \epsilon_i) + i\eta
        } 
    \right]  
    \label{eq:chi_imp_para}
\end{align}
where, $n_i$ and $n_a$ are $1$ and $0$ for occupied and virtual (Vir.) molecular spatial orbitals, respectively.
After simple manipulations, the matrix elements of $\chi^{\text{imp}}$ have a simplified form
\begin{align}
    \chi^\text{imp}_{kn}(\omega) = 
    \lim_{\eta \rightarrow 0^{+}}
    \sum_{\substack{ i\in \text{Occ.} \\ a \in \text{Vir.} }}
        \frac{2\Omega_{ia}} {(\omega+i\eta)^2 - \Omega_{ia}^2} 
        \langle \phi_i | g_k | \phi_a \rangle 
        \langle \phi_a | g_n | \phi_i \rangle.
    \label{eq:chi_imp_para_dirac}
\end{align}
where $\Omega_{ia} = \epsilon_a - \epsilon_i$ and the Dirac notation is used.
For closed-shell molecules, a factor four is applied instead of two in Eq.~\eqref{eq:chi_imp_para_dirac} and the sum is restricted to spin-up orbitals only.

\subsection{Dynamic polarizability and related properties}
\label{subsec:prop_calc}
The dynamic linear-response properties, such as dipole polarizabilities, can be derived from the frequency-dependent response matrix.
Moreover, with a proper choice of density and potential basis functions, distributed polarizabilities \cite{Stone1985} are also available.

The potential basis set is constructed as 
\begin{align}
    g_{n(a,\ell, m)}(\bm{r}) = w_a(\bm{r}) R_{\ell}^{m}(\bm{r}-\bm{R}_a)
\end{align}
where $R_{\ell}^{m}(\bm{r}-\bm{R}_a)$ is a real solid harmonic with angular quantum number $\ell$ and magnetic quantum number $m$, using the position of nucleus $a$ as origin. 
The functions $w_a(\bm{r})$ ($0 \leq w_a(\bm{r}) \leq 1$) are so-called atom-in-molecule (AIM) weights functions, determining which proportion of the molecular ground-state density is assigned to atom $a$. \cite{Angyan1994} The rationale is that the integral of the product of such a potential basis function and a (response) density can be interpreted as distributed multipole moment.

The density basis set can be organized similarly, using atom-centered functions with unit monopole, dipole, etc. moments.
One may use Gaussian basis sets, 
\cite{Gutlein2019} or construct a basis that is bi-orthogonal to the potential basis set, \cite{Verstraelen2014} which is also the approach followed in this work.

Both basis sets are truncated by discarding all solid harmonics for which $\ell > \ell_\text{max}$, where $\ell_\text{max}$ is a user-specified threshold.
For instance, $\ell_\text{max}=0$ and $\ell_\text{max}=1$ corresponds to fluctuating charges only and charges+dipoles, respectively.
The total basis size is therefore $2 (\ell_\text{max}+1)^2$ times the number of atoms.

For the remainder of this section, it is convenient to change the index notation and introduce a compound index $t$ (or $u$) for the components of multipole moments, instead of separate indices $\ell$ and $m$. With these, we can define a distributed polarizability as:
\begin{align}
    \alpha^{ab}_{tu} = - \chi_{k(a,t)\,n(b,u)}
\end{align}
where $\chi_{k(a,t)\,n(b,u)}$ are elements of the response matrix defined in Eq.~\eqref{eq:acsk2_interacting_response}.
When considering the case $\ell_\text{max}=1$, distributed polarizability comprises several physically distinct blocks: the distributed charge-flow, $\alpha_{qq}^{ab}$, charge-dipole, $\alpha_{q \beta}^{ab}$, dipole-charge, $\alpha_{\alpha q}^{ab}$, and dipole-dipole, $\alpha_{\alpha \beta}^{ab}$, components.
The subscript $q$ is used to denote the monopole block and $\alpha$($\beta$) refers to dipole blocks, whereby
both $\alpha$ and $\beta$ could be $x$, $y$, or $z$.
The total molecular dipole polarizability can be recovered as follows:
\cite{Angyan1994}
\begin{equation}
    \label{eq:tot_alpha_3c}
    \alpha_{\alpha \beta} = \sum_{a,b}
    r_{\alpha}^a \alpha_{qq}^{ab} r_{\beta}^b 
    + r_{\alpha}^a \alpha_{q\beta}^{ab} 
    + \alpha_{\alpha q}^{ab} r_{\beta}^b 
    + \alpha_{\alpha \beta}^{ab}
\end{equation}
where $r_{\alpha}^a$ and $r_{\beta}^b$ are corresponding atomic coordinates.
The isotropic dipole polarizability of molecules is formally given as:
\begin{equation}
\label{eq:ave_alpha}
    \overline{\alpha} =\frac{1}{3} (\alpha_{xx} + \alpha_{yy} + \alpha_{zz}) 
\end{equation}
Eqs.~\eqref{eq:tot_alpha_3c}--\eqref{eq:ave_alpha} remain valid for frequency-dependent polarizabilities. When a real frequency is used, $\overline{\alpha}$ becomes complex (when $i\eta\neq 0$). The dipole strength function $S(\omega)$ can be computed using the imaginary part of $\overline{\alpha}$ with \cite{martin2020electronic}
\begin{align}
    S(\omega) = \frac{2 \omega}{\pi \hbar^2} \Im[\overline{\alpha}(\omega)].
\end{align}
The excitation energies are thus determined from the positions of the peaks, and the area under the peaks represents the oscillator strength of the corresponding transition.
In addition, the isotropic $C_6$ dispersion coefficient can be computed with imaginary frequencies ($\omega = iu$) using Casimir-Polder equation:\cite{Zaremba1976}
\begin{equation}
\label{eq:C6}
    C_6 = \frac{3}{\pi} \int_0^{\infty} \overline{\alpha}^A(iu) \overline{\alpha}^B(iu) d u,
\end{equation}
where, superscripts $A$ and $B$ denote different molecules.

\subsection{Two-site ACKS2$\omega$ model}
Before studying the absorption spectra of molecular systems, it is instructive to start with a simple model system to illustrate the workings of the ACKS2$\omega$ approach in analogy to the TDDFT method.
We consider a two-site system of an electric dipole along $x$-axis with coordinates of -1 and 1, and only s-type density and potential functions are considered.
The two-site ACKS2$\omega$ problem can be solved analytically after making the following assumptions:
\begin{enumerate}
    \item The hardness matrix $\eta$ is $\begin{bmatrix} a & b \\ b & a \end{bmatrix}$ where non-negative numbers $a$ and $b$ denotes atomic hardness and the classical Coulomb interaction between the density functions, respectively. 
    The xc contribution to the hardness is neglected entirely, i.e., the RPA is used.
    \item The overlap matrix is $\begin{bmatrix} c & 0 \\ 0 & c \end{bmatrix}$
    where $c$ is the overlap between the potential and density functions on the same site. Any overlap between functions on different sites is neglected for simplicity.
    Furthermore, we assume that the density basis functions are normalized, \textit{i.e.} $D_k = 1$ for all $k$ in Eq.~\eqref{eq:charge_vec}.
    \item The non-interacting response matrix is $\begin{bmatrix} f & -f \\ -f & f \end{bmatrix}$
    with 
    \begin{align}
        f=\frac{2 \Omega d}{(\omega+i\eta)^2 - \Omega^2} 
    \end{align}
    where $\Omega_{12}$ is the excitation energy between the ground state and the first excited state, determining the pole position of the non-interacting response function. 
    In addition, an arbitrary amplifier $d$ is added to create a more general non-interacting response matrix.
\end{enumerate}
One may solve the interacting response of this model analytically, by working out Eq.~\eqref{eq:acsk2_interacting_response}.
Doing so leads to an interacting response matrix of the form $\begin{bmatrix} g & -g \\ -g & g \end{bmatrix}$ with 
\begin{align}
    g = \frac{2 \Omega d}{[(\omega + i\eta)^2 -\Omega^2] c^2 -4\Omega(a-b)d}.
\end{align}
Therefore, the pole of the interacting response function is 
\begin{align}
    \omega = \sqrt{\Omega^2 + \frac{4 \Omega (a-b)d}{c^2}} > \Omega
\end{align}
where the condition $a>b$ is generally satisfied with the RPA approximation, because the Hartree kernel is positive definite.
This inequality implies that the pole of the interacting response is shifted to higher frequencies compared to the non-interacting response, when using RPA.

\begin{figure}[htbp]
\centering
    \includegraphics[scale=1.0]{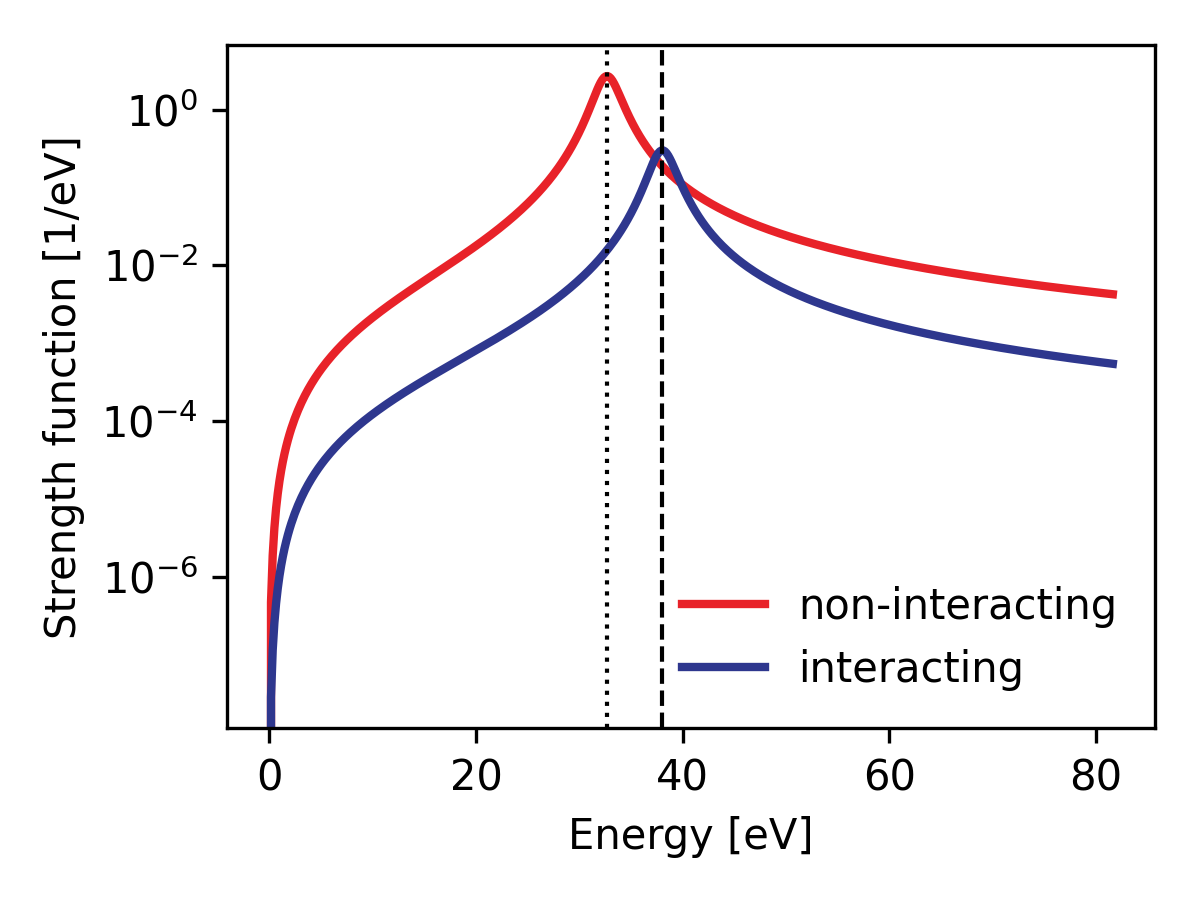}
    \caption{
    Spectra of a finite dipole along $x$-axis with a single excited state.
    The pole of the non-interacting (interacting) system is indicated by a dotted (dashed) line.
    Numerical parameters are given in the main text.}
    \label{fig:spectrum_two_sites}
\end{figure}

Figure \ref{fig:spectrum_two_sites} gives the absorption spectra computed by both non-interacting and interacting response functions, where the parameters $a$, $b$, $c$, $d$, $\Omega$, $\eta$ are assigned 1.0, 0.2, 3.0, 1.2, 1.2, and 0.05 (all in atomic units), respectively.
The position of the pole is depicted with dotted (dashed) lines for the interacting (non-interacting) response function, overlapping with the peak position in the corresponding absorption spectrum.
It can be seen that the pole of the interacting response function shifts to a higher frequency compared to the non-interacting case.

\section{Computational Details}
\label{sec:cp_details}
This section presents the details of the numerical validation of ACKS2$\omega$ using the TS42 database and its TS27 subset.
As discussed in Section \ref{subsec:basis_sets}, both potential and density basis functions are frequency-independent. 
Here, the basis sets are constructed in the same way as in one of our previous works on ACKS2 \cite{Verstraelen2014}.
The MBIS method \cite{Verstraelen2016} was used to define the AIM weight functions, $w_a(\bm{r})$.
Using the same procedure as in Ref.~\onlinecite{Verstraelen2014}, the density basis set consists of non-interacting responses to each potential basis function, when used as a perturbation. This set is augmented with one Fukui function, to have at least one basis function that is not norm-preserving. Later, a linear transformation is applied to the density basis functions, such that they become bi-orthogonal to the potential basis, enabling one to label them as s-type, p-type, etc.
This definition of the density basis set is relatively expensive, and is geared towards an accurate description of linear response functions.
The goal of this work is to test the validity of the ACKS2$\omega$ theory, justifying the use of carefully constructed basis functions.
The bi-orthogonality of the density and potential basis functions also simplifies the elements of ACKS2$\omega$ working matrix in Eq.~\eqref{eq:acks2w_P}.
More specifically, the overlap matrix $\bm{O}$ becomes an identity matrix, and all non-zero elements of the vector $\bm{D}$ are equal to 1, corresponding to integrals from s-type density functions.
Moreover, when the adiabatic approximation is used, the hardness matrix becomes frequency-independent and takes the same form as in the ACKS2 paper.\cite{Verstraelen2014}
More specifically, integrals including the hardness kernel of the explicit functional were evaluated numerically using a pruned Becke-Levedev integration grid.\cite{Becke1988}
Integrals over the Hartree kernel were implemented with a Becke-Poisson solver.\cite{BeckeDiskson1988}
Although the analytic expressions for LDA and PBE kernels are already provided in LibXC, integrals involving the exchange-correlation kernel were evaluated with the finite difference method used to be consistent with the previous work.\cite{Verstraelen2014} 
The only frequency-dependent parameters of ACKS2$\omega$ are found in the non-interacting response tensor, which can be evaluated using Eq.~\eqref{eq:chi_imp_para} on the same numerical grids.

The workflow of the ACKS2$\omega$ parameterization in this work is analogous to our earlier assessment of the ACKS2 theory: \cite{Verstraelen2014}
\begin{enumerate}
\item[(1)] The molecular structures listed in the T42 data set are optimized using DFT at B3LYP/aug-cc-pVDZ level. (The optimized geometries are available in the Supplementary Material.)
\item[(2)] A new DFT calculation with LDA or PBE functional is employed to obtain ground-state information used in ACKS2$\omega$ parameter evaluations, e.g., unperturbed electron density and Kohn-Sham orbitals.
All DFT calculations were performed in the quantum chemistry program GAUSSIAN16 \cite{g16}.
\item[(3)] The Kohn-Sham density and orbitals are then imported into ACKS2$\omega$ implemented in the Horton library,\cite{horton} where the MIBS  partitioning scheme \cite{Verstraelen2016} is applied to generate the weight function $w_a(\bm{r})$ for each atom. 
In ACKS2$\omega$, all parameters are evaluated on numerical Becke-Lebedev grids \cite{Becke1988,lebedev1999quadrature} implemented in Horton,\cite{horton} providing six user options from cheap to expensive: \texttt{coarse}, \texttt{medium}, \texttt{fine}, \texttt{veryfine}, \texttt{ultrafine}, and \texttt{insane}. 
\texttt{medium} grids are employed for all ACKS2$\omega$ calculations, unless noted elsewhere.
\item[(4)] Frequency-dependent distributed polarizabilities are evaluated up to charges ($\ell_{max}=0$) and charges+dipoles ($\ell_{max} = 1$), respectively.
An LDA or PBE hardness kernel is used, consistently with the functional in step (2).
\item[(5)] Consequently, dynamic dipole polarizabilities can be constructed using the distributed polarizabilities from Eq.~\eqref{eq:tot_alpha_3c}.
\end{enumerate}

To demonstrate the accuracy of ACKS2$\omega$ on dynamic linear-response properties, we first studied absorption spectra of 42 molecules from the TS42 data set.
We only estimate the absorption spectra at the PBE/aug-cc-pVDZ level to save computational resources.
300 real frequencies ranging from 0.2 to 0.5 a.u. (i.e., 5.4$\sim$13.6 eV) are selected for absorption spectra, and the parameter $\eta$ in Eq.~\eqref{eq:chi_imp_para} is set to 0.001.
Then, the isotropic $C_6$ coefficients (in a.u.) are investigated for 903 molecular pairs constructed from the 42 molecules.
The effect of xc functional is investigated on the $C_6$ coefficient by comparing two types of functional, i.e., LDA and PBE.
The Slater exchange functional \cite{PhysRev.136.B864, PhysRev.140.A1133, slater1974quantum} and the VWN5 correlation functional \cite{Vosko1980} are used for the LDA functional in this work.
Moreover, four different Dunning basis sets, aug-cc-pVDZ, aug-cc-pVTZ, d-aug-cc-pVDZ, and d-aug-cc-pVTZ, are utilized to study the impact of the basis size.
The integral in Eq.~\eqref{eq:C6} is evaluated using the Gaussian-Legendre quadrature with 12 imaginary frequencies.
Furthermore, for comparison, we also evaluated the linear-response properties using linear-response TDDFT (LrTDDFT) with the LDA (PBE) functional, referred to as LrTDLDA (LrTDPBE).
All LrTDDFT calculations are carried out in the quantum chemistry program Dalton. \cite{Aidas2014,Jorgensen1988,Olsen1989}

In the following context, we use the notation ACKS2$\omega$@\textit{X} to specify the ACKS2$\omega$ model with parameters estimated using functional \textit{X}, for instance ACKS2$\omega$@LDA indicates that LDA functional is used. 
The ACKS2$\omega$@\textit{X} evaluated with both $\ell_{max}=0$ and $\ell_{max}=1$ are called s-type and sp-type ACKS2$\omega$@\textit{X}, respectively.
As an exception to this nomenclature, ACKS2$\omega$@LDAx means that the ground-state DFT  uses full LDA xc functional, just as ACKS2$\omega$@LDA, but that the correlation contribution to the hardness kernel is neglected. (Only Hartree and exchange are included.)

In this work, all parameters in ACKS2$\omega$ are computed as expectation values of an electronic wavefunction, which is the most time-consuming step.
For applications of ACKS2$\omega$ to larger systems, this step should be replaced by a simpler empirical model for the ACKS2$\omega$, analogous to general PFFs, such that the calculation of all matrix elements becomes fast, with a quadratic complexity with the number of atoms in the system.
In the long run, we hope that screening approximations and advanced Poisson solvers may further reduce the scaling of setting up and solving the relevant part of the equations to $O(N_\text{atom}\log N_\text{atom})$.
    
In comparison, the bottleneck of TDDFT is the transformation of electron repulsion integrals (ERI) from atomic orbitals (AO) to molecular orbitals (MO), which cannot be applied in large-scale systems due to its significantly expensive $O(N_\text{occ} N_\text{AO}^4)$ complexity from the conventional transformation or $O(N_\text{occ}^2 N_\text{vir}^2 N_\text{aux})$ from density fitting procedures where $N_\text{AO}$ is the size of the basis set and $N_\text{occ}$, $N_\text{vir}$ and $N_\text{aux}\approx 3N_\text{AO}$ are the number of occupied, virtual orbitals and the size of auxiliary basis sets, respectively.\cite{Limaye1994,Tang1970,Bender1972,Sherrill2010}
    
Finally, for the matrix inversion to calculate the interacting response matrix, a similar linear-algebra technology can be employed for both ACKS2$\omega$ and TDDFT with a semi-local exchange-correlation functional.
However, the dimension of ACKS2$\omega$ working matrix with $l_\text{max} = 1$ is $4N_\text{atom}$, in practice much less than the TDDFT counterpart, i.e., $N_\text{nocc} \times N_\text{nvir}$.
    
Because our current implementation is merely a prototype, we only illustrate timings of the different steps in the computational workflow, for the case of benzene with an Aug-cc-pVTZ basis set, on a 4-core AMD EPYC 7552 processor (AMD Zen2 microarchitecture). The following numbers may change with future software and hardware improvements. The calculation of the hardness matrix and 13 (twelve frequency-dependent and one static) Kohn-Sham response matrices take 101~s in this case. Once the matrix elements are available, the ACKS2$\omega$@LDA response calculation is trivial and has a much lower runtime (1.97~ms) compared to the corresponding LrTDLDA calculation (290~s), excluding the walltime of ground-state DFT and ERI transformation.

\section{Results and Discussion}
\label{sec:results}

\subsection{Absorption spectra of molecules in the TS42 set}
\label{subsubsec:42mol_supectrum}
\begin{figure}[htbp]
\centering
    \includegraphics[width=12cm]{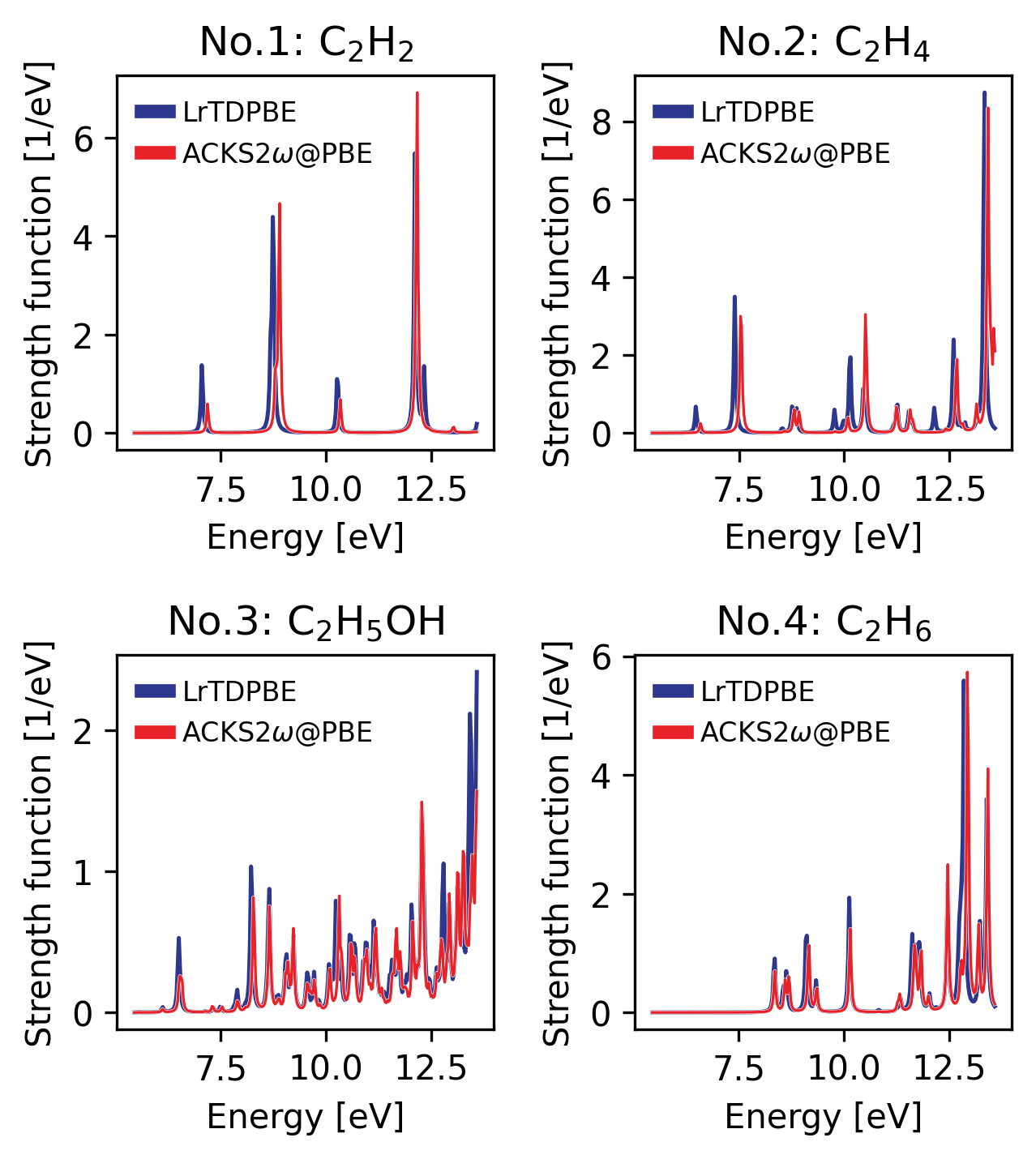}
    \caption{
    Comparison of optical absorption spectra, computed with sp-type ACKS2$\omega$@PBE and LrTDPBE methods for \ce{C2H2}, \ce{C6H4}, \ce{C2H5OH} and \ce{C2H6} molecules, using the aug-cc-pVDZ basis set in the DFT calculations.
    }
    \label{fig:spectrum_examples}
\end{figure}

Figure. \ref{fig:spectrum_examples} shows the results computed by sp-type ACKS2$\omega$ for four example molecules, i.e., \ce{C2H2}, \ce{C2H4}, \ce{C2H5OH} and \ce{C2H6}, while the other spectra can be found in Figures S1-S5 of the Supplementary Material.
From the figures, we can see that the position of peaks obtained by the ACKS2$\omega$ model overlaps almost perfectly with the LrTDPBE data, suggesting that the ACKS2$\omega$ model is a faithful approximation of its TDDFT reference.
At higher excitation energies, beyond 13.6 eV, we observe a similar correspondence of the spectra, which is not included in the figures for the sake of visual clarity.
It should be noted that a near-quantitative reproduction of the reference can be achieved with the sp-type basis functions only, i.e., charges+dipoles functions, which agrees well with the ACKS2 model. \cite{Verstraelen2014}
The small deviation between ACKS2$\omega$ and TDDFT is potentially due to the atom-condensed basis functions and the numerical integration errors in the calculation of the parameters.

\subsection{Comparison of ACKS2$\omega$ and TDDFT $C_6$ coefficients}
\label{subsec:c6_vs_tddft}
\begin{figure}[htbp]
\centering
    \includegraphics[scale=1.0]{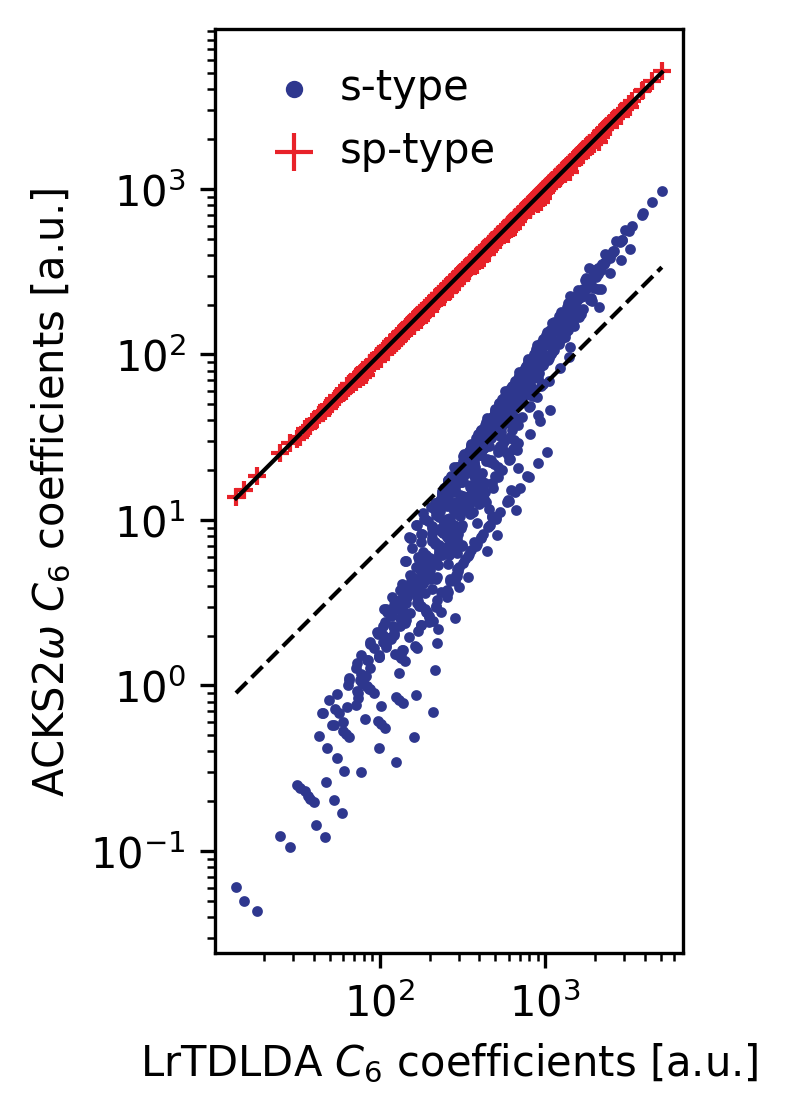} 
    \caption{
    Parity plot of the $C_6$ coefficients (in a.u.) of all 903 molecular dimers computed by ACKS2$\omega$@LDA compared to the LrTDLDA reference values.
    Two datasets are included: the lowest set (blue) is generated by s-type ACKS2$\omega$@LDA (see text), while the higher set (red) is generated by sp-type ACKS2$\omega$@LDA.
    The first bisector is plotted as a solid black line, and a dashed line parallel to the bisector (factor 0.07 lower) is representative of the MAPE of the s-type ACKS2$\omega$ model.
    }
\label{fig:c6_acks2w_vs_tddft}
\end{figure}

This section presents a comparison of $C_6$ coefficients obtained with ACKS2$\omega$ and TDDFT, to validate the ACKS2$\omega$ formalism.
Figure \ref{fig:c6_acks2w_vs_tddft} shows a parity plot of $C_6$ dispersion coefficients for 903 molecular pairs, based on the TS42 data set, computed with s-type and sp-type ACKS2$\omega$ models. The ACKS2$\omega$ parameters are derived from LDA/aug-cc-pVDZ calculations and LrTDLDA/aug-cc-pVDZ $C_6$ coefficients are used as a reference.
All s-type ACKS2$\omega$ parametrizations significantly underestimate $C_6$ coefficients, averaging around 93\% with respect to the reference values.
The reason for this large error is that the dipole polarizability is significantly underestimated due to less complete basis sets in s-type ACKS2,\cite{Verstraelen2014}.
This effect is amplified in the isotropic $C_6$ coefficient, because it scales quadratically with the dipole polarizability. 
In contrast, very small errors can already be obtained with the sp-type ACKS2$\omega$ model, \textit{i.e.}\ just considering fluctuating atomic charges and dipoles.

Table \ref{tbl:acks2w_performance_basis} presents the mean percentage errors (MPE) and mean absolute percentage errors (MAPE) over all molecule pairs, between LrTDLDA reference values and sp-type ACKS2$\omega$@LDA models, for different orbital basis sets.
For all tested basis sets, the sp-type ACKS2$\omega$@LDA has a negative MPE, indicating that ACKS2$\omega$@LDA slightly underestimates $C_6$ with respect to the TDDFT reference.
Moreover, in all cases, sp-type ACKS2$\omega$@LDA has a MAPE below 3\%, which numerically confirms the validity of ACKS2$\omega$ and its ability to construct faithful approximations of its TDDFT reference.
The absolute value of the MPE is clearly smaller than the MAPE, suggesting that the underestimation of $C_6$ by ACKS2$\omega$ is not systematic.
Overall, the MPE and MAPE are not sensitive to the orbital basis, safe for a slight increase in error for larger orbital basis sets.
This is to be expected, since the TDDFT calculations with larger basis sets have richer response functions, which are harder to reproduce with a simple sp-type ACKS2$\omega$ model.

\begin{table}[htpb]
\caption{
    Comparison of the performance of sp-type ACKS2$\omega$@LDA with different basis sets used in ground-state DFT calculations.
    MPE and MAPE stand for the mean percentage error and mean absolute percentage error between $C_6$ coefficients of LrTDLDA references and the corresponding ACKS2$\omega$ results for the 903 intermolecular pairs from the TS42 data set.
    }
    \centering
    \begin{tabular}{ccc}
        \hline\hline
        Basis sets & MPE (\%) & MAPE (\%)  \\
        \hline
        aug-cc-pVDZ & -1.62 & 2.35 \\
        aug-cc-pVTZ & -1.94 & 2.60  \\
        d-aug-cc-pVDZ & -1.41 & 2.31  \\
        d-aug-cc-pVTZ & -1.81 & 2.55  \\ 
        \hline\hline
    \end{tabular}
    \label{tbl:acks2w_performance_basis}
\end{table}

The non-zero MPEs and MAPEs in table \ref{tbl:acks2w_performance_basis} are due to the finite sp basis used in ACKS2$\omega$.
This is unavoidably a smaller basis compared to the TDDFT reference calculations, in which the Casida equations are solved in the product space of the occupied and virtual Kohn-Sham orbitals.\cite{CASIDA1995}
One may reduce the error of ACKS2$\omega$ by systematically including higher-order atomic multipoles, shown in table~\ref{tbl:acks2w_performance}.
It should be noted that \texttt{veryfine} numerical grids were needed to obtain numerically stable results for higher values of $\ell_\text{max}$.
This also results in a small deviation from $\ell_\text{max}=1$ compared to table~\ref{tbl:acks2w_performance_basis}, for which \texttt{medium} grids were used.
Increasing $\ell_\text{max}$ beyond 4 is expected to lower the MAPE even further, provided that even larger integration grids are used, and in the limit of a complete basis set, we expect the MAPE to vanish.
The near zero ($-4.88\times 10^{-4}\%$) MPE obtained with monopoles and dipoles is likely coincidental, as the MPE is higher for  $\ell_\text{max} > 1$.

\begin{table}[htpb]
\caption{
    Comparison of the performance of ACKS2$\omega$@LDA with the aug-cc-pVDZ basis set derived using the different bi-orthogonal basis sets (defined by $\ell_\text{max}$, see text).
    MAPE (MPE) stand for the mean absolute percentage error (mean percentage error) between $C_6$ coefficients of LrTDLDA references and the corresponding ACKS2$\omega$ results for the 903 intermolecular pairs from the TS42 data set.
    }
    \centering
    \begin{tabular}{ccc}
        \hline\hline
        $\ell_\text{max}$ & MAPE (\%) & MPE (\%)  \\
        \hline
        0 & 93.35 & $2.18 \times 10^{-1}$\\
        1 & 2.39  & $-4.88 \times 10^{-4}$\\
        2 & 2.12  & $1.32 \times 10^{-3}$\\
        3 & 1.79  & $1.27 \times 10^{-3}$\\ 
        4 & 1.59  & $3.63 \times 10^{-4}$\\ 
        \hline\hline
    \end{tabular}
    \label{tbl:acks2w_performance}
\end{table}

\subsection{Comparison to experimental $C_6$ coefficients}
\label{subsec:c6_vs_exp}

\begin{figure}[htbp]
\centering
    \includegraphics[scale=1.0]{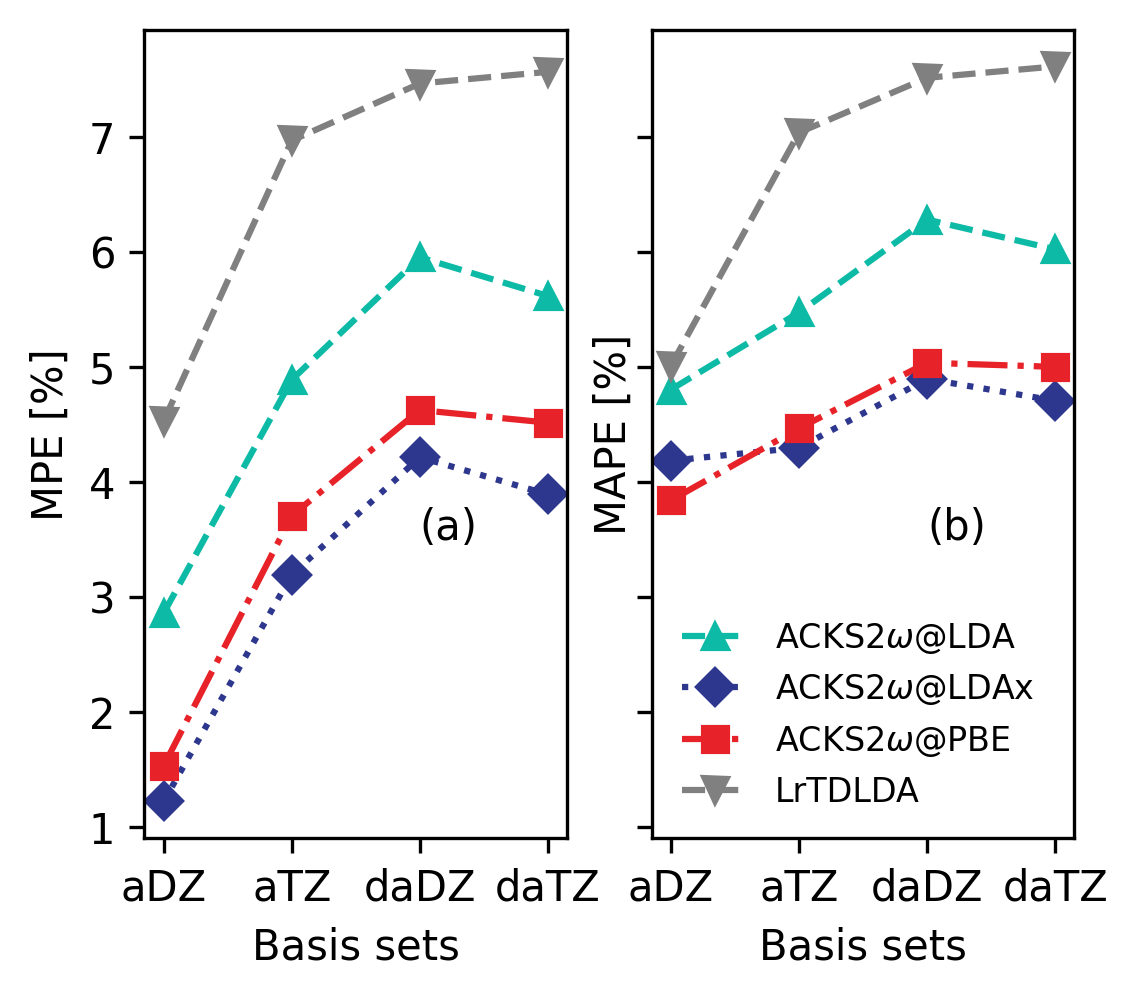}
    \caption{
    The impact of the size of the basis set on the error of different methods with respect to experimental reference data: (a) MPE and (b) MAPE, where `aDZ', `aTZ', `daDZ', and `daTZ' denote aug-cc-pVDZ, aug-cc-pVTZ, d-aug-cc-pVDZ, and d-aug-cc-pVTZ basis sets, respectively.
    }
\label{fig:comparison_diff_bs}
\end{figure}

Figure \ref{fig:comparison_diff_bs} shows the MPE and MAPE of $C_6$ coefficients computed by ACKS2$\omega$ and LrTDLDA from experimentally-derived reference taken from Ref.~\onlinecite{Tkatchenko2009}.
We only consider results computed by sp-type ACKS2$\omega$ due to the large errors made by s-type ACKS2$\omega$ discussed in Section \ref{subsec:c6_vs_tddft}.
ACKS2$\omega$ parameters were derived using different xc functionals to study the impact of the functional.
Some interesting observations can be made:
\begin{enumerate}
\item[(1)] The positive MPE implies that all computational models in this work slightly overestimate $C_6$ values to some extent.
\item[(2)] For LrTDLDA, the MPE almost equals MAPE, except when using the aug-cc-pVDZ basis. This suggests LrTDLDA systematically overestimates $C_6$ values when extensive basis sets are used.
The MAPE (and MPE) increases with the size of the basis, and it is found to be converged at around 7\%.
\item[(3)] The MAPE (MPE) of ACKS2$\omega$ has a similar trend as LrTDLDA, but two significant differences should be pointed out.
First, the ACKS2$\omega$ model has a larger MAPE than MPE, indicating that the overestimation is not systematic in the ACKS2 model, i.e., for some molecules, ACKS2$\omega$ underestimates $C_6$ coefficients.
Second, ACKS2$\omega$@LDA has a lower MAPE (MPE) than the corresponding LrTDLDA for all cases, implying that the improvement compared to LrTDDFT is somewhat systematic.
One possible expansion is that TDDFT always gives an overestimated $C_6$ with LDA functional, and the incomplete basis in ACKS2$\omega$ compensates for the overestimation of TDDFT to some extent, leading to a lower MAPE.
This type of error compensation is obviously fortuitous, and should therefore not be relied upon blindly.
\item[(4)] The comparison between ACKS2$\omega$@LDA and ACKS2$\omega$@PBE shows a systematic improvement when using PBE instead of LDA.
One possible reason could be that the contribution of correlation functional can be described more precisely with PBE than LDA.
This inspired us to also test the ACKS2$\omega$@LDAx model, which entirely ignores the correlation contribution in the hardness kernel, compared to ACKS2$\omega$@LDA.
As expected, it shows a significant deviation from the ACKS2$\omega$@LDA model.
Interestingly, ACKS2$\omega$@LDAx performs slightly better than ACKS2$\omega$@PBE, except for the aug-cc-pVDZ basis, where the latter has a lower MAPE (3.84\%).
Besides the potentially beneficial error compensation, these results also show that ACKS2$\omega$ is quite sensitive to the choice of functional, because the TDDFT reference also exhibits this sensitivity.
\end{enumerate}

\subsection{Comparison of ACKS2$\omega$ and range-separated TDDFT $C_6$ coefficients}

Toulouse \textit{et al.} calculated $C_6$ coefficients using range-separated hybrid TDDFT at the LDA/d-aug-cc-pVTZ level (TDRSHLDA) for 27 homodimers from the TS27 data set. \cite{Toulouse2013}
The spirit of the TDRSHLDA method is to improve the quality of exchange kernel in TDDFT, by incorporating Hartree-Fock exchange at longer ranges, thereby suppressing systematic errors of pure exchange functionals.
Table \ref{tbl:ts27_c6_comparison} reports the results of $C_6$ coefficients computed by the ACKS2$\omega$ model for the TS27 data set, where effects of different Hartree-exchange-correlation kernels are also presented.
The literature results for TDRSHLDA \cite{Toulouse2013} and experimental data \cite{Tkatchenko2009} are also presented.
We use a similar notation as Ref.~\onlinecite{Toulouse2013}, e.g., the bare LDA indicates no hardness kernel is applied, while RPA@LDA, TDLDAx, and TDLDA mean that the response kernel is evaluated considering only Hartree, Hartree + LDA exchange, and whole LDA Hartree-exchange-correlation kernel, respectively. 
Corresponding results with PBE functional are also reported.

As we can see from Table \ref{tbl:ts27_c6_comparison}, bare LDA and PBE largely overestimate the $C_6$ coefficients by as much as 138.96\% and 134.92\%, respectively, as one could expect. \cite{Toulouse2013}
However, RPA@LDA and RPA@PBE, by making use of the Hartree kernel only, underestimate the coefficients by 12.64\% and 13.59\%, respectively.
LDAx, LDA, and PBE give $C_6$ coefficients with overall comparable accuracy, LDAx having a slightly smaller MAPE of 5.29\% in comparison to the MAPE of LDA and PBE, 6.18\% and 5.38\%, respectively.
The MAPE of ACKS2$\omega$@LDAx is lower than LrTDLDA (6.3\%),\cite{Toulouse2013} but greater than TDRSHLDA (5.17\%),\cite{Toulouse2013}, demonstrating that when correlation is entirely ignored in response calculations, ACKS2$\omega$@LDAx gives a better performance, close to the TDRSHLDA method.
Note that ACKS2$\omega$@LDAx has a positive MPE but TDRSHLDA negative, indicating that $C_6$ coefficients obtained by ACKS2$\omega$@LDAx and TDRSHLDA are in general overestimated and underestimated, respectively.
As discussed above, LrTDLDA overestimates $C_6$ coefficients and ACKS2$\omega$@LDA slightly compensates for the overestimation due to the less complete bio-orthogonal basis sets.
However, the compensation is too small to offer a significant improvement compared to  TDRSHLDA.
Interestingly, the results are improved and even better than PBE by removing the electron correlation contribution in response calculations, which again demonstrates that the effects of xc functional on $C_6$ coefficients with ACKS2$\omega$ are of significant importance.

\begin{table}[htbp]
\begin{threeparttable}
    \caption{
        Isotropic $C_6$ coefficients (in a.u.) for the TS27 data set obtained by the ACKS2$\omega$ model with bare LDA, RPA@LDA, LDAx, LDA, PBE, RPA@PBE, and PBE kernels, with d-aug-cc-pVTZ basis sets.
        Computational values of TDRSHLDA \cite{Toulouse2013} are presented, as well as the experimental reference data compiled from Ref.~\onlinecite{Tkatchenko2009}.
        The geometry was optimized at the B3LYP/aug-cc-pVDZ level. 
        Mean percentage error (MPE) and mean absolute percentage errors (MAPE) over all molecules with respect to reference values are given.
    }
\footnotesize
\renewcommand*{\arraystretch}{0.7}
\begin{tabular}{lrrrrrrrrr}
\hline\hline
            & \multicolumn{7}{c}{ACKS2$\omega$} & TDRSHLDA\tnote{a} & Exp.\tnote{b} \\
            \cline{2-8}
            &  Bare LDA &  RPA@LDA &   LDAx &  LDA &  Bare PBE &  RPA@PBE &   PBE & & \\
\hline
\ce{H2}      &     20.11 &    11.01 &   14.03 &   14.45 &     18.88 &    10.41 &   13.65 &     12.70 &      12.10 \\
\ce{HF}      &     32.94 &    19.14 &   22.20 &   22.58 &     33.17 &    19.29 &   22.65 &     19.20 &      19.00 \\
\ce{H2O}     &     84.09 &    43.17 &   51.07 &   52.09 &     83.95 &    43.17 &   51.92 &     43.40 &      45.30 \\
\ce{N2}      &    179.25 &    63.11 &   72.21 &   73.16 &    178.47 &    63.16 &   73.22 &     72.70 &      73.30 \\
\ce{CO}      &    182.42 &    65.87 &   76.60 &   77.73 &    180.90 &    65.88 &   77.91 &     77.10 &      81.40 \\
\ce{NH3}     &    166.17 &    79.17 &   94.82 &   96.88 &    163.92 &    78.34 &   95.78 &     80.80 &      89.00 \\
\ce{CH4}     &    241.97 &   111.97 &  134.34 &  137.05 &    234.09 &   109.11 &  134.05 &    121.20 &     129.70 \\
\ce{HCl}     &    295.43 &   107.80 &  127.46 &  129.76 &    290.76 &   106.49 &  128.33 &    122.90 &     130.40 \\
\ce{CO2}     &    392.45 &   137.70 &  155.13 &  156.97 &    390.03 &   138.01 &  157.54 &    150.90 &     158.70 \\
\ce{H2CO}    &    314.90 &   128.76 &  150.73 &  153.25 &    311.55 &   128.13 &  152.87 &    138.40 &     165.20 \\
\ce{N2O}     &    581.17 &   156.22 &  175.71 &  177.72 &    578.60 &   156.75 &  178.37 &    179.80 &     184.90 \\
\ce{C2H2}    &    497.75 &   180.45 &  212.02 &  216.01 &    494.75 &   180.27 &  216.33 &    198.90 &     204.10 \\
\ce{HBr}     &    524.77 &   182.43 &  216.97 &  221.24 &    520.74 &   181.76 &  221.56 &    205.50 &     216.60 \\
\ce{H2S}     &    543.41 &   182.32 &  219.14 &  223.72 &    529.72 &   178.96 &  220.21 &    209.00 &     216.80 \\
\ce{CH3OH}   &    426.69 &   196.89 &  231.62 &  235.60 &    418.51 &   194.35 &  233.70 &    205.00 &     222.00 \\
\ce{SO2}     &    961.52 &   263.01 &  298.19 &  301.89 &    960.93 &   264.22 &  304.05 &    295.30 &     294.00 \\
\ce{C2H4}    &    651.83 &   259.34 &  307.48 &  313.29 &    640.10 &   256.29 &  311.24 &    287.30 &     300.20 \\
\ce{CH3NH2}  &    601.55 &   271.57 &  320.30 &  326.20 &    588.09 &   267.27 &  322.48 &    279.60 &     303.80 \\
\ce{SiH4}    &    777.20 &   302.19 &  369.60 &  378.07 &    725.68 &   288.91 &  364.81 &    329.60 &     343.90 \\
\ce{C2H6}    &    750.74 &   332.35 &  392.56 &  399.72 &    727.40 &   324.66 &  392.87 &    352.70 &     381.90 \\
\ce{Cl2}     &   1095.68 &   315.22 &  366.12 &  371.26 &   1077.08 &   312.46 &  369.93 &    385.40 &     389.20 \\
\ce{CH3CHO}  &    923.40 &   375.91 &  434.97 &  441.78 &    904.54 &   371.51 &  438.12 &    386.60 &     401.70 \\
\ce{COS}     &   1415.93 &   358.36 &  409.47 &  415.10 &   1399.21 &   357.42 &  415.24 &    425.40 &     402.20 \\
\ce{CH3OCH3} &   1091.09 &   485.48 &  566.65 &  576.14 &   1061.04 &   476.27 &  568.58 &    496.10 &     534.10 \\
\ce{C3H6}    &   1461.23 &   583.62 &  683.18 &  694.79 &   1423.72 &   573.31 &  686.75 &    622.00 &     662.10 \\
\ce{CS2}     &   3759.03 &   733.53 &  841.33 &  852.85 &   3710.36 &   730.39 &  852.50 &    923.00 &     871.10 \\
\ce{CCl4}    &   5918.94 &  1723.00 & 1949.94 & 1971.80 &   5787.63 &  1703.58 & 1962.53 &   1924.90 &    2024.10 \\
\hline
    MPE      &    138.96 &   -12.64 &    2.42 &    4.20 &    134.92 &   -13.59 &    3.32 &     -3.83              \\
    MAPE     &    138.96 &    12.69 &    5.29 &    6.18 &    134.92 &    13.70 &    5.38 &      5.17              \\ 
\hline\hline
\end{tabular}
\begin{tablenotes}
\item[a] From Ref.~\onlinecite{Toulouse2013}, computed by range-separated hybrid TDDFT at the LDA/d-aug-cc-pVTZ level.
\item[b] From Ref.~\onlinecite{Tkatchenko2009}, obtained from experimental dipole oscillator strength distribution data.
\end{tablenotes}
\label{tbl:ts27_c6_comparison}
\end{threeparttable}
\end{table}

\section{Conclusions and Outlook}
\label{sec:summary}
A frequency-dependent ACKS2 model, referred to as ACKS2$\omega$, is proposed using the quasi-energy formalism.
This allows us to approach ACKS2$\omega$ directly using the variational principle, in the same way as the time-independent ACKS2 derivation.
For the xc contribution to the hardness, an adiabatic approximation is applied, and the hardness is thereby frequency-independent, just as in the ACKS2 model.
For the frequency-dependent non-interacting response matrix, a similar Lehmann representation is employed as in the static case.
Given the hardness and non-interacting response matrix, the interacting response matrix can be reproduced in analogy to the procedure used in the ACKS2 model.

The ACKS2$\omega$ equations are validated with several numerical assessments. Absorption spectra obtained from the strength function are evaluated for all 42 molecular monomers from the TS42 data set.
The results agree well with the LrTDDFT reference data, indicating that the ACKS2$\omega$ model is reasonably accurate for linear-response property calculations of finite systems.
Furthermore, we test $C_6$ coefficients for the TS42 data set to validate the new model, including 903 organic and inorganic molecule pairs.
These tests confirm that ACKS2$\omega$ can reproduce TDDFT $C_6$ coefficients with only a small systematic error, when at least fluctuating atomic charges and dipoles are used.

A comparison of ACKS2$\omega$, using the LDA and PBE kernels, to experimental data and results obtained with range-separated calculations, provides some additional insights.
The deviation of the ACKS2$\omega$ $C_6$ coefficients from the experiment is quite sensitive to the choice of the functional, with PBE and LDAx (correlation omitted from the hardness kernel), being favorable choices. The best ACKS2$\omega$ model, ACKS2$\omega$@PBE with the aug-cc-pVDZ basis, gives $C_6$ coefficients of molecules with an MAPE of 3.84\%, a slight improvement over the TS method (6.3\%).
This improvement is significant because it potentially provides a better way to describe non-local dispersion energy between molecular dimers, which is a topic of ongoing research.
In addition, the ACKS2$\omega$@LDAx model has only a slightly inferior accuracy of $C_6$ coefficients than TDRSHLDA response calculations, with MAPEs of 5.29\% and 5.17\%, respectively, for 27 homodimers from the TS27 database.

The strength of the ACKS2$\omega$ model is that, once parameterized, calculations involving response kernels have a low computational cost, comparable in complexity to conventional polarizable force fields.
In this work, the parametrization is relatively expensive to obtain accurate predictions, thereby showing that ACKS2$\omega$ is capable of reproducing its LrTDDFT reference.
In future work, simpler force-field like parametrizations could be used instead, to overcome the computational bottleneck of the current parametrization.
We anticipate that this approach will be beneficial, e.g. for calculations of long-range correlation energies or optical properties of extended systems, for which DFT would be infeasible.

\section*{Supplementary Material}

Additional display items showing spectra for the remaining molecules of the TS42 set, not included in Fig.~\ref{fig:spectrum_examples} are included in a Supplementary PDF file.
Optimized geometries for the TS42 molecules are stored in XYZ format in a Supplementary ZIP file.

\begin{acknowledgments}
Y.C. and T.V. acknowledge the Foundation of Scientific Research - Flanders (FWO, file number G0A9717N) and the Research Board of Ghent University (BOF) for their financial support.
The resources and services used in this work were provided by the VSC (Flemish Supercomputer Center), funded by the Research Foundation - Flanders (FWO) and the Flemish Government.
We thank Dr. Jelle Vekeman for helpful comments on the manuscript.
\end{acknowledgments}

\section*{Data Availability Statement}

The data that supports the findings of this study are available within the article and its supplementary material.

\clearpage
\bibliography{reference}

\begin{thebibliography}{111}%
\makeatletter
\providecommand \@ifxundefined [1]{%
 \@ifx{#1\undefined}
}%
\providecommand \@ifnum [1]{%
 \ifnum #1\expandafter \@firstoftwo
 \else \expandafter \@secondoftwo
 \fi
}%
\providecommand \@ifx [1]{%
 \ifx #1\expandafter \@firstoftwo
 \else \expandafter \@secondoftwo
 \fi
}%
\providecommand \natexlab [1]{#1}%
\providecommand \enquote  [1]{``#1''}%
\providecommand \bibnamefont  [1]{#1}%
\providecommand \bibfnamefont [1]{#1}%
\providecommand \citenamefont [1]{#1}%
\providecommand \href@noop [0]{\@secondoftwo}%
\providecommand \href [0]{\begingroup \@sanitize@url \@href}%
\providecommand \@href[1]{\@@startlink{#1}\@@href}%
\providecommand \@@href[1]{\endgroup#1\@@endlink}%
\providecommand \@sanitize@url [0]{\catcode `\\12\catcode `\$12\catcode
  `\&12\catcode `\#12\catcode `\^12\catcode `\_12\catcode `\%12\relax}%
\providecommand \@@startlink[1]{}%
\providecommand \@@endlink[0]{}%
\providecommand \url  [0]{\begingroup\@sanitize@url \@url }%
\providecommand \@url [1]{\endgroup\@href {#1}{\urlprefix }}%
\providecommand \urlprefix  [0]{URL }%
\providecommand \Eprint [0]{\href }%
\providecommand \doibase [0]{https://doi.org/}%
\providecommand \selectlanguage [0]{\@gobble}%
\providecommand \bibinfo  [0]{\@secondoftwo}%
\providecommand \bibfield  [0]{\@secondoftwo}%
\providecommand \translation [1]{[#1]}%
\providecommand \BibitemOpen [0]{}%
\providecommand \bibitemStop [0]{}%
\providecommand \bibitemNoStop [0]{.\EOS\space}%
\providecommand \EOS [0]{\spacefactor3000\relax}%
\providecommand \BibitemShut  [1]{\csname bibitem#1\endcsname}%
\let\auto@bib@innerbib\@empty
\bibitem [{\citenamefont {Warshel}\ and\ \citenamefont
  {Levitt}(1976)}]{Warshel1976}%
  \BibitemOpen
  \bibfield  {author} {\bibinfo {author} {\bibfnamefont {A.}~\bibnamefont
  {Warshel}}\ and\ \bibinfo {author} {\bibfnamefont {M.}~\bibnamefont
  {Levitt}},\ }\href {https://doi.org/10.1016/0022-2836(76)90311-9} {\bibfield
  {journal} {\bibinfo  {journal} {J. Mol. Biol.}\ }\textbf {\bibinfo {volume}
  {103}},\ \bibinfo {pages} {227} (\bibinfo {year} {1976})}\BibitemShut
  {NoStop}%
\bibitem [{\citenamefont {Kaminski}\ \emph {et~al.}(2002)\citenamefont
  {Kaminski}, \citenamefont {Stern}, \citenamefont {Berne}, \citenamefont
  {Friesner}, \citenamefont {Cao}, \citenamefont {Murphy}, \citenamefont
  {Zhou},\ and\ \citenamefont {Halgren}}]{Kaminski2002}%
  \BibitemOpen
  \bibfield  {author} {\bibinfo {author} {\bibfnamefont {G.~A.}\ \bibnamefont
  {Kaminski}}, \bibinfo {author} {\bibfnamefont {H.~A.}\ \bibnamefont {Stern}},
  \bibinfo {author} {\bibfnamefont {B.~J.}\ \bibnamefont {Berne}}, \bibinfo
  {author} {\bibfnamefont {R.~A.}\ \bibnamefont {Friesner}}, \bibinfo {author}
  {\bibfnamefont {Y.~X.}\ \bibnamefont {Cao}}, \bibinfo {author} {\bibfnamefont
  {R.~B.}\ \bibnamefont {Murphy}}, \bibinfo {author} {\bibfnamefont
  {R.}~\bibnamefont {Zhou}},\ and\ \bibinfo {author} {\bibfnamefont {T.~A.}\
  \bibnamefont {Halgren}},\ }\href {https://doi.org/10.1002/jcc.10125}
  {\bibfield  {journal} {\bibinfo  {journal} {J. Comput. Chem.}\ }\textbf
  {\bibinfo {volume} {23}},\ \bibinfo {pages} {1515} (\bibinfo {year}
  {2002})}\BibitemShut {NoStop}%
\bibitem [{\citenamefont {Ren}\ and\ \citenamefont {Ponder}(2002)}]{Ren2002}%
  \BibitemOpen
  \bibfield  {author} {\bibinfo {author} {\bibfnamefont {P.}~\bibnamefont
  {Ren}}\ and\ \bibinfo {author} {\bibfnamefont {J.~W.}\ \bibnamefont
  {Ponder}},\ }\href {https://doi.org/10.1002/jcc.10127} {\bibfield  {journal}
  {\bibinfo  {journal} {J. Comput. Chem.}\ }\textbf {\bibinfo {volume} {23}},\
  \bibinfo {pages} {1497} (\bibinfo {year} {2002})}\BibitemShut {NoStop}%
\bibitem [{\citenamefont {Jorgensen}\ \emph {et~al.}(2007)\citenamefont
  {Jorgensen}, \citenamefont {Jensen},\ and\ \citenamefont
  {Alexandrova}}]{Jorgensen2007}%
  \BibitemOpen
  \bibfield  {author} {\bibinfo {author} {\bibfnamefont {W.~L.}\ \bibnamefont
  {Jorgensen}}, \bibinfo {author} {\bibfnamefont {K.~P.}\ \bibnamefont
  {Jensen}},\ and\ \bibinfo {author} {\bibfnamefont {A.~N.}\ \bibnamefont
  {Alexandrova}},\ }\href {https://doi.org/10.1021/ct7001754} {\bibfield
  {journal} {\bibinfo  {journal} {J. Chem. Theory Comput.}\ }\textbf {\bibinfo
  {volume} {3}},\ \bibinfo {pages} {1987} (\bibinfo {year} {2007})}\BibitemShut
  {NoStop}%
\bibitem [{\citenamefont {Shi}\ \emph {et~al.}(2013)\citenamefont {Shi},
  \citenamefont {Xia}, \citenamefont {Zhang}, \citenamefont {Best},
  \citenamefont {Wu}, \citenamefont {Ponder},\ and\ \citenamefont
  {Ren}}]{Shi2013}%
  \BibitemOpen
  \bibfield  {author} {\bibinfo {author} {\bibfnamefont {Y.}~\bibnamefont
  {Shi}}, \bibinfo {author} {\bibfnamefont {Z.}~\bibnamefont {Xia}}, \bibinfo
  {author} {\bibfnamefont {J.}~\bibnamefont {Zhang}}, \bibinfo {author}
  {\bibfnamefont {R.}~\bibnamefont {Best}}, \bibinfo {author} {\bibfnamefont
  {C.}~\bibnamefont {Wu}}, \bibinfo {author} {\bibfnamefont {J.~W.}\
  \bibnamefont {Ponder}},\ and\ \bibinfo {author} {\bibfnamefont
  {P.}~\bibnamefont {Ren}},\ }\href {https://doi.org/10.1021/ct4003702}
  {\bibfield  {journal} {\bibinfo  {journal} {J. Chem. Theory Comput.}\
  }\textbf {\bibinfo {volume} {9}},\ \bibinfo {pages} {4046} (\bibinfo {year}
  {2013})}\BibitemShut {NoStop}%
\bibitem [{\citenamefont {Lamoureux}\ \emph {et~al.}(2003)\citenamefont
  {Lamoureux}, \citenamefont {MacKerell},\ and\ \citenamefont
  {Roux}}]{Lamoureux2003}%
  \BibitemOpen
  \bibfield  {author} {\bibinfo {author} {\bibfnamefont {G.}~\bibnamefont
  {Lamoureux}}, \bibinfo {author} {\bibfnamefont {A.~D.}\ \bibnamefont
  {MacKerell}},\ and\ \bibinfo {author} {\bibfnamefont {B.}~\bibnamefont
  {Roux}},\ }\href {https://doi.org/10.1063/1.1598191} {\bibfield  {journal}
  {\bibinfo  {journal} {J. Chem. Phys.}\ }\textbf {\bibinfo {volume} {119}},\
  \bibinfo {pages} {5185} (\bibinfo {year} {2003})}\BibitemShut {NoStop}%
\bibitem [{\citenamefont {Yu}\ \emph {et~al.}(2003)\citenamefont {Yu},
  \citenamefont {Hansson},\ and\ \citenamefont {{Van Gunsteren}}}]{Yu2003}%
  \BibitemOpen
  \bibfield  {author} {\bibinfo {author} {\bibfnamefont {H.}~\bibnamefont
  {Yu}}, \bibinfo {author} {\bibfnamefont {T.}~\bibnamefont {Hansson}},\ and\
  \bibinfo {author} {\bibfnamefont {W.~F.}\ \bibnamefont {{Van Gunsteren}}},\
  }\href {https://doi.org/10.1063/1.1523915} {\bibfield  {journal} {\bibinfo
  {journal} {J. Chem. Phys.}\ }\textbf {\bibinfo {volume} {118}},\ \bibinfo
  {pages} {221} (\bibinfo {year} {2003})}\BibitemShut {NoStop}%
\bibitem [{\citenamefont {Lemkul}\ \emph {et~al.}(2016)\citenamefont {Lemkul},
  \citenamefont {Huang}, \citenamefont {Roux},\ and\ \citenamefont
  {Mackerell}}]{Lemkul2016}%
  \BibitemOpen
  \bibfield  {author} {\bibinfo {author} {\bibfnamefont {J.~A.}\ \bibnamefont
  {Lemkul}}, \bibinfo {author} {\bibfnamefont {J.}~\bibnamefont {Huang}},
  \bibinfo {author} {\bibfnamefont {B.}~\bibnamefont {Roux}},\ and\ \bibinfo
  {author} {\bibfnamefont {A.~D.}\ \bibnamefont {Mackerell}},\ }\href
  {https://doi.org/10.1021/acs.chemrev.5b00505} {\bibfield  {journal} {\bibinfo
   {journal} {Chem. Rev.}\ }\textbf {\bibinfo {volume} {116}},\ \bibinfo
  {pages} {4983} (\bibinfo {year} {2016})}\BibitemShut {NoStop}%
\bibitem [{\citenamefont {Mortier}\ \emph {et~al.}(1985)\citenamefont
  {Mortier}, \citenamefont {Genechten},\ and\ \citenamefont
  {Gasteiger}}]{Mortier1985}%
  \BibitemOpen
  \bibfield  {author} {\bibinfo {author} {\bibfnamefont {W.~J.}\ \bibnamefont
  {Mortier}}, \bibinfo {author} {\bibfnamefont {K.~V.}\ \bibnamefont
  {Genechten}},\ and\ \bibinfo {author} {\bibfnamefont {J.}~\bibnamefont
  {Gasteiger}},\ }\href {https://doi.org/10.1021/ja00290a017} {\bibfield
  {journal} {\bibinfo  {journal} {J. Am. Chem. Soc.}\ }\textbf {\bibinfo
  {volume} {107}},\ \bibinfo {pages} {829} (\bibinfo {year}
  {1985})}\BibitemShut {NoStop}%
\bibitem [{\citenamefont {Mortier}\ \emph {et~al.}(1986)\citenamefont
  {Mortier}, \citenamefont {Ghosh},\ and\ \citenamefont
  {Shankar}}]{Mortier1986}%
  \BibitemOpen
  \bibfield  {author} {\bibinfo {author} {\bibfnamefont {W.~J.}\ \bibnamefont
  {Mortier}}, \bibinfo {author} {\bibfnamefont {S.~K.}\ \bibnamefont {Ghosh}},\
  and\ \bibinfo {author} {\bibfnamefont {S.}~\bibnamefont {Shankar}},\ }\href
  {https://doi.org/10.1021/ja00275a013} {\bibfield  {journal} {\bibinfo
  {journal} {J. Am. Chem. Soc.}\ }\textbf {\bibinfo {volume} {108}},\ \bibinfo
  {pages} {4315} (\bibinfo {year} {1986})}\BibitemShut {NoStop}%
\bibitem [{\citenamefont {Bai}\ \emph {et~al.}(2017)\citenamefont {Bai},
  \citenamefont {Kale},\ and\ \citenamefont {Herzfeld}}]{C7SC01181D}%
  \BibitemOpen
  \bibfield  {author} {\bibinfo {author} {\bibfnamefont {C.}~\bibnamefont
  {Bai}}, \bibinfo {author} {\bibfnamefont {S.}~\bibnamefont {Kale}},\ and\
  \bibinfo {author} {\bibfnamefont {J.}~\bibnamefont {Herzfeld}},\ }\href
  {https://doi.org/10.1039/C7SC01181D} {\bibfield  {journal} {\bibinfo
  {journal} {Chem. Sci.}\ }\textbf {\bibinfo {volume} {8}},\ \bibinfo {pages}
  {4203} (\bibinfo {year} {2017})}\BibitemShut {NoStop}%
\bibitem [{\citenamefont {Cools-Ceuppens}\ \emph {et~al.}(2022)\citenamefont
  {Cools-Ceuppens}, \citenamefont {Dambre},\ and\ \citenamefont
  {Verstraelen}}]{Cools-Ceuppens2022}%
  \BibitemOpen
  \bibfield  {author} {\bibinfo {author} {\bibfnamefont {M.}~\bibnamefont
  {Cools-Ceuppens}}, \bibinfo {author} {\bibfnamefont {J.}~\bibnamefont
  {Dambre}},\ and\ \bibinfo {author} {\bibfnamefont {T.}~\bibnamefont
  {Verstraelen}},\ }\href {https://doi.org/10.1021/acs.jctc.1c00978} {\bibfield
   {journal} {\bibinfo  {journal} {J. Chem. Theory Comput.}\ }\textbf {\bibinfo
  {volume} {18}},\ \bibinfo {pages} {1672} (\bibinfo {year} {2022})},\ \Eprint
  {https://arxiv.org/abs/2109.13111} {arXiv:2109.13111} \BibitemShut {NoStop}%
\bibitem [{\citenamefont {{Van Maaren}}\ and\ \citenamefont {{Van
  Spoel}}(2001)}]{VanMaaren2001}%
  \BibitemOpen
  \bibfield  {author} {\bibinfo {author} {\bibfnamefont {P.~J.}\ \bibnamefont
  {{Van Maaren}}}\ and\ \bibinfo {author} {\bibfnamefont {D.~D.}\ \bibnamefont
  {{Van Spoel}}},\ }\href {https://doi.org/10.1021/jp003843l} {\bibfield
  {journal} {\bibinfo  {journal} {J. Phys. Chem. B}\ }\textbf {\bibinfo
  {volume} {105}},\ \bibinfo {pages} {2618} (\bibinfo {year}
  {2001})}\BibitemShut {NoStop}%
\bibitem [{\citenamefont {Kunz}\ and\ \citenamefont {{Van
  Gunsteren}}(2009)}]{Kunz2009}%
  \BibitemOpen
  \bibfield  {author} {\bibinfo {author} {\bibfnamefont {A.~P.~E.}\
  \bibnamefont {Kunz}}\ and\ \bibinfo {author} {\bibfnamefont {W.~F.}\
  \bibnamefont {{Van Gunsteren}}},\ }\href {https://doi.org/10.1021/jp903164s}
  {\bibfield  {journal} {\bibinfo  {journal} {J. Phys. Chem. A}\ }\textbf
  {\bibinfo {volume} {113}},\ \bibinfo {pages} {11570} (\bibinfo {year}
  {2009})}\BibitemShut {NoStop}%
\bibitem [{\citenamefont {Rick}\ \emph {et~al.}(1994)\citenamefont {Rick},
  \citenamefont {Stuart},\ and\ \citenamefont {Berne}}]{Rick1994}%
  \BibitemOpen
  \bibfield  {author} {\bibinfo {author} {\bibfnamefont {S.~W.}\ \bibnamefont
  {Rick}}, \bibinfo {author} {\bibfnamefont {S.~J.}\ \bibnamefont {Stuart}},\
  and\ \bibinfo {author} {\bibfnamefont {B.~J.}\ \bibnamefont {Berne}},\ }\href
  {https://doi.org/10.1063/1.468398} {\bibfield  {journal} {\bibinfo  {journal}
  {J. Chem. Phys.}\ }\textbf {\bibinfo {volume} {101}},\ \bibinfo {pages}
  {6141} (\bibinfo {year} {1994})},\ \Eprint {https://arxiv.org/abs/9406002}
  {arXiv:9406002 [chem-ph]} \BibitemShut {NoStop}%
\bibitem [{\citenamefont {Rick}\ and\ \citenamefont {Berne}(1996)}]{Rick1996}%
  \BibitemOpen
  \bibfield  {author} {\bibinfo {author} {\bibfnamefont {S.~W.}\ \bibnamefont
  {Rick}}\ and\ \bibinfo {author} {\bibfnamefont {B.~J.}\ \bibnamefont
  {Berne}},\ }\href {https://doi.org/10.1021/ja952535b} {\bibfield  {journal}
  {\bibinfo  {journal} {J. Am. Chem. Soc.}\ }\textbf {\bibinfo {volume}
  {118}},\ \bibinfo {pages} {672} (\bibinfo {year} {1996})}\BibitemShut
  {NoStop}%
\bibitem [{\citenamefont {Stern}\ \emph {et~al.}(2001)\citenamefont {Stern},
  \citenamefont {Rittner}, \citenamefont {Berne},\ and\ \citenamefont
  {Friesner}}]{Stern2001}%
  \BibitemOpen
  \bibfield  {author} {\bibinfo {author} {\bibfnamefont {H.~A.}\ \bibnamefont
  {Stern}}, \bibinfo {author} {\bibfnamefont {F.}~\bibnamefont {Rittner}},
  \bibinfo {author} {\bibfnamefont {B.~J.}\ \bibnamefont {Berne}},\ and\
  \bibinfo {author} {\bibfnamefont {R.~A.}\ \bibnamefont {Friesner}},\ }\href
  {https://doi.org/10.1063/1.1376165} {\bibfield  {journal} {\bibinfo
  {journal} {J. Chem. Phys.}\ }\textbf {\bibinfo {volume} {115}},\ \bibinfo
  {pages} {2237} (\bibinfo {year} {2001})}\BibitemShut {NoStop}%
\bibitem [{\citenamefont {Patel}\ and\ \citenamefont
  {Brooks}(2004)}]{Patel2004}%
  \BibitemOpen
  \bibfield  {author} {\bibinfo {author} {\bibfnamefont {S.}~\bibnamefont
  {Patel}}\ and\ \bibinfo {author} {\bibfnamefont {C.~L.}\ \bibnamefont
  {Brooks}},\ }\href {https://doi.org/10.1002/jcc.10355} {\bibfield  {journal}
  {\bibinfo  {journal} {J. Comput. Chem.}\ }\textbf {\bibinfo {volume} {25}},\
  \bibinfo {pages} {1} (\bibinfo {year} {2004})}\BibitemShut {NoStop}%
\bibitem [{\citenamefont {Patel}\ \emph {et~al.}(2004)\citenamefont {Patel},
  \citenamefont {Mackerell},\ and\ \citenamefont {Brooks}}]{Patel2004a}%
  \BibitemOpen
  \bibfield  {author} {\bibinfo {author} {\bibfnamefont {S.}~\bibnamefont
  {Patel}}, \bibinfo {author} {\bibfnamefont {A.~D.}\ \bibnamefont
  {Mackerell}},\ and\ \bibinfo {author} {\bibfnamefont {C.~L.}\ \bibnamefont
  {Brooks}},\ }\href {https://doi.org/10.1002/jcc.20077} {\bibfield  {journal}
  {\bibinfo  {journal} {J. Comput. Chem.}\ }\textbf {\bibinfo {volume} {25}},\
  \bibinfo {pages} {1504} (\bibinfo {year} {2004})}\BibitemShut {NoStop}%
\bibitem [{\citenamefont {Rapp{\'{e}}}\ and\ \citenamefont
  {Goddard}(1991)}]{Rappe1991}%
  \BibitemOpen
  \bibfield  {author} {\bibinfo {author} {\bibfnamefont {A.~K.}\ \bibnamefont
  {Rapp{\'{e}}}}\ and\ \bibinfo {author} {\bibfnamefont {W.~A.}\ \bibnamefont
  {Goddard}},\ }\href {https://doi.org/10.1021/j100161a070} {\bibfield
  {journal} {\bibinfo  {journal} {J. Phys. Chem.}\ }\textbf {\bibinfo {volume}
  {95}},\ \bibinfo {pages} {3358} (\bibinfo {year} {1991})}\BibitemShut
  {NoStop}%
\bibitem [{\citenamefont {Zhong}\ and\ \citenamefont
  {Patel}(2010)}]{Zhong2010}%
  \BibitemOpen
  \bibfield  {author} {\bibinfo {author} {\bibfnamefont {Y.}~\bibnamefont
  {Zhong}}\ and\ \bibinfo {author} {\bibfnamefont {S.}~\bibnamefont {Patel}},\
  }\href {https://doi.org/10.1021/jp101597r} {\bibfield  {journal} {\bibinfo
  {journal} {J. Phys. Chem. B}\ }\textbf {\bibinfo {volume} {114}},\ \bibinfo
  {pages} {11076} (\bibinfo {year} {2010})}\BibitemShut {NoStop}%
\bibitem [{\citenamefont {York}\ and\ \citenamefont {Yang}(1996)}]{York1996}%
  \BibitemOpen
  \bibfield  {author} {\bibinfo {author} {\bibfnamefont {D.~M.}\ \bibnamefont
  {York}}\ and\ \bibinfo {author} {\bibfnamefont {W.}~\bibnamefont {Yang}},\
  }\href {https://doi.org/10.1063/1.470886} {\bibfield  {journal} {\bibinfo
  {journal} {J. Chem. Phys.}\ }\textbf {\bibinfo {volume} {104}},\ \bibinfo
  {pages} {159} (\bibinfo {year} {1996})}\BibitemShut {NoStop}%
\bibitem [{\citenamefont {Devereux}\ \emph {et~al.}(2014)\citenamefont
  {Devereux}, \citenamefont {Raghunathan}, \citenamefont {Fedorov},\ and\
  \citenamefont {Meuwly}}]{Devereux2014}%
  \BibitemOpen
  \bibfield  {author} {\bibinfo {author} {\bibfnamefont {M.}~\bibnamefont
  {Devereux}}, \bibinfo {author} {\bibfnamefont {S.}~\bibnamefont
  {Raghunathan}}, \bibinfo {author} {\bibfnamefont {D.~G.}\ \bibnamefont
  {Fedorov}},\ and\ \bibinfo {author} {\bibfnamefont {M.}~\bibnamefont
  {Meuwly}},\ }\href {https://doi.org/10.1021/ct500511t} {\bibfield  {journal}
  {\bibinfo  {journal} {J. Chem. Theory Comput.}\ }\textbf {\bibinfo {volume}
  {10}},\ \bibinfo {pages} {4229} (\bibinfo {year} {2014})}\BibitemShut
  {NoStop}%
\bibitem [{\citenamefont {Unke}\ \emph {et~al.}(2017)\citenamefont {Unke},
  \citenamefont {Devereux},\ and\ \citenamefont {Meuwly}}]{Unke2017}%
  \BibitemOpen
  \bibfield  {author} {\bibinfo {author} {\bibfnamefont {O.~T.}\ \bibnamefont
  {Unke}}, \bibinfo {author} {\bibfnamefont {M.}~\bibnamefont {Devereux}},\
  and\ \bibinfo {author} {\bibfnamefont {M.}~\bibnamefont {Meuwly}},\
  }\bibfield  {journal} {\bibinfo  {journal} {J. Chem. Phys.}\ }\textbf
  {\bibinfo {volume} {147}},\ \href {https://doi.org/10.1063/1.4993424}
  {10.1063/1.4993424} (\bibinfo {year} {2017})\BibitemShut {NoStop}%
\bibitem [{\citenamefont {Devereux}\ \emph {et~al.}(2020)\citenamefont
  {Devereux}, \citenamefont {Pezzella}, \citenamefont {Raghunathan},\ and\
  \citenamefont {Meuwly}}]{Devereux2020}%
  \BibitemOpen
  \bibfield  {author} {\bibinfo {author} {\bibfnamefont {M.}~\bibnamefont
  {Devereux}}, \bibinfo {author} {\bibfnamefont {M.}~\bibnamefont {Pezzella}},
  \bibinfo {author} {\bibfnamefont {S.}~\bibnamefont {Raghunathan}},\ and\
  \bibinfo {author} {\bibfnamefont {M.}~\bibnamefont {Meuwly}},\ }\href
  {https://doi.org/10.1021/acs.jctc.0c00883} {\bibfield  {journal} {\bibinfo
  {journal} {J. Chem. Theory Comput.}\ }\textbf {\bibinfo {volume} {16}},\
  \bibinfo {pages} {7267} (\bibinfo {year} {2020})}\BibitemShut {NoStop}%
\bibitem [{\citenamefont {Sanderson}(1951)}]{Sanderson1951}%
  \BibitemOpen
  \bibfield  {author} {\bibinfo {author} {\bibfnamefont {R.~T.}\ \bibnamefont
  {Sanderson}},\ }\href {https://doi.org/10.1126/science.114.2973.670}
  {\bibfield  {journal} {\bibinfo  {journal} {Science (80-. ).}\ }\textbf
  {\bibinfo {volume} {114}},\ \bibinfo {pages} {670} (\bibinfo {year}
  {1951})}\BibitemShut {NoStop}%
\bibitem [{\citenamefont {{Van Duin}}\ \emph {et~al.}(2003)\citenamefont {{Van
  Duin}}, \citenamefont {Strachan}, \citenamefont {Stewman}, \citenamefont
  {Zhang}, \citenamefont {Xu},\ and\ \citenamefont {Goddard}}]{VanDuin2003}%
  \BibitemOpen
  \bibfield  {author} {\bibinfo {author} {\bibfnamefont {A.~C.}\ \bibnamefont
  {{Van Duin}}}, \bibinfo {author} {\bibfnamefont {A.}~\bibnamefont
  {Strachan}}, \bibinfo {author} {\bibfnamefont {S.}~\bibnamefont {Stewman}},
  \bibinfo {author} {\bibfnamefont {Q.}~\bibnamefont {Zhang}}, \bibinfo
  {author} {\bibfnamefont {X.}~\bibnamefont {Xu}},\ and\ \bibinfo {author}
  {\bibfnamefont {W.~A.}\ \bibnamefont {Goddard}},\ }\href
  {https://doi.org/10.1021/jp0276303} {\bibfield  {journal} {\bibinfo
  {journal} {J. Phys. Chem. A}\ }\textbf {\bibinfo {volume} {107}},\ \bibinfo
  {pages} {3803} (\bibinfo {year} {2003})}\BibitemShut {NoStop}%
\bibitem [{\citenamefont {Smirnov}\ and\ \citenamefont
  {Bougeard}(2003)}]{Smirnov2003}%
  \BibitemOpen
  \bibfield  {author} {\bibinfo {author} {\bibfnamefont {K.~S.}\ \bibnamefont
  {Smirnov}}\ and\ \bibinfo {author} {\bibfnamefont {D.}~\bibnamefont
  {Bougeard}},\ }\href {https://doi.org/10.1016/S0301-0104(03)00275-1}
  {\bibfield  {journal} {\bibinfo  {journal} {Chem. Phys.}\ }\textbf {\bibinfo
  {volume} {292}},\ \bibinfo {pages} {53} (\bibinfo {year} {2003})}\BibitemShut
  {NoStop}%
\bibitem [{\citenamefont {Hallil}\ \emph {et~al.}(2006)\citenamefont {Hallil},
  \citenamefont {T{\'{e}}tot}, \citenamefont {Berthier}, \citenamefont
  {Braems},\ and\ \citenamefont {Creuze}}]{Hallil2006}%
  \BibitemOpen
  \bibfield  {author} {\bibinfo {author} {\bibfnamefont {A.}~\bibnamefont
  {Hallil}}, \bibinfo {author} {\bibfnamefont {R.}~\bibnamefont {T{\'{e}}tot}},
  \bibinfo {author} {\bibfnamefont {F.}~\bibnamefont {Berthier}}, \bibinfo
  {author} {\bibfnamefont {I.}~\bibnamefont {Braems}},\ and\ \bibinfo {author}
  {\bibfnamefont {J.}~\bibnamefont {Creuze}},\ }\href
  {https://doi.org/10.1103/PhysRevB.73.165406} {\bibfield  {journal} {\bibinfo
  {journal} {Phys. Rev. B - Condens. Matter Mater. Phys.}\ }\textbf {\bibinfo
  {volume} {73}},\ \bibinfo {pages} {1} (\bibinfo {year} {2006})}\BibitemShut
  {NoStop}%
\bibitem [{\citenamefont {Bultinck}\ \emph {et~al.}(2002)\citenamefont
  {Bultinck}, \citenamefont {Langenaeker}, \citenamefont {Lahorte},
  \citenamefont {{De Proft}}, \citenamefont {Geerlings}, \citenamefont
  {Waroquier},\ and\ \citenamefont {Tollenaere}}]{Bultinck2002}%
  \BibitemOpen
  \bibfield  {author} {\bibinfo {author} {\bibfnamefont {P.}~\bibnamefont
  {Bultinck}}, \bibinfo {author} {\bibfnamefont {W.}~\bibnamefont
  {Langenaeker}}, \bibinfo {author} {\bibfnamefont {P.}~\bibnamefont
  {Lahorte}}, \bibinfo {author} {\bibfnamefont {F.}~\bibnamefont {{De Proft}}},
  \bibinfo {author} {\bibfnamefont {P.}~\bibnamefont {Geerlings}}, \bibinfo
  {author} {\bibfnamefont {M.}~\bibnamefont {Waroquier}},\ and\ \bibinfo
  {author} {\bibfnamefont {J.~P.}\ \bibnamefont {Tollenaere}},\ }\href
  {https://doi.org/10.1021/jp0205463} {\bibfield  {journal} {\bibinfo
  {journal} {J. Phys. Chem. A}\ }\textbf {\bibinfo {volume} {106}},\ \bibinfo
  {pages} {7887} (\bibinfo {year} {2002})}\BibitemShut {NoStop}%
\bibitem [{\citenamefont {{Van Duin}}\ \emph {et~al.}(2001)\citenamefont {{Van
  Duin}}, \citenamefont {Dasgupta}, \citenamefont {Lorant},\ and\ \citenamefont
  {Goddard}}]{VanDuin2001}%
  \BibitemOpen
  \bibfield  {author} {\bibinfo {author} {\bibfnamefont {A.~C.}\ \bibnamefont
  {{Van Duin}}}, \bibinfo {author} {\bibfnamefont {S.}~\bibnamefont
  {Dasgupta}}, \bibinfo {author} {\bibfnamefont {F.}~\bibnamefont {Lorant}},\
  and\ \bibinfo {author} {\bibfnamefont {W.~A.}\ \bibnamefont {Goddard}},\
  }\href {https://doi.org/10.1021/jp004368u} {\bibfield  {journal} {\bibinfo
  {journal} {J. Phys. Chem. A}\ }\textbf {\bibinfo {volume} {105}},\ \bibinfo
  {pages} {9396} (\bibinfo {year} {2001})}\BibitemShut {NoStop}%
\bibitem [{\citenamefont {Bultinck}\ \emph {et~al.}(2004)\citenamefont
  {Bultinck}, \citenamefont {Vanholme}, \citenamefont {Popelier}, \citenamefont
  {{De Proft}},\ and\ \citenamefont {Geerlings}}]{Bultinck2004}%
  \BibitemOpen
  \bibfield  {author} {\bibinfo {author} {\bibfnamefont {P.}~\bibnamefont
  {Bultinck}}, \bibinfo {author} {\bibfnamefont {R.}~\bibnamefont {Vanholme}},
  \bibinfo {author} {\bibfnamefont {P.~L.}\ \bibnamefont {Popelier}}, \bibinfo
  {author} {\bibfnamefont {F.}~\bibnamefont {{De Proft}}},\ and\ \bibinfo
  {author} {\bibfnamefont {P.}~\bibnamefont {Geerlings}},\ }\href
  {https://doi.org/10.1021/jp0469281} {\bibfield  {journal} {\bibinfo
  {journal} {J. Phys. Chem. A}\ }\textbf {\bibinfo {volume} {108}},\ \bibinfo
  {pages} {10359} (\bibinfo {year} {2004})}\BibitemShut {NoStop}%
\bibitem [{\citenamefont {Yang}\ and\ \citenamefont {Sharp}(2006)}]{Yang2006}%
  \BibitemOpen
  \bibfield  {author} {\bibinfo {author} {\bibfnamefont {Q.}~\bibnamefont
  {Yang}}\ and\ \bibinfo {author} {\bibfnamefont {K.~A.}\ \bibnamefont
  {Sharp}},\ }\href {https://doi.org/10.1021/ct060009c} {\bibfield  {journal}
  {\bibinfo  {journal} {J. Chem. Theory Comput.}\ }\textbf {\bibinfo {volume}
  {2}},\ \bibinfo {pages} {1152} (\bibinfo {year} {2006})}\BibitemShut
  {NoStop}%
\bibitem [{\citenamefont {Verstraelen}\ \emph {et~al.}(2012)\citenamefont
  {Verstraelen}, \citenamefont {Pauwels}, \citenamefont {{De Proft}},
  \citenamefont {{Van Speybroeck}}, \citenamefont {Geerlings},\ and\
  \citenamefont {Waroquier}}]{Verstraelen2012}%
  \BibitemOpen
  \bibfield  {author} {\bibinfo {author} {\bibfnamefont {T.}~\bibnamefont
  {Verstraelen}}, \bibinfo {author} {\bibfnamefont {E.}~\bibnamefont
  {Pauwels}}, \bibinfo {author} {\bibfnamefont {F.}~\bibnamefont {{De Proft}}},
  \bibinfo {author} {\bibfnamefont {V.}~\bibnamefont {{Van Speybroeck}}},
  \bibinfo {author} {\bibfnamefont {P.}~\bibnamefont {Geerlings}},\ and\
  \bibinfo {author} {\bibfnamefont {M.}~\bibnamefont {Waroquier}},\ }\href
  {https://doi.org/10.1021/ct200512e} {\bibfield  {journal} {\bibinfo
  {journal} {J. Chem. Theory Comput.}\ }\textbf {\bibinfo {volume} {8}},\
  \bibinfo {pages} {661} (\bibinfo {year} {2012})}\BibitemShut {NoStop}%
\bibitem [{\citenamefont {Ionescu}\ \emph {et~al.}(2013)\citenamefont
  {Ionescu}, \citenamefont {Geidl}, \citenamefont {Svobodová~Vařeková},\
  and\ \citenamefont {Koča}}]{Ionescu2013}%
  \BibitemOpen
  \bibfield  {author} {\bibinfo {author} {\bibfnamefont {C.-M.}\ \bibnamefont
  {Ionescu}}, \bibinfo {author} {\bibfnamefont {S.}~\bibnamefont {Geidl}},
  \bibinfo {author} {\bibfnamefont {R.}~\bibnamefont {Svobodová~Vařeková}},\
  and\ \bibinfo {author} {\bibfnamefont {J.}~\bibnamefont {Koča}},\ }\href
  {https://doi.org/10.1021/ci400448n} {\bibfield  {journal} {\bibinfo
  {journal} {J. Chem. Inf. Model.}\ }\textbf {\bibinfo {volume} {53}},\
  \bibinfo {pages} {2548} (\bibinfo {year} {2013})}\BibitemShut {NoStop}%
\bibitem [{\citenamefont {Chelli}\ \emph {et~al.}(1999)\citenamefont {Chelli},
  \citenamefont {Procacci}, \citenamefont {Righini},\ and\ \citenamefont
  {Califano}}]{chelli1999}%
  \BibitemOpen
  \bibfield  {author} {\bibinfo {author} {\bibfnamefont {R.}~\bibnamefont
  {Chelli}}, \bibinfo {author} {\bibfnamefont {P.}~\bibnamefont {Procacci}},
  \bibinfo {author} {\bibfnamefont {R.}~\bibnamefont {Righini}},\ and\ \bibinfo
  {author} {\bibfnamefont {S.}~\bibnamefont {Califano}},\ }\href
  {https://doi.org/10.1063/1.480198} {\bibfield  {journal} {\bibinfo  {journal}
  {J. Chem. Phys.}\ }\textbf {\bibinfo {volume} {111}},\ \bibinfo {pages}
  {8569} (\bibinfo {year} {1999})},\ \bibinfo {note} {number: 18}\BibitemShut
  {NoStop}%
\bibitem [{\citenamefont {Nistor}\ \emph {et~al.}(2006)\citenamefont {Nistor},
  \citenamefont {Polihronov}, \citenamefont {Müser},\ and\ \citenamefont
  {Mosey}}]{nistor2006}%
  \BibitemOpen
  \bibfield  {author} {\bibinfo {author} {\bibfnamefont {R.~a.}\ \bibnamefont
  {Nistor}}, \bibinfo {author} {\bibfnamefont {J.~G.}\ \bibnamefont
  {Polihronov}}, \bibinfo {author} {\bibfnamefont {M.~H.}\ \bibnamefont
  {Müser}},\ and\ \bibinfo {author} {\bibfnamefont {N.~J.}\ \bibnamefont
  {Mosey}},\ }\href {https://doi.org/10.1063/1.2346671} {\bibfield  {journal}
  {\bibinfo  {journal} {J. Chem. Phys.}\ }\textbf {\bibinfo {volume} {125}},\
  \bibinfo {pages} {94108} (\bibinfo {year} {2006})}\BibitemShut {NoStop}%
\bibitem [{\citenamefont {Warren}\ \emph {et~al.}(2008)\citenamefont {Warren},
  \citenamefont {Davis},\ and\ \citenamefont {Patel}}]{warren2008}%
  \BibitemOpen
  \bibfield  {author} {\bibinfo {author} {\bibfnamefont {L.}~\bibnamefont
  {Warren}}, \bibinfo {author} {\bibfnamefont {J.}~\bibnamefont {Davis}},\ and\
  \bibinfo {author} {\bibfnamefont {S.}~\bibnamefont {Patel}},\ }\href
  {https://doi.org/10.1063/1.2872603} {\bibfield  {journal} {\bibinfo
  {journal} {J. Chem. Phys.}\ }\textbf {\bibinfo {volume} {128}},\ \bibinfo
  {pages} {144110} (\bibinfo {year} {2008})}\BibitemShut {NoStop}%
\bibitem [{\citenamefont {Nistor}\ and\ \citenamefont
  {Müser}(2009)}]{nistor2009}%
  \BibitemOpen
  \bibfield  {author} {\bibinfo {author} {\bibfnamefont {R.}~\bibnamefont
  {Nistor}}\ and\ \bibinfo {author} {\bibfnamefont {M.}~\bibnamefont
  {Müser}},\ }\href {https://doi.org/10.1103/physrevb.79.104303} {\bibfield
  {journal} {\bibinfo  {journal} {Phys. Rev. B}\ }\textbf {\bibinfo {volume}
  {79}},\ \bibinfo {pages} {104303} (\bibinfo {year} {2009})}\BibitemShut
  {NoStop}%
\bibitem [{\citenamefont {Verstraelen}\ \emph {et~al.}(2013)\citenamefont
  {Verstraelen}, \citenamefont {Ayers}, \citenamefont {{Van Speybroeck}},\ and\
  \citenamefont {Waroquier}}]{Verstraelen2013}%
  \BibitemOpen
  \bibfield  {author} {\bibinfo {author} {\bibfnamefont {T.}~\bibnamefont
  {Verstraelen}}, \bibinfo {author} {\bibfnamefont {P.~W.}\ \bibnamefont
  {Ayers}}, \bibinfo {author} {\bibfnamefont {V.}~\bibnamefont {{Van
  Speybroeck}}},\ and\ \bibinfo {author} {\bibfnamefont {M.}~\bibnamefont
  {Waroquier}},\ }\bibfield  {journal} {\bibinfo  {journal} {Journal of
  Chemical Physics}\ }\textbf {\bibinfo {volume} {138}},\ \href
  {https://doi.org/10.1063/1.4791569} {10.1063/1.4791569} (\bibinfo {year}
  {2013})\BibitemShut {NoStop}%
\bibitem [{\citenamefont {Chelli}\ \emph {et~al.}(2005)\citenamefont {Chelli},
  \citenamefont {Pagliai}, \citenamefont {Procacci}, \citenamefont {Cardini},\
  and\ \citenamefont {Schettino}}]{Chelli2005}%
  \BibitemOpen
  \bibfield  {author} {\bibinfo {author} {\bibfnamefont {R.}~\bibnamefont
  {Chelli}}, \bibinfo {author} {\bibfnamefont {M.}~\bibnamefont {Pagliai}},
  \bibinfo {author} {\bibfnamefont {P.}~\bibnamefont {Procacci}}, \bibinfo
  {author} {\bibfnamefont {G.}~\bibnamefont {Cardini}},\ and\ \bibinfo {author}
  {\bibfnamefont {V.}~\bibnamefont {Schettino}},\ }\bibfield  {journal}
  {\bibinfo  {journal} {J. Chem. Phys.}\ }\textbf {\bibinfo {volume} {122}},\
  \href {https://doi.org/10.1063/1.1851504} {10.1063/1.1851504} (\bibinfo
  {year} {2005})\BibitemShut {NoStop}%
\bibitem [{\citenamefont {Chen}\ and\ \citenamefont
  {Mart{\'{i}}nez}(2007)}]{Chen2007}%
  \BibitemOpen
  \bibfield  {author} {\bibinfo {author} {\bibfnamefont {J.}~\bibnamefont
  {Chen}}\ and\ \bibinfo {author} {\bibfnamefont {T.~J.}\ \bibnamefont
  {Mart{\'{i}}nez}},\ }\href {https://doi.org/10.1016/j.cplett.2007.02.065}
  {\bibfield  {journal} {\bibinfo  {journal} {Chem. Phys. Lett.}\ }\textbf
  {\bibinfo {volume} {438}},\ \bibinfo {pages} {315} (\bibinfo {year}
  {2007})}\BibitemShut {NoStop}%
\bibitem [{\citenamefont {Verstraelen}\ \emph {et~al.}(2014)\citenamefont
  {Verstraelen}, \citenamefont {Vandenbrande},\ and\ \citenamefont
  {Ayers}}]{Verstraelen2014}%
  \BibitemOpen
  \bibfield  {author} {\bibinfo {author} {\bibfnamefont {T.}~\bibnamefont
  {Verstraelen}}, \bibinfo {author} {\bibfnamefont {S.}~\bibnamefont
  {Vandenbrande}},\ and\ \bibinfo {author} {\bibfnamefont {P.~W.}\ \bibnamefont
  {Ayers}},\ }\bibfield  {journal} {\bibinfo  {journal} {Journal of Chemical
  Physics}\ }\textbf {\bibinfo {volume} {141}},\ \href
  {https://doi.org/10.1063/1.4901513} {10.1063/1.4901513} (\bibinfo {year}
  {2014})\BibitemShut {NoStop}%
\bibitem [{\citenamefont {Islam}\ \emph {et~al.}(2016)\citenamefont {Islam},
  \citenamefont {Kolesov}, \citenamefont {Verstraelen}, \citenamefont
  {Kaxiras},\ and\ \citenamefont {van Duin}}]{Islam2016}%
  \BibitemOpen
  \bibfield  {author} {\bibinfo {author} {\bibfnamefont {M.~M.}\ \bibnamefont
  {Islam}}, \bibinfo {author} {\bibfnamefont {G.}~\bibnamefont {Kolesov}},
  \bibinfo {author} {\bibfnamefont {T.}~\bibnamefont {Verstraelen}}, \bibinfo
  {author} {\bibfnamefont {E.}~\bibnamefont {Kaxiras}},\ and\ \bibinfo {author}
  {\bibfnamefont {A.~C.~T.}\ \bibnamefont {van Duin}},\ }\href
  {https://doi.org/10.1021/acs.jctc.6b00432} {\bibfield  {journal} {\bibinfo
  {journal} {J. Chem. Theory Comput.}\ }\textbf {\bibinfo {volume} {12}},\
  \bibinfo {pages} {3463} (\bibinfo {year} {2016})}\BibitemShut {NoStop}%
\bibitem [{\citenamefont {G{\"{u}}tlein}\ \emph {et~al.}(2019)\citenamefont
  {G{\"{u}}tlein}, \citenamefont {Lang}, \citenamefont {Reuter}, \citenamefont
  {Blumberger},\ and\ \citenamefont {Oberhofer}}]{Gutlein2019}%
  \BibitemOpen
  \bibfield  {author} {\bibinfo {author} {\bibfnamefont {P.}~\bibnamefont
  {G{\"{u}}tlein}}, \bibinfo {author} {\bibfnamefont {L.}~\bibnamefont {Lang}},
  \bibinfo {author} {\bibfnamefont {K.}~\bibnamefont {Reuter}}, \bibinfo
  {author} {\bibfnamefont {J.}~\bibnamefont {Blumberger}},\ and\ \bibinfo
  {author} {\bibfnamefont {H.}~\bibnamefont {Oberhofer}},\ }\href
  {https://doi.org/10.1021/acs.jctc.9b00415} {\bibfield  {journal} {\bibinfo
  {journal} {Journal of Chemical Theory and Computation}\ }\textbf {\bibinfo
  {volume} {15}},\ \bibinfo {pages} {4516} (\bibinfo {year}
  {2019})}\BibitemShut {NoStop}%
\bibitem [{\citenamefont {G{\"{u}}tlein}\ \emph {et~al.}(2020)\citenamefont
  {G{\"{u}}tlein}, \citenamefont {Blumberger},\ and\ \citenamefont
  {Oberhofer}}]{Gutlein2020}%
  \BibitemOpen
  \bibfield  {author} {\bibinfo {author} {\bibfnamefont {P.}~\bibnamefont
  {G{\"{u}}tlein}}, \bibinfo {author} {\bibfnamefont {J.}~\bibnamefont
  {Blumberger}},\ and\ \bibinfo {author} {\bibfnamefont {H.}~\bibnamefont
  {Oberhofer}},\ }\href {https://doi.org/10.1021/acs.jctc.0c00151} {\bibfield
  {journal} {\bibinfo  {journal} {Journal of chemical theory and computation}\
  }\textbf {\bibinfo {volume} {16}},\ \bibinfo {pages} {5723} (\bibinfo {year}
  {2020})}\BibitemShut {NoStop}%
\bibitem [{\citenamefont {Stone}\ and\ \citenamefont {Tong}(1989)}]{Stone1989}%
  \BibitemOpen
  \bibfield  {author} {\bibinfo {author} {\bibfnamefont {A.}~\bibnamefont
  {Stone}}\ and\ \bibinfo {author} {\bibfnamefont {C.-S.}\ \bibnamefont
  {Tong}},\ }\href {https://doi.org/10.1016/0301-0104(89)87098-3} {\bibfield
  {journal} {\bibinfo  {journal} {Chemical Physics}\ }\textbf {\bibinfo
  {volume} {137}},\ \bibinfo {pages} {121} (\bibinfo {year}
  {1989})}\BibitemShut {NoStop}%
\bibitem [{\citenamefont {Dobson}\ \emph {et~al.}(2006)\citenamefont {Dobson},
  \citenamefont {White},\ and\ \citenamefont {Rubio}}]{Dobson2006}%
  \BibitemOpen
  \bibfield  {author} {\bibinfo {author} {\bibfnamefont {J.~F.}\ \bibnamefont
  {Dobson}}, \bibinfo {author} {\bibfnamefont {A.}~\bibnamefont {White}},\ and\
  \bibinfo {author} {\bibfnamefont {A.}~\bibnamefont {Rubio}},\ }\href
  {https://doi.org/10.1103/PhysRevLett.96.073201} {\bibfield  {journal}
  {\bibinfo  {journal} {Phys. Rev. Lett.}\ }\textbf {\bibinfo {volume} {96}},\
  \bibinfo {pages} {4} (\bibinfo {year} {2006})}\BibitemShut {NoStop}%
\bibitem [{\citenamefont {Hermann}\ \emph {et~al.}(2017)\citenamefont
  {Hermann}, \citenamefont {Alf{\`{e}}},\ and\ \citenamefont
  {Tkatchenko}}]{Hermann2017}%
  \BibitemOpen
  \bibfield  {author} {\bibinfo {author} {\bibfnamefont {J.}~\bibnamefont
  {Hermann}}, \bibinfo {author} {\bibfnamefont {D.}~\bibnamefont
  {Alf{\`{e}}}},\ and\ \bibinfo {author} {\bibfnamefont {A.}~\bibnamefont
  {Tkatchenko}},\ }\bibfield  {journal} {\bibinfo  {journal} {Nat. Commun.}\
  }\textbf {\bibinfo {volume} {8}},\ \href
  {https://doi.org/10.1038/ncomms14052} {10.1038/ncomms14052} (\bibinfo {year}
  {2017})\BibitemShut {NoStop}%
\bibitem [{\citenamefont {Jackson}\ \emph {et~al.}(2007)\citenamefont
  {Jackson}, \citenamefont {Yang},\ and\ \citenamefont
  {Jellinek}}]{Jackson2007}%
  \BibitemOpen
  \bibfield  {author} {\bibinfo {author} {\bibfnamefont {K.}~\bibnamefont
  {Jackson}}, \bibinfo {author} {\bibfnamefont {M.}~\bibnamefont {Yang}},\ and\
  \bibinfo {author} {\bibfnamefont {J.}~\bibnamefont {Jellinek}},\ }\href
  {https://doi.org/10.1021/jp0719457} {\bibfield  {journal} {\bibinfo
  {journal} {J. Phys. Chem. C}\ }\textbf {\bibinfo {volume} {111}},\ \bibinfo
  {pages} {17952} (\bibinfo {year} {2007})}\BibitemShut {NoStop}%
\bibitem [{\citenamefont {Jackson}\ and\ \citenamefont
  {Jellinek}(2016)}]{Jackson2016}%
  \BibitemOpen
  \bibfield  {author} {\bibinfo {author} {\bibfnamefont {K.}~\bibnamefont
  {Jackson}}\ and\ \bibinfo {author} {\bibfnamefont {J.}~\bibnamefont
  {Jellinek}},\ }\href {https://doi.org/10.1063/1.4972813} {\bibfield
  {journal} {\bibinfo  {journal} {J. Chem. Phys.}\ }\textbf {\bibinfo {volume}
  {145}},\ \bibinfo {pages} {244302} (\bibinfo {year} {2016})}\BibitemShut
  {NoStop}%
\bibitem [{\citenamefont {Misquitta}\ \emph {et~al.}(2010)\citenamefont
  {Misquitta}, \citenamefont {Spencer}, \citenamefont {Stone},\ and\
  \citenamefont {Alavi}}]{Misquitta2010}%
  \BibitemOpen
  \bibfield  {author} {\bibinfo {author} {\bibfnamefont {A.~J.}\ \bibnamefont
  {Misquitta}}, \bibinfo {author} {\bibfnamefont {J.}~\bibnamefont {Spencer}},
  \bibinfo {author} {\bibfnamefont {A.~J.}\ \bibnamefont {Stone}},\ and\
  \bibinfo {author} {\bibfnamefont {A.}~\bibnamefont {Alavi}},\ }\href
  {https://doi.org/10.1103/PhysRevB.82.075312} {\bibfield  {journal} {\bibinfo
  {journal} {Phys. Rev. B - Condens. Matter Mater. Phys.}\ }\textbf {\bibinfo
  {volume} {82}},\ \bibinfo {pages} {1} (\bibinfo {year} {2010})}\BibitemShut
  {NoStop}%
\bibitem [{\citenamefont {Dobson}\ and\ \citenamefont
  {Gould}(2012)}]{Dobson2012}%
  \BibitemOpen
  \bibfield  {author} {\bibinfo {author} {\bibfnamefont {J.~F.}\ \bibnamefont
  {Dobson}}\ and\ \bibinfo {author} {\bibfnamefont {T.}~\bibnamefont {Gould}},\
  }\bibfield  {journal} {\bibinfo  {journal} {J. Phys. Condens. Matter}\
  }\textbf {\bibinfo {volume} {24}},\ \href
  {https://doi.org/10.1088/0953-8984/24/7/073201}
  {10.1088/0953-8984/24/7/073201} (\bibinfo {year} {2012})\BibitemShut
  {NoStop}%
\bibitem [{\citenamefont {Misquitta}\ \emph {et~al.}(2014)\citenamefont
  {Misquitta}, \citenamefont {Maezono}, \citenamefont {Drummond}, \citenamefont
  {Stone},\ and\ \citenamefont {Needs}}]{Misquitta2014}%
  \BibitemOpen
  \bibfield  {author} {\bibinfo {author} {\bibfnamefont {A.~J.}\ \bibnamefont
  {Misquitta}}, \bibinfo {author} {\bibfnamefont {R.}~\bibnamefont {Maezono}},
  \bibinfo {author} {\bibfnamefont {N.~D.}\ \bibnamefont {Drummond}}, \bibinfo
  {author} {\bibfnamefont {A.~J.}\ \bibnamefont {Stone}},\ and\ \bibinfo
  {author} {\bibfnamefont {R.~J.}\ \bibnamefont {Needs}},\ }\href
  {https://doi.org/10.1103/PhysRevB.89.045140} {\bibfield  {journal} {\bibinfo
  {journal} {Phys. Rev. B - Condens. Matter Mater. Phys.}\ }\textbf {\bibinfo
  {volume} {89}},\ \bibinfo {pages} {1} (\bibinfo {year} {2014})}\BibitemShut
  {NoStop}%
\bibitem [{\citenamefont {Runge}\ and\ \citenamefont
  {Gross}(1984)}]{Runge1984}%
  \BibitemOpen
  \bibfield  {author} {\bibinfo {author} {\bibfnamefont {E.}~\bibnamefont
  {Runge}}\ and\ \bibinfo {author} {\bibfnamefont {E.~K.~U.}\ \bibnamefont
  {Gross}},\ }\href {https://doi.org/10.1103/PhysRevLett.52.997} {\bibfield
  {journal} {\bibinfo  {journal} {Physical Review Letters}\ }\textbf {\bibinfo
  {volume} {52}},\ \bibinfo {pages} {997} (\bibinfo {year} {1984})}\BibitemShut
  {NoStop}%
\bibitem [{\citenamefont {Ullrich}(2011)}]{ullrich2011time}%
  \BibitemOpen
  \bibfield  {author} {\bibinfo {author} {\bibfnamefont {C.~A.}\ \bibnamefont
  {Ullrich}},\ }\href@noop {} {\emph {\bibinfo {title} {Time-dependent
  density-functional theory: concepts and applications}}}\ (\bibinfo
  {publisher} {OUP Oxford},\ \bibinfo {year} {2011})\BibitemShut {NoStop}%
\bibitem [{\citenamefont {Marques}\ \emph {et~al.}(2012)\citenamefont
  {Marques}, \citenamefont {Maitra}, \citenamefont {Nogueira}, \citenamefont
  {Gross},\ and\ \citenamefont {Rubio}}]{marques2012fundamentals}%
  \BibitemOpen
  \bibfield  {author} {\bibinfo {author} {\bibfnamefont {M.~A.}\ \bibnamefont
  {Marques}}, \bibinfo {author} {\bibfnamefont {N.~T.}\ \bibnamefont {Maitra}},
  \bibinfo {author} {\bibfnamefont {F.~M.}\ \bibnamefont {Nogueira}}, \bibinfo
  {author} {\bibfnamefont {E.~K.}\ \bibnamefont {Gross}},\ and\ \bibinfo
  {author} {\bibfnamefont {A.}~\bibnamefont {Rubio}},\ }\href@noop {} {\emph
  {\bibinfo {title} {Fundamentals of time-dependent density functional
  theory}}},\ Vol.\ \bibinfo {volume} {837}\ (\bibinfo  {publisher}
  {Springer},\ \bibinfo {year} {2012})\BibitemShut {NoStop}%
\bibitem [{\citenamefont {Burke}\ \emph {et~al.}(2005)\citenamefont {Burke},
  \citenamefont {Werschnik},\ and\ \citenamefont {Gross}}]{Burke2005}%
  \BibitemOpen
  \bibfield  {author} {\bibinfo {author} {\bibfnamefont {K.}~\bibnamefont
  {Burke}}, \bibinfo {author} {\bibfnamefont {J.}~\bibnamefont {Werschnik}},\
  and\ \bibinfo {author} {\bibfnamefont {E.~K.}\ \bibnamefont {Gross}},\
  }\bibfield  {journal} {\bibinfo  {journal} {J. Chem. Phys.}\ }\textbf
  {\bibinfo {volume} {123}},\ \href {https://doi.org/10.1063/1.1904586}
  {10.1063/1.1904586} (\bibinfo {year} {2005}),\ \Eprint
  {https://arxiv.org/abs/0410362} {arXiv:0410362 [cond-mat]} \BibitemShut
  {NoStop}%
\bibitem [{\citenamefont {Casida}\ and\ \citenamefont
  {Huix-Rotllant}(2012)}]{Casida2012}%
  \BibitemOpen
  \bibfield  {author} {\bibinfo {author} {\bibfnamefont {M.~E.}\ \bibnamefont
  {Casida}}\ and\ \bibinfo {author} {\bibfnamefont {M.}~\bibnamefont
  {Huix-Rotllant}},\ }\href
  {https://doi.org/10.1146/annurev-physchem-032511-143803} {\bibfield
  {journal} {\bibinfo  {journal} {Annu. Rev. Phys. Chem.}\ }\textbf {\bibinfo
  {volume} {63}},\ \bibinfo {pages} {287} (\bibinfo {year} {2012})},\ \Eprint
  {https://arxiv.org/abs/1108.0611} {arXiv:1108.0611} \BibitemShut {NoStop}%
\bibitem [{\citenamefont {Marques}\ and\ \citenamefont
  {Gross}(2004)}]{Marques2004}%
  \BibitemOpen
  \bibfield  {author} {\bibinfo {author} {\bibfnamefont {M.~A.}\ \bibnamefont
  {Marques}}\ and\ \bibinfo {author} {\bibfnamefont {E.~K.}\ \bibnamefont
  {Gross}},\ }\href {https://doi.org/10.1146/annurev.physchem.55.091602.094449}
  {\bibfield  {journal} {\bibinfo  {journal} {Annu. Rev. Phys. Chem.}\ }\textbf
  {\bibinfo {volume} {55}},\ \bibinfo {pages} {427} (\bibinfo {year}
  {2004})}\BibitemShut {NoStop}%
\bibitem [{\citenamefont {Laurent}\ and\ \citenamefont
  {Jacquemin}(2013)}]{Laurent2013}%
  \BibitemOpen
  \bibfield  {author} {\bibinfo {author} {\bibfnamefont {A.~D.}\ \bibnamefont
  {Laurent}}\ and\ \bibinfo {author} {\bibfnamefont {D.}~\bibnamefont
  {Jacquemin}},\ }\href {https://doi.org/10.1002/qua.24438} {\bibfield
  {journal} {\bibinfo  {journal} {Int. J. Quantum Chem.}\ }\textbf {\bibinfo
  {volume} {113}},\ \bibinfo {pages} {2019} (\bibinfo {year}
  {2013})}\BibitemShut {NoStop}%
\bibitem [{\citenamefont {Kjellgren}\ \emph {et~al.}(2019)\citenamefont
  {Kjellgren}, \citenamefont {Hedeg{\aa}rd},\ and\ \citenamefont
  {Jensen}}]{Kjellgren2019}%
  \BibitemOpen
  \bibfield  {author} {\bibinfo {author} {\bibfnamefont {E.~R.}\ \bibnamefont
  {Kjellgren}}, \bibinfo {author} {\bibfnamefont {E.~D.}\ \bibnamefont
  {Hedeg{\aa}rd}},\ and\ \bibinfo {author} {\bibfnamefont {H.~J.~A.}\
  \bibnamefont {Jensen}},\ }\bibfield  {journal} {\bibinfo  {journal} {J. Chem.
  Phys.}\ }\textbf {\bibinfo {volume} {151}},\ \href
  {https://doi.org/10.1063/1.5119312} {10.1063/1.5119312} (\bibinfo {year}
  {2019})\BibitemShut {NoStop}%
\bibitem [{\citenamefont {Osinga}\ \emph {et~al.}(1997)\citenamefont {Osinga},
  \citenamefont {{Van Gisbergen}}, \citenamefont {Snijders},\ and\
  \citenamefont {Baerends}}]{Osinga1997}%
  \BibitemOpen
  \bibfield  {author} {\bibinfo {author} {\bibfnamefont {V.~P.}\ \bibnamefont
  {Osinga}}, \bibinfo {author} {\bibfnamefont {S.~J.}\ \bibnamefont {{Van
  Gisbergen}}}, \bibinfo {author} {\bibfnamefont {J.~G.}\ \bibnamefont
  {Snijders}},\ and\ \bibinfo {author} {\bibfnamefont {E.~J.}\ \bibnamefont
  {Baerends}},\ }\href {https://doi.org/10.1063/1.473555} {\bibfield  {journal}
  {\bibinfo  {journal} {J. Chem. Phys.}\ }\textbf {\bibinfo {volume} {106}},\
  \bibinfo {pages} {5091} (\bibinfo {year} {1997})}\BibitemShut {NoStop}%
\bibitem [{\citenamefont {Aiga}\ \emph {et~al.}(1999)\citenamefont {Aiga},
  \citenamefont {Tada},\ and\ \citenamefont {Yoshimura}}]{Aiga1999}%
  \BibitemOpen
  \bibfield  {author} {\bibinfo {author} {\bibfnamefont {F.}~\bibnamefont
  {Aiga}}, \bibinfo {author} {\bibfnamefont {T.}~\bibnamefont {Tada}},\ and\
  \bibinfo {author} {\bibfnamefont {R.}~\bibnamefont {Yoshimura}},\ }\href
  {https://doi.org/10.1063/1.479570} {\bibfield  {journal} {\bibinfo  {journal}
  {J. Chem. Phys.}\ }\textbf {\bibinfo {volume} {111}},\ \bibinfo {pages}
  {2878} (\bibinfo {year} {1999})}\BibitemShut {NoStop}%
\bibitem [{\citenamefont {Castro}\ \emph {et~al.}(2004)\citenamefont {Castro},
  \citenamefont {Marques}, \citenamefont {Alonso}, \citenamefont {Bertsch},\
  and\ \citenamefont {Rubio}}]{Castro2004}%
  \BibitemOpen
  \bibfield  {author} {\bibinfo {author} {\bibfnamefont {A.}~\bibnamefont
  {Castro}}, \bibinfo {author} {\bibfnamefont {M.~A.}\ \bibnamefont {Marques}},
  \bibinfo {author} {\bibfnamefont {J.~A.}\ \bibnamefont {Alonso}}, \bibinfo
  {author} {\bibfnamefont {G.~F.}\ \bibnamefont {Bertsch}},\ and\ \bibinfo
  {author} {\bibfnamefont {A.}~\bibnamefont {Rubio}},\ }\href
  {https://doi.org/10.1140/epjd/e2003-00306-3} {\bibfield  {journal} {\bibinfo
  {journal} {Eur. Phys. J. D}\ }\textbf {\bibinfo {volume} {28}},\ \bibinfo
  {pages} {211} (\bibinfo {year} {2004})},\ \Eprint
  {https://arxiv.org/abs/0206307} {arXiv:0206307 [cond-mat]} \BibitemShut
  {NoStop}%
\bibitem [{\citenamefont {Hedeg{\aa}rd}\ \emph {et~al.}(2013)\citenamefont
  {Hedeg{\aa}rd}, \citenamefont {Heiden}, \citenamefont {Knecht}, \citenamefont
  {Fromager},\ and\ \citenamefont {Jensen}}]{Hedegard2013}%
  \BibitemOpen
  \bibfield  {author} {\bibinfo {author} {\bibfnamefont {E.~D.}\ \bibnamefont
  {Hedeg{\aa}rd}}, \bibinfo {author} {\bibfnamefont {F.}~\bibnamefont
  {Heiden}}, \bibinfo {author} {\bibfnamefont {S.}~\bibnamefont {Knecht}},
  \bibinfo {author} {\bibfnamefont {E.}~\bibnamefont {Fromager}},\ and\
  \bibinfo {author} {\bibfnamefont {H.~J.~A.}\ \bibnamefont {Jensen}},\ }\href
  {https://doi.org/10.1063/1.4826533} {\bibfield  {journal} {\bibinfo
  {journal} {J. Chem. Phys.}\ }\textbf {\bibinfo {volume} {139}},\ \bibinfo
  {pages} {1} (\bibinfo {year} {2013})}\BibitemShut {NoStop}%
\bibitem [{\citenamefont {Fromager}\ \emph {et~al.}(2013)\citenamefont
  {Fromager}, \citenamefont {Knecht},\ and\ \citenamefont
  {Jensen}}]{Fromager2013}%
  \BibitemOpen
  \bibfield  {author} {\bibinfo {author} {\bibfnamefont {E.}~\bibnamefont
  {Fromager}}, \bibinfo {author} {\bibfnamefont {S.}~\bibnamefont {Knecht}},\
  and\ \bibinfo {author} {\bibfnamefont {H.~J.~A.}\ \bibnamefont {Jensen}},\
  }\bibfield  {journal} {\bibinfo  {journal} {J. Chem. Phys.}\ }\textbf
  {\bibinfo {volume} {138}},\ \href {https://doi.org/10.1063/1.4792199}
  {10.1063/1.4792199} (\bibinfo {year} {2013}),\ \Eprint
  {https://arxiv.org/abs/1211.4829} {arXiv:1211.4829} \BibitemShut {NoStop}%
\bibitem [{\citenamefont {Sa{\l}ek}\ \emph {et~al.}(2002)\citenamefont
  {Sa{\l}ek}, \citenamefont {Vahtras}, \citenamefont {Helgaker},\ and\
  \citenamefont {{\AA}gren}}]{Saek2002}%
  \BibitemOpen
  \bibfield  {author} {\bibinfo {author} {\bibfnamefont {P.}~\bibnamefont
  {Sa{\l}ek}}, \bibinfo {author} {\bibfnamefont {O.}~\bibnamefont {Vahtras}},
  \bibinfo {author} {\bibfnamefont {T.}~\bibnamefont {Helgaker}},\ and\
  \bibinfo {author} {\bibfnamefont {H.}~\bibnamefont {{\AA}gren}},\ }\href
  {https://doi.org/10.1063/1.1516805} {\bibfield  {journal} {\bibinfo
  {journal} {J. Chem. Phys.}\ }\textbf {\bibinfo {volume} {117}},\ \bibinfo
  {pages} {9630} (\bibinfo {year} {2002})}\BibitemShut {NoStop}%
\bibitem [{\citenamefont {Tunell}\ \emph {et~al.}(2003)\citenamefont {Tunell},
  \citenamefont {Rinkevicius}, \citenamefont {Vahtras}, \citenamefont {Salek},
  \citenamefont {Helgaker},\ and\ \citenamefont {Agren}}]{Tunell2003}%
  \BibitemOpen
  \bibfield  {author} {\bibinfo {author} {\bibfnamefont {I.}~\bibnamefont
  {Tunell}}, \bibinfo {author} {\bibfnamefont {Z.}~\bibnamefont {Rinkevicius}},
  \bibinfo {author} {\bibfnamefont {O.}~\bibnamefont {Vahtras}}, \bibinfo
  {author} {\bibfnamefont {P.}~\bibnamefont {Salek}}, \bibinfo {author}
  {\bibfnamefont {T.}~\bibnamefont {Helgaker}},\ and\ \bibinfo {author}
  {\bibfnamefont {H.}~\bibnamefont {Agren}},\ }\href
  {https://doi.org/10.1063/1.1622926} {\bibfield  {journal} {\bibinfo
  {journal} {J. Chem. Phys.}\ }\textbf {\bibinfo {volume} {119}},\ \bibinfo
  {pages} {11024} (\bibinfo {year} {2003})}\BibitemShut {NoStop}%
\bibitem [{\citenamefont {Christiansen}\ \emph {et~al.}(1998)\citenamefont
  {Christiansen}, \citenamefont {J{\o}rgensen},\ and\ \citenamefont
  {H{\"{a}}ttig}}]{Christiansen1998}%
  \BibitemOpen
  \bibfield  {author} {\bibinfo {author} {\bibfnamefont {O.}~\bibnamefont
  {Christiansen}}, \bibinfo {author} {\bibfnamefont {P.}~\bibnamefont
  {J{\o}rgensen}},\ and\ \bibinfo {author} {\bibfnamefont {C.}~\bibnamefont
  {H{\"{a}}ttig}},\ }\href
  {https://doi.org/10.1002/(SICI)1097-461X(1998)68:1<1::AID-QUA1>3.0.CO;2-Z}
  {\bibfield  {journal} {\bibinfo  {journal} {International Journal of Quantum
  Chemistry}\ }\textbf {\bibinfo {volume} {68}},\ \bibinfo {pages} {1}
  (\bibinfo {year} {1998})}\BibitemShut {NoStop}%
\bibitem [{\citenamefont {Telnov}\ and\ \citenamefont
  {Chu}(1997)}]{Telnov1997}%
  \BibitemOpen
  \bibfield  {author} {\bibinfo {author} {\bibfnamefont {D.~A.}\ \bibnamefont
  {Telnov}}\ and\ \bibinfo {author} {\bibfnamefont {S.~I.}\ \bibnamefont
  {Chu}},\ }\href {https://doi.org/10.1016/S0009-2614(96)01370-X} {\bibfield
  {journal} {\bibinfo  {journal} {Chemical Physics Letters}\ }\textbf {\bibinfo
  {volume} {264}},\ \bibinfo {pages} {466} (\bibinfo {year}
  {1997})}\BibitemShut {NoStop}%
\bibitem [{\citenamefont {Salek}\ \emph {et~al.}(2005)\citenamefont {Salek},
  \citenamefont {Helgaker},\ and\ \citenamefont {Saue}}]{Salek2005}%
  \BibitemOpen
  \bibfield  {author} {\bibinfo {author} {\bibfnamefont {P.}~\bibnamefont
  {Salek}}, \bibinfo {author} {\bibfnamefont {T.}~\bibnamefont {Helgaker}},\
  and\ \bibinfo {author} {\bibfnamefont {T.}~\bibnamefont {Saue}},\ }\href
  {https://doi.org/10.1016/j.chemphys.2004.10.011} {\bibfield  {journal}
  {\bibinfo  {journal} {Chemical Physics}\ }\textbf {\bibinfo {volume} {311}},\
  \bibinfo {pages} {187} (\bibinfo {year} {2005})}\BibitemShut {NoStop}%
\bibitem [{\citenamefont {Deb}\ and\ \citenamefont {Ghosh}(1982)}]{Deb1982}%
  \BibitemOpen
  \bibfield  {author} {\bibinfo {author} {\bibfnamefont {B.~M.}\ \bibnamefont
  {Deb}}\ and\ \bibinfo {author} {\bibfnamefont {S.~K.}\ \bibnamefont
  {Ghosh}},\ }\href {https://doi.org/10.1063/1.443611} {\bibfield  {journal}
  {\bibinfo  {journal} {The Journal of Chemical Physics}\ }\textbf {\bibinfo
  {volume} {77}},\ \bibinfo {pages} {342} (\bibinfo {year} {1982})}\BibitemShut
  {NoStop}%
\bibitem [{\citenamefont {Maitra}\ and\ \citenamefont
  {Burke}(2002)}]{Maitra2002}%
  \BibitemOpen
  \bibfield  {author} {\bibinfo {author} {\bibfnamefont {N.~T.}\ \bibnamefont
  {Maitra}}\ and\ \bibinfo {author} {\bibfnamefont {K.}~\bibnamefont {Burke}},\
  }\href {https://doi.org/10.1016/S0009-2614(02)00586-9} {\bibfield  {journal}
  {\bibinfo  {journal} {Chem. Phys. Lett.}\ }\textbf {\bibinfo {volume}
  {359}},\ \bibinfo {pages} {237} (\bibinfo {year} {2002})}\BibitemShut
  {NoStop}%
\bibitem [{\citenamefont {Samal}\ and\ \citenamefont
  {Harbola}(2006)}]{Samal2006}%
  \BibitemOpen
  \bibfield  {author} {\bibinfo {author} {\bibfnamefont {P.}~\bibnamefont
  {Samal}}\ and\ \bibinfo {author} {\bibfnamefont {M.~K.}\ \bibnamefont
  {Harbola}},\ }\href {https://doi.org/10.1016/j.cplett.2006.11.026} {\bibfield
   {journal} {\bibinfo  {journal} {Chem. Phys. Lett.}\ }\textbf {\bibinfo
  {volume} {433}},\ \bibinfo {pages} {204} (\bibinfo {year} {2006})},\ \Eprint
  {https://arxiv.org/abs/0611428} {arXiv:0611428 [cond-mat]} \BibitemShut
  {NoStop}%
\bibitem [{\citenamefont {Maitra}\ and\ \citenamefont
  {Burke}(2007)}]{Maitra2007}%
  \BibitemOpen
  \bibfield  {author} {\bibinfo {author} {\bibfnamefont {N.~T.}\ \bibnamefont
  {Maitra}}\ and\ \bibinfo {author} {\bibfnamefont {K.}~\bibnamefont {Burke}},\
  }\href {https://doi.org/10.1016/j.cplett.2007.04.091} {\bibfield  {journal}
  {\bibinfo  {journal} {Chem. Phys. Lett.}\ }\textbf {\bibinfo {volume}
  {441}},\ \bibinfo {pages} {167} (\bibinfo {year} {2007})},\ \Eprint
  {https://arxiv.org/abs/0704.2084} {arXiv:0704.2084} \BibitemShut {NoStop}%
\bibitem [{\citenamefont {Kapoor}\ \emph {et~al.}(2013)\citenamefont {Kapoor},
  \citenamefont {Ruggenthaler},\ and\ \citenamefont {Bauer}}]{Kapoor2013}%
  \BibitemOpen
  \bibfield  {author} {\bibinfo {author} {\bibfnamefont {V.}~\bibnamefont
  {Kapoor}}, \bibinfo {author} {\bibfnamefont {M.}~\bibnamefont
  {Ruggenthaler}},\ and\ \bibinfo {author} {\bibfnamefont {D.}~\bibnamefont
  {Bauer}},\ }\href {https://doi.org/10.1103/PhysRevA.87.042521} {\bibfield
  {journal} {\bibinfo  {journal} {Phys. Rev. A - At. Mol. Opt. Phys.}\ }\textbf
  {\bibinfo {volume} {87}},\ \bibinfo {pages} {1} (\bibinfo {year}
  {2013})}\BibitemShut {NoStop}%
\bibitem [{\citenamefont {Langhoff}\ \emph {et~al.}(1972)\citenamefont
  {Langhoff}, \citenamefont {Epstein},\ and\ \citenamefont
  {Karplus}}]{Langhoff1972}%
  \BibitemOpen
  \bibfield  {author} {\bibinfo {author} {\bibfnamefont {P.~W.}\ \bibnamefont
  {Langhoff}}, \bibinfo {author} {\bibfnamefont {S.~T.}\ \bibnamefont
  {Epstein}},\ and\ \bibinfo {author} {\bibfnamefont {M.}~\bibnamefont
  {Karplus}},\ }\href {https://doi.org/10.1103/RevModPhys.44.602} {\bibfield
  {journal} {\bibinfo  {journal} {Reviews of Modern Physics}\ }\textbf
  {\bibinfo {volume} {44}},\ \bibinfo {pages} {602} (\bibinfo {year}
  {1972})}\BibitemShut {NoStop}%
\bibitem [{\citenamefont {Mayer}\ \emph {et~al.}(2008)\citenamefont {Mayer},
  \citenamefont {Lambin},\ and\ \citenamefont
  {Åstrand}}]{mayer_electrostatic_2008}%
  \BibitemOpen
  \bibfield  {author} {\bibinfo {author} {\bibfnamefont {A.}~\bibnamefont
  {Mayer}}, \bibinfo {author} {\bibfnamefont {P.}~\bibnamefont {Lambin}},\ and\
  \bibinfo {author} {\bibfnamefont {P.-O.}\ \bibnamefont {Åstrand}},\ }\href
  {https://doi.org/10.1088/0957-4484/19/02/025203} {\bibfield  {journal}
  {\bibinfo  {journal} {Nanotechnology}\ }\textbf {\bibinfo {volume} {19}},\
  \bibinfo {pages} {025203} (\bibinfo {year} {2008})}\BibitemShut {NoStop}%
\bibitem [{\citenamefont {Smalø}\ \emph {et~al.}(2013)\citenamefont {Smalø},
  \citenamefont {Åstrand},\ and\ \citenamefont {Mayer}}]{smalo_combined_2013}%
  \BibitemOpen
  \bibfield  {author} {\bibinfo {author} {\bibfnamefont {H.~S.}\ \bibnamefont
  {Smalø}}, \bibinfo {author} {\bibfnamefont {P.-O.}\ \bibnamefont
  {Åstrand}},\ and\ \bibinfo {author} {\bibfnamefont {A.}~\bibnamefont
  {Mayer}},\ }\href {https://doi.org/10.1080/00268976.2013.797116} {\bibfield
  {journal} {\bibinfo  {journal} {Mol. Phys.}\ }\textbf {\bibinfo {volume}
  {111}},\ \bibinfo {pages} {1470} (\bibinfo {year} {2013})}\BibitemShut
  {NoStop}%
\bibitem [{\citenamefont {Haghdani}\ \emph {et~al.}(2014)\citenamefont
  {Haghdani}, \citenamefont {Davari}, \citenamefont {Sandnes},\ and\
  \citenamefont {Åstrand}}]{haghdani_complex_2014}%
  \BibitemOpen
  \bibfield  {author} {\bibinfo {author} {\bibfnamefont {S.}~\bibnamefont
  {Haghdani}}, \bibinfo {author} {\bibfnamefont {N.}~\bibnamefont {Davari}},
  \bibinfo {author} {\bibfnamefont {R.}~\bibnamefont {Sandnes}},\ and\ \bibinfo
  {author} {\bibfnamefont {P.-O.}\ \bibnamefont {Åstrand}},\ }\href
  {https://doi.org/10.1021/jp507639z} {\bibfield  {journal} {\bibinfo
  {journal} {J. Phys. Chem. A}\ }\textbf {\bibinfo {volume} {118}},\ \bibinfo
  {pages} {11282} (\bibinfo {year} {2014})}\BibitemShut {NoStop}%
\bibitem [{\citenamefont {Hermann}\ and\ \citenamefont
  {Tkatchenko}(2020)}]{Hermann2020}%
  \BibitemOpen
  \bibfield  {author} {\bibinfo {author} {\bibfnamefont {J.}~\bibnamefont
  {Hermann}}\ and\ \bibinfo {author} {\bibfnamefont {A.}~\bibnamefont
  {Tkatchenko}},\ }\href {https://doi.org/10.1103/PhysRevLett.124.146401}
  {\bibfield  {journal} {\bibinfo  {journal} {Phys. Rev. Lett.}\ }\textbf
  {\bibinfo {volume} {124}},\ \bibinfo {pages} {146401} (\bibinfo {year}
  {2020})}\BibitemShut {NoStop}%
\bibitem [{\citenamefont {Ambrosetti}\ \emph {et~al.}(2014)\citenamefont
  {Ambrosetti}, \citenamefont {Reilly}, \citenamefont {Distasio},\ and\
  \citenamefont {Tkatchenko}}]{ambrosetti_long-range_2014}%
  \BibitemOpen
  \bibfield  {author} {\bibinfo {author} {\bibfnamefont {A.}~\bibnamefont
  {Ambrosetti}}, \bibinfo {author} {\bibfnamefont {A.~M.}\ \bibnamefont
  {Reilly}}, \bibinfo {author} {\bibfnamefont {R.~A.}\ \bibnamefont
  {Distasio}},\ and\ \bibinfo {author} {\bibfnamefont {A.}~\bibnamefont
  {Tkatchenko}},\ }\bibfield  {journal} {\bibinfo  {journal} {J. Chem. Phys.}\
  }\textbf {\bibinfo {volume} {140}},\ \href
  {https://doi.org/10.1063/1.4865104} {10.1063/1.4865104} (\bibinfo {year}
  {2014})\BibitemShut {NoStop}%
\bibitem [{\citenamefont {Wildman}\ \emph {et~al.}(2019)\citenamefont
  {Wildman}, \citenamefont {Donati}, \citenamefont {Lipparini}, \citenamefont
  {Mennucci},\ and\ \citenamefont {Li}}]{wildman_nonequilibrium_2019}%
  \BibitemOpen
  \bibfield  {author} {\bibinfo {author} {\bibfnamefont {A.}~\bibnamefont
  {Wildman}}, \bibinfo {author} {\bibfnamefont {G.}~\bibnamefont {Donati}},
  \bibinfo {author} {\bibfnamefont {F.}~\bibnamefont {Lipparini}}, \bibinfo
  {author} {\bibfnamefont {B.}~\bibnamefont {Mennucci}},\ and\ \bibinfo
  {author} {\bibfnamefont {X.}~\bibnamefont {Li}},\ }\href
  {https://doi.org/10.1021/acs.jctc.8b00836} {\bibfield  {journal} {\bibinfo
  {journal} {J. Chem. Theory Comput.}\ }\textbf {\bibinfo {volume} {15}},\
  \bibinfo {pages} {43} (\bibinfo {year} {2019})}\BibitemShut {NoStop}%
\bibitem [{\citenamefont {Misquitta}\ and\ \citenamefont
  {Stone}(2018)}]{misquitta_isa-pol_2018}%
  \BibitemOpen
  \bibfield  {author} {\bibinfo {author} {\bibfnamefont {A.~J.}\ \bibnamefont
  {Misquitta}}\ and\ \bibinfo {author} {\bibfnamefont {A.~J.}\ \bibnamefont
  {Stone}},\ }\href {https://doi.org/10.1007/s00214-018-2371-4} {\bibfield
  {journal} {\bibinfo  {journal} {Theor. Chem. Acc.}\ }\textbf {\bibinfo
  {volume} {137}},\ \bibinfo {pages} {153} (\bibinfo {year}
  {2018})}\BibitemShut {NoStop}%
\bibitem [{\citenamefont {Schriber}\ \emph {et~al.}(2021)\citenamefont
  {Schriber}, \citenamefont {Nascimento}, \citenamefont {Koutsoukas},
  \citenamefont {Spronk}, \citenamefont {Cheney},\ and\ \citenamefont
  {Sherrill}}]{schriber_cliff_2021}%
  \BibitemOpen
  \bibfield  {author} {\bibinfo {author} {\bibfnamefont {J.~B.}\ \bibnamefont
  {Schriber}}, \bibinfo {author} {\bibfnamefont {D.~R.}\ \bibnamefont
  {Nascimento}}, \bibinfo {author} {\bibfnamefont {A.}~\bibnamefont
  {Koutsoukas}}, \bibinfo {author} {\bibfnamefont {S.~A.}\ \bibnamefont
  {Spronk}}, \bibinfo {author} {\bibfnamefont {D.~L.}\ \bibnamefont {Cheney}},\
  and\ \bibinfo {author} {\bibfnamefont {C.~D.}\ \bibnamefont {Sherrill}},\
  }\href {https://doi.org/10.1063/5.0042989} {\bibfield  {journal} {\bibinfo
  {journal} {J. Chem. Phys.}\ }\textbf {\bibinfo {volume} {154}},\ \bibinfo
  {pages} {184110} (\bibinfo {year} {2021})}\BibitemShut {NoStop}%
\bibitem [{\citenamefont {Tkatchenko}\ and\ \citenamefont
  {Scheffler}(2009)}]{Tkatchenko2009}%
  \BibitemOpen
  \bibfield  {author} {\bibinfo {author} {\bibfnamefont {A.}~\bibnamefont
  {Tkatchenko}}\ and\ \bibinfo {author} {\bibfnamefont {M.}~\bibnamefont
  {Scheffler}},\ }\href {https://doi.org/10.1103/PhysRevLett.102.073005}
  {\bibfield  {journal} {\bibinfo  {journal} {Phys. Rev. Lett.}\ }\textbf
  {\bibinfo {volume} {102}},\ \bibinfo {pages} {73005} (\bibinfo {year}
  {2009})}\BibitemShut {NoStop}%
\bibitem [{\citenamefont {Toulouse}\ \emph {et~al.}(2013)\citenamefont
  {Toulouse}, \citenamefont {Rebolini}, \citenamefont {Gould}, \citenamefont
  {Dobson}, \citenamefont {Seal},\ and\ \citenamefont
  {{\'{A}}ngy{\'{a}}n}}]{Toulouse2013}%
  \BibitemOpen
  \bibfield  {author} {\bibinfo {author} {\bibfnamefont {J.}~\bibnamefont
  {Toulouse}}, \bibinfo {author} {\bibfnamefont {E.}~\bibnamefont {Rebolini}},
  \bibinfo {author} {\bibfnamefont {T.}~\bibnamefont {Gould}}, \bibinfo
  {author} {\bibfnamefont {J.~F.}\ \bibnamefont {Dobson}}, \bibinfo {author}
  {\bibfnamefont {P.}~\bibnamefont {Seal}},\ and\ \bibinfo {author}
  {\bibfnamefont {J.~G.}\ \bibnamefont {{\'{A}}ngy{\'{a}}n}},\ }\bibfield
  {journal} {\bibinfo  {journal} {J. Chem. Phys.}\ }\textbf {\bibinfo {volume}
  {138}},\ \href {https://doi.org/10.1063/1.4804981} {10.1063/1.4804981}
  (\bibinfo {year} {2013})\BibitemShut {NoStop}%
\bibitem [{\citenamefont {Rice}\ and\ \citenamefont {Handy}(1991)}]{Rice1991}%
  \BibitemOpen
  \bibfield  {author} {\bibinfo {author} {\bibfnamefont {J.~E.}\ \bibnamefont
  {Rice}}\ and\ \bibinfo {author} {\bibfnamefont {N.~C.}\ \bibnamefont
  {Handy}},\ }\href {https://doi.org/10.1063/1.460558} {\bibfield  {journal}
  {\bibinfo  {journal} {The Journal of Chemical Physics}\ }\textbf {\bibinfo
  {volume} {94}},\ \bibinfo {pages} {4959} (\bibinfo {year}
  {1991})}\BibitemShut {NoStop}%
\bibitem [{\citenamefont {Stone}(1985)}]{Stone1985}%
  \BibitemOpen
  \bibfield  {author} {\bibinfo {author} {\bibfnamefont {A.}~\bibnamefont
  {Stone}},\ }\href {https://doi.org/10.1080/00268978500102901} {\bibfield
  {journal} {\bibinfo  {journal} {Molecular Physics}\ }\textbf {\bibinfo
  {volume} {56}},\ \bibinfo {pages} {1065} (\bibinfo {year} {1985})},\ \Eprint
  {https://arxiv.org/abs/https://doi.org/10.1080/00268978500102901}
  {https://doi.org/10.1080/00268978500102901} \BibitemShut {NoStop}%
\bibitem [{\citenamefont {{\'{A}}ngy{\'{a}}n}\ \emph
  {et~al.}(1994)\citenamefont {{\'{A}}ngy{\'{a}}n}, \citenamefont {Jansen},
  \citenamefont {Loss}, \citenamefont {H{\"{a}}ttig},\ and\ \citenamefont
  {He{\ss}}}]{Angyan1994}%
  \BibitemOpen
  \bibfield  {author} {\bibinfo {author} {\bibfnamefont {J.~G.}\ \bibnamefont
  {{\'{A}}ngy{\'{a}}n}}, \bibinfo {author} {\bibfnamefont {G.}~\bibnamefont
  {Jansen}}, \bibinfo {author} {\bibfnamefont {M.}~\bibnamefont {Loss}},
  \bibinfo {author} {\bibfnamefont {C.}~\bibnamefont {H{\"{a}}ttig}},\ and\
  \bibinfo {author} {\bibfnamefont {B.~A.}\ \bibnamefont {He{\ss}}},\ }\href
  {https://doi.org/10.1016/0009-2614(94)87056-X} {\bibfield  {journal}
  {\bibinfo  {journal} {Chemical Physics Letters}\ }\textbf {\bibinfo {volume}
  {219}},\ \bibinfo {pages} {267} (\bibinfo {year} {1994})}\BibitemShut
  {NoStop}%
\bibitem [{\citenamefont {Martin}(2020)}]{martin2020electronic}%
  \BibitemOpen
  \bibfield  {author} {\bibinfo {author} {\bibfnamefont {R.~M.}\ \bibnamefont
  {Martin}},\ }\href@noop {} {\emph {\bibinfo {title} {Electronic structure:
  basic theory and practical methods}}}\ (\bibinfo  {publisher} {Cambridge
  university press},\ \bibinfo {year} {2020})\BibitemShut {NoStop}%
\bibitem [{\citenamefont {Zaremba}\ and\ \citenamefont
  {Kohn}(1976)}]{Zaremba1976}%
  \BibitemOpen
  \bibfield  {author} {\bibinfo {author} {\bibfnamefont {E.}~\bibnamefont
  {Zaremba}}\ and\ \bibinfo {author} {\bibfnamefont {W.}~\bibnamefont {Kohn}},\
  }\href {https://doi.org/10.1103/PhysRevB.13.2270} {\bibfield  {journal}
  {\bibinfo  {journal} {Phys. Rev. B}\ }\textbf {\bibinfo {volume} {13}},\
  \bibinfo {pages} {2270} (\bibinfo {year} {1976})},\ \Eprint
  {https://arxiv.org/abs/arXiv:1011.1669v3} {arXiv:arXiv:1011.1669v3}
  \BibitemShut {NoStop}%
\bibitem [{\citenamefont {Verstraelen}\ \emph {et~al.}(2016)\citenamefont
  {Verstraelen}, \citenamefont {Vandenbrande}, \citenamefont {Heidar-Zadeh},
  \citenamefont {Vanduyfhuys}, \citenamefont {{Van Speybroeck}}, \citenamefont
  {Waroquier},\ and\ \citenamefont {Ayers}}]{Verstraelen2016}%
  \BibitemOpen
  \bibfield  {author} {\bibinfo {author} {\bibfnamefont {T.}~\bibnamefont
  {Verstraelen}}, \bibinfo {author} {\bibfnamefont {S.}~\bibnamefont
  {Vandenbrande}}, \bibinfo {author} {\bibfnamefont {F.}~\bibnamefont
  {Heidar-Zadeh}}, \bibinfo {author} {\bibfnamefont {L.}~\bibnamefont
  {Vanduyfhuys}}, \bibinfo {author} {\bibfnamefont {V.}~\bibnamefont {{Van
  Speybroeck}}}, \bibinfo {author} {\bibfnamefont {M.}~\bibnamefont
  {Waroquier}},\ and\ \bibinfo {author} {\bibfnamefont {P.~W.}\ \bibnamefont
  {Ayers}},\ }\href {https://doi.org/10.1021/acs.jctc.6b00456} {\bibfield
  {journal} {\bibinfo  {journal} {J. Chem. Theory Comput.}\ }\textbf {\bibinfo
  {volume} {12}},\ \bibinfo {pages} {3894} (\bibinfo {year}
  {2016})}\BibitemShut {NoStop}%
\bibitem [{\citenamefont {Becke}(1988)}]{Becke1988}%
  \BibitemOpen
  \bibfield  {author} {\bibinfo {author} {\bibfnamefont {A.~D.}\ \bibnamefont
  {Becke}},\ }\href {https://doi.org/10.1063/1.454033} {\bibfield  {journal}
  {\bibinfo  {journal} {J. Chem. Phys.}\ }\textbf {\bibinfo {volume} {88}},\
  \bibinfo {pages} {2547} (\bibinfo {year} {1988})}\BibitemShut {NoStop}%
\bibitem [{\citenamefont {Becke}\ and\ \citenamefont
  {Dickson}(1988)}]{BeckeDiskson1988}%
  \BibitemOpen
  \bibfield  {author} {\bibinfo {author} {\bibfnamefont {A.~D.}\ \bibnamefont
  {Becke}}\ and\ \bibinfo {author} {\bibfnamefont {R.~M.}\ \bibnamefont
  {Dickson}},\ }\href {https://doi.org/10.1063/1.455005} {\bibfield  {journal}
  {\bibinfo  {journal} {J. Chem. Phys.}\ }\textbf {\bibinfo {volume} {89}},\
  \bibinfo {pages} {2993} (\bibinfo {year} {1988})}\BibitemShut {NoStop}%
\bibitem [{\citenamefont {Frisch}\ \emph {et~al.}(2016)\citenamefont {Frisch},
  \citenamefont {Trucks}, \citenamefont {Schlegel}, \citenamefont {Scuseria},
  \citenamefont {Robb}, \citenamefont {Cheeseman}, \citenamefont {Scalmani},
  \citenamefont {Barone}, \citenamefont {Petersson}, \citenamefont {Nakatsuji},
  \citenamefont {Li}, \citenamefont {Caricato}, \citenamefont {Marenich},
  \citenamefont {Bloino}, \citenamefont {Janesko}, \citenamefont {Gomperts},
  \citenamefont {Mennucci}, \citenamefont {Hratchian}, \citenamefont {Ortiz},
  \citenamefont {Izmaylov}, \citenamefont {Sonnenberg}, \citenamefont
  {Williams-Young}, \citenamefont {Ding}, \citenamefont {Lipparini},
  \citenamefont {Egidi}, \citenamefont {Goings}, \citenamefont {Peng},
  \citenamefont {Petrone}, \citenamefont {Henderson}, \citenamefont
  {Ranasinghe}, \citenamefont {Zakrzewski}, \citenamefont {Gao}, \citenamefont
  {Rega}, \citenamefont {Zheng}, \citenamefont {Liang}, \citenamefont {Hada},
  \citenamefont {Ehara}, \citenamefont {Toyota}, \citenamefont {Fukuda},
  \citenamefont {Hasegawa}, \citenamefont {Ishida}, \citenamefont {Nakajima},
  \citenamefont {Honda}, \citenamefont {Kitao}, \citenamefont {Nakai},
  \citenamefont {Vreven}, \citenamefont {Throssell}, \citenamefont
  {Montgomery}, \citenamefont {Peralta}, \citenamefont {Ogliaro}, \citenamefont
  {Bearpark}, \citenamefont {Heyd}, \citenamefont {Brothers}, \citenamefont
  {Kudin}, \citenamefont {Staroverov}, \citenamefont {Keith}, \citenamefont
  {Kobayashi}, \citenamefont {Normand}, \citenamefont {Raghavachari},
  \citenamefont {Rendell}, \citenamefont {Burant}, \citenamefont {Iyengar},
  \citenamefont {Tomasi}, \citenamefont {Cossi}, \citenamefont {Millam},
  \citenamefont {Klene}, \citenamefont {Adamo}, \citenamefont {Cammi},
  \citenamefont {Ochterski}, \citenamefont {Martin}, \citenamefont {Morokuma},
  \citenamefont {Farkas}, \citenamefont {Foresman},\ and\ \citenamefont
  {Fox}}]{g16}%
  \BibitemOpen
  \bibfield  {author} {\bibinfo {author} {\bibfnamefont {M.~J.}\ \bibnamefont
  {Frisch}}, \bibinfo {author} {\bibfnamefont {G.~W.}\ \bibnamefont {Trucks}},
  \bibinfo {author} {\bibfnamefont {H.~B.}\ \bibnamefont {Schlegel}}, \bibinfo
  {author} {\bibfnamefont {G.~E.}\ \bibnamefont {Scuseria}}, \bibinfo {author}
  {\bibfnamefont {M.~A.}\ \bibnamefont {Robb}}, \bibinfo {author}
  {\bibfnamefont {J.~R.}\ \bibnamefont {Cheeseman}}, \bibinfo {author}
  {\bibfnamefont {G.}~\bibnamefont {Scalmani}}, \bibinfo {author}
  {\bibfnamefont {V.}~\bibnamefont {Barone}}, \bibinfo {author} {\bibfnamefont
  {G.~A.}\ \bibnamefont {Petersson}}, \bibinfo {author} {\bibfnamefont
  {H.}~\bibnamefont {Nakatsuji}}, \bibinfo {author} {\bibfnamefont
  {X.}~\bibnamefont {Li}}, \bibinfo {author} {\bibfnamefont {M.}~\bibnamefont
  {Caricato}}, \bibinfo {author} {\bibfnamefont {A.~V.}\ \bibnamefont
  {Marenich}}, \bibinfo {author} {\bibfnamefont {J.}~\bibnamefont {Bloino}},
  \bibinfo {author} {\bibfnamefont {B.~G.}\ \bibnamefont {Janesko}}, \bibinfo
  {author} {\bibfnamefont {R.}~\bibnamefont {Gomperts}}, \bibinfo {author}
  {\bibfnamefont {B.}~\bibnamefont {Mennucci}}, \bibinfo {author}
  {\bibfnamefont {H.~P.}\ \bibnamefont {Hratchian}}, \bibinfo {author}
  {\bibfnamefont {J.~V.}\ \bibnamefont {Ortiz}}, \bibinfo {author}
  {\bibfnamefont {A.~F.}\ \bibnamefont {Izmaylov}}, \bibinfo {author}
  {\bibfnamefont {J.~L.}\ \bibnamefont {Sonnenberg}}, \bibinfo {author}
  {\bibfnamefont {D.}~\bibnamefont {Williams-Young}}, \bibinfo {author}
  {\bibfnamefont {F.}~\bibnamefont {Ding}}, \bibinfo {author} {\bibfnamefont
  {F.}~\bibnamefont {Lipparini}}, \bibinfo {author} {\bibfnamefont
  {F.}~\bibnamefont {Egidi}}, \bibinfo {author} {\bibfnamefont
  {J.}~\bibnamefont {Goings}}, \bibinfo {author} {\bibfnamefont
  {B.}~\bibnamefont {Peng}}, \bibinfo {author} {\bibfnamefont {A.}~\bibnamefont
  {Petrone}}, \bibinfo {author} {\bibfnamefont {T.}~\bibnamefont {Henderson}},
  \bibinfo {author} {\bibfnamefont {D.}~\bibnamefont {Ranasinghe}}, \bibinfo
  {author} {\bibfnamefont {V.~G.}\ \bibnamefont {Zakrzewski}}, \bibinfo
  {author} {\bibfnamefont {J.}~\bibnamefont {Gao}}, \bibinfo {author}
  {\bibfnamefont {N.}~\bibnamefont {Rega}}, \bibinfo {author} {\bibfnamefont
  {G.}~\bibnamefont {Zheng}}, \bibinfo {author} {\bibfnamefont
  {W.}~\bibnamefont {Liang}}, \bibinfo {author} {\bibfnamefont
  {M.}~\bibnamefont {Hada}}, \bibinfo {author} {\bibfnamefont {M.}~\bibnamefont
  {Ehara}}, \bibinfo {author} {\bibfnamefont {K.}~\bibnamefont {Toyota}},
  \bibinfo {author} {\bibfnamefont {R.}~\bibnamefont {Fukuda}}, \bibinfo
  {author} {\bibfnamefont {J.}~\bibnamefont {Hasegawa}}, \bibinfo {author}
  {\bibfnamefont {M.}~\bibnamefont {Ishida}}, \bibinfo {author} {\bibfnamefont
  {T.}~\bibnamefont {Nakajima}}, \bibinfo {author} {\bibfnamefont
  {Y.}~\bibnamefont {Honda}}, \bibinfo {author} {\bibfnamefont
  {O.}~\bibnamefont {Kitao}}, \bibinfo {author} {\bibfnamefont
  {H.}~\bibnamefont {Nakai}}, \bibinfo {author} {\bibfnamefont
  {T.}~\bibnamefont {Vreven}}, \bibinfo {author} {\bibfnamefont
  {K.}~\bibnamefont {Throssell}}, \bibinfo {author} {\bibfnamefont {J.~A.}\
  \bibnamefont {Montgomery}, \bibfnamefont {{Jr.}}}, \bibinfo {author}
  {\bibfnamefont {J.~E.}\ \bibnamefont {Peralta}}, \bibinfo {author}
  {\bibfnamefont {F.}~\bibnamefont {Ogliaro}}, \bibinfo {author} {\bibfnamefont
  {M.~J.}\ \bibnamefont {Bearpark}}, \bibinfo {author} {\bibfnamefont {J.~J.}\
  \bibnamefont {Heyd}}, \bibinfo {author} {\bibfnamefont {E.~N.}\ \bibnamefont
  {Brothers}}, \bibinfo {author} {\bibfnamefont {K.~N.}\ \bibnamefont {Kudin}},
  \bibinfo {author} {\bibfnamefont {V.~N.}\ \bibnamefont {Staroverov}},
  \bibinfo {author} {\bibfnamefont {T.~A.}\ \bibnamefont {Keith}}, \bibinfo
  {author} {\bibfnamefont {R.}~\bibnamefont {Kobayashi}}, \bibinfo {author}
  {\bibfnamefont {J.}~\bibnamefont {Normand}}, \bibinfo {author} {\bibfnamefont
  {K.}~\bibnamefont {Raghavachari}}, \bibinfo {author} {\bibfnamefont {A.~P.}\
  \bibnamefont {Rendell}}, \bibinfo {author} {\bibfnamefont {J.~C.}\
  \bibnamefont {Burant}}, \bibinfo {author} {\bibfnamefont {S.~S.}\
  \bibnamefont {Iyengar}}, \bibinfo {author} {\bibfnamefont {J.}~\bibnamefont
  {Tomasi}}, \bibinfo {author} {\bibfnamefont {M.}~\bibnamefont {Cossi}},
  \bibinfo {author} {\bibfnamefont {J.~M.}\ \bibnamefont {Millam}}, \bibinfo
  {author} {\bibfnamefont {M.}~\bibnamefont {Klene}}, \bibinfo {author}
  {\bibfnamefont {C.}~\bibnamefont {Adamo}}, \bibinfo {author} {\bibfnamefont
  {R.}~\bibnamefont {Cammi}}, \bibinfo {author} {\bibfnamefont {J.~W.}\
  \bibnamefont {Ochterski}}, \bibinfo {author} {\bibfnamefont {R.~L.}\
  \bibnamefont {Martin}}, \bibinfo {author} {\bibfnamefont {K.}~\bibnamefont
  {Morokuma}}, \bibinfo {author} {\bibfnamefont {O.}~\bibnamefont {Farkas}},
  \bibinfo {author} {\bibfnamefont {J.~B.}\ \bibnamefont {Foresman}},\ and\
  \bibinfo {author} {\bibfnamefont {D.~J.}\ \bibnamefont {Fox}},\ }\href@noop
  {} {\bibinfo {title} {Gaussian˜16 {R}evision {C}.01}} (\bibinfo {year}
  {2016}),\ \bibinfo {note} {gaussian Inc. Wallingford CT}\BibitemShut
  {NoStop}%
\bibitem [{\citenamefont {Verstraelen}\ \emph {et~al.}(2017)\citenamefont
  {Verstraelen}, \citenamefont {Tecmer}, \citenamefont {Heidar-Zadeh},
  \citenamefont {González-Espinoza}, \citenamefont {Chan}, \citenamefont
  {Kim}, \citenamefont {Boguslawski}, \citenamefont {Fias}, \citenamefont
  {Vandenbrande}, \citenamefont {Berrocal},\ and\ \citenamefont
  {Ayers}}]{horton}%
  \BibitemOpen
  \bibfield  {author} {\bibinfo {author} {\bibfnamefont {T.}~\bibnamefont
  {Verstraelen}}, \bibinfo {author} {\bibfnamefont {P.}~\bibnamefont {Tecmer}},
  \bibinfo {author} {\bibfnamefont {F.}~\bibnamefont {Heidar-Zadeh}}, \bibinfo
  {author} {\bibfnamefont {C.~E.}\ \bibnamefont {González-Espinoza}}, \bibinfo
  {author} {\bibfnamefont {M.}~\bibnamefont {Chan}}, \bibinfo {author}
  {\bibfnamefont {T.~D.}\ \bibnamefont {Kim}}, \bibinfo {author} {\bibfnamefont
  {K.}~\bibnamefont {Boguslawski}}, \bibinfo {author} {\bibfnamefont
  {S.}~\bibnamefont {Fias}}, \bibinfo {author} {\bibfnamefont {S.}~\bibnamefont
  {Vandenbrande}}, \bibinfo {author} {\bibfnamefont {D.}~\bibnamefont
  {Berrocal}},\ and\ \bibinfo {author} {\bibfnamefont {P.~W.}\ \bibnamefont
  {Ayers}},\ }\href {http://theochem.github.com/horton/} {\bibinfo {title}
  {Horton 2.1.1}} (\bibinfo {year} {2017})\BibitemShut {NoStop}%
\bibitem [{\citenamefont {Lebedev}\ and\ \citenamefont
  {Laikov}(1999)}]{lebedev1999quadrature}%
  \BibitemOpen
  \bibfield  {author} {\bibinfo {author} {\bibfnamefont {V.~I.}\ \bibnamefont
  {Lebedev}}\ and\ \bibinfo {author} {\bibfnamefont {D.}~\bibnamefont
  {Laikov}},\ }\href@noop {} {\bibfield  {journal} {\bibinfo  {journal}
  {Doklady Mathematics}\ }\textbf {\bibinfo {volume} {59}},\ \bibinfo {pages}
  {477} (\bibinfo {year} {1999})}\BibitemShut {NoStop}%
\bibitem [{\citenamefont {Hohenberg}\ and\ \citenamefont
  {Kohn}(1964)}]{PhysRev.136.B864}%
  \BibitemOpen
  \bibfield  {author} {\bibinfo {author} {\bibfnamefont {P.}~\bibnamefont
  {Hohenberg}}\ and\ \bibinfo {author} {\bibfnamefont {W.}~\bibnamefont
  {Kohn}},\ }\href {https://doi.org/10.1103/PhysRev.136.B864} {\bibfield
  {journal} {\bibinfo  {journal} {Phys. Rev.}\ }\textbf {\bibinfo {volume}
  {136}},\ \bibinfo {pages} {B864} (\bibinfo {year} {1964})}\BibitemShut
  {NoStop}%
\bibitem [{\citenamefont {Kohn}\ and\ \citenamefont
  {Sham}(1965)}]{PhysRev.140.A1133}%
  \BibitemOpen
  \bibfield  {author} {\bibinfo {author} {\bibfnamefont {W.}~\bibnamefont
  {Kohn}}\ and\ \bibinfo {author} {\bibfnamefont {L.~J.}\ \bibnamefont
  {Sham}},\ }\href {https://doi.org/10.1103/PhysRev.140.A1133} {\bibfield
  {journal} {\bibinfo  {journal} {Phys. Rev.}\ }\textbf {\bibinfo {volume}
  {140}},\ \bibinfo {pages} {A1133} (\bibinfo {year} {1965})}\BibitemShut
  {NoStop}%
\bibitem [{\citenamefont {Slater}(1974)}]{slater1974quantum}%
  \BibitemOpen
  \bibfield  {author} {\bibinfo {author} {\bibfnamefont {J.~C.}\ \bibnamefont
  {Slater}},\ }\href@noop {} {\bibfield  {journal} {\bibinfo  {journal} {The
  self-Consistent Field for Molecular and solids}\ }\textbf {\bibinfo {volume}
  {4}} (\bibinfo {year} {1974})}\BibitemShut {NoStop}%
\bibitem [{\citenamefont {Vosko}\ \emph {et~al.}(1980)\citenamefont {Vosko},
  \citenamefont {Wilk},\ and\ \citenamefont {Nusair}}]{Vosko1980}%
  \BibitemOpen
  \bibfield  {author} {\bibinfo {author} {\bibfnamefont {S.~H.}\ \bibnamefont
  {Vosko}}, \bibinfo {author} {\bibfnamefont {L.}~\bibnamefont {Wilk}},\ and\
  \bibinfo {author} {\bibfnamefont {M.}~\bibnamefont {Nusair}},\ }\href
  {https://doi.org/10.1139/p80-159} {\bibfield  {journal} {\bibinfo  {journal}
  {Can. J. Phys.}\ }\textbf {\bibinfo {volume} {58}},\ \bibinfo {pages} {1200}
  (\bibinfo {year} {1980})}\BibitemShut {NoStop}%
\bibitem [{\citenamefont {Aidas}\ \emph {et~al.}(2014)\citenamefont {Aidas},
  \citenamefont {Angeli}, \citenamefont {Bak}, \citenamefont {Bakken},
  \citenamefont {Bast}, \citenamefont {Boman}, \citenamefont {Christiansen},
  \citenamefont {Cimiraglia}, \citenamefont {Coriani}, \citenamefont {Dahle},
  \citenamefont {Dalskov}, \citenamefont {Ekstr{\"{o}}m}, \citenamefont
  {Enevoldsen}, \citenamefont {Eriksen}, \citenamefont {Ettenhuber},
  \citenamefont {Fern{\'{a}}ndez}, \citenamefont {Ferrighi}, \citenamefont
  {Fliegl}, \citenamefont {Frediani}, \citenamefont {Hald}, \citenamefont
  {Halkier}, \citenamefont {H{\"{a}}ttig}, \citenamefont {Heiberg},
  \citenamefont {Helgaker}, \citenamefont {Hennum}, \citenamefont {Hettema},
  \citenamefont {Hjerten{\ae}s}, \citenamefont {H{\o}st}, \citenamefont
  {H{\o}yvik}, \citenamefont {Iozzi}, \citenamefont {Jans{\'{i}}k},
  \citenamefont {Jensen}, \citenamefont {Jonsson}, \citenamefont
  {J{\o}rgensen}, \citenamefont {Kauczor}, \citenamefont {Kirpekar},
  \citenamefont {Kj{\ae}rgaard}, \citenamefont {Klopper}, \citenamefont
  {Knecht}, \citenamefont {Kobayashi}, \citenamefont {Koch}, \citenamefont
  {Kongsted}, \citenamefont {Krapp}, \citenamefont {Kristensen}, \citenamefont
  {Ligabue}, \citenamefont {Lutn{\ae}s}, \citenamefont {Melo}, \citenamefont
  {Mikkelsen}, \citenamefont {Myhre}, \citenamefont {Neiss}, \citenamefont
  {Nielsen}, \citenamefont {Norman}, \citenamefont {Olsen}, \citenamefont
  {Olsen}, \citenamefont {Osted}, \citenamefont {Packer}, \citenamefont
  {Pawlowski}, \citenamefont {Pedersen}, \citenamefont {Provasi}, \citenamefont
  {Reine}, \citenamefont {Rinkevicius}, \citenamefont {Ruden}, \citenamefont
  {Ruud}, \citenamefont {Rybkin}, \citenamefont {Sa{\l}ek}, \citenamefont
  {Samson}, \citenamefont {de~Mer{\'{a}}s}, \citenamefont {Saue}, \citenamefont
  {Sauer}, \citenamefont {Schimmelpfennig}, \citenamefont {Sneskov},
  \citenamefont {Steindal}, \citenamefont {Sylvester-Hvid}, \citenamefont
  {Taylor}, \citenamefont {Teale}, \citenamefont {Tellgren}, \citenamefont
  {Tew}, \citenamefont {Thorvaldsen}, \citenamefont {Th{\o}gersen},
  \citenamefont {Vahtras}, \citenamefont {Watson}, \citenamefont {Wilson},
  \citenamefont {Ziolkowski},\ and\ \citenamefont {{\AA}gren}}]{Aidas2014}%
  \BibitemOpen
  \bibfield  {author} {\bibinfo {author} {\bibfnamefont {K.}~\bibnamefont
  {Aidas}}, \bibinfo {author} {\bibfnamefont {C.}~\bibnamefont {Angeli}},
  \bibinfo {author} {\bibfnamefont {K.~L.}\ \bibnamefont {Bak}}, \bibinfo
  {author} {\bibfnamefont {V.}~\bibnamefont {Bakken}}, \bibinfo {author}
  {\bibfnamefont {R.}~\bibnamefont {Bast}}, \bibinfo {author} {\bibfnamefont
  {L.}~\bibnamefont {Boman}}, \bibinfo {author} {\bibfnamefont
  {O.}~\bibnamefont {Christiansen}}, \bibinfo {author} {\bibfnamefont
  {R.}~\bibnamefont {Cimiraglia}}, \bibinfo {author} {\bibfnamefont
  {S.}~\bibnamefont {Coriani}}, \bibinfo {author} {\bibfnamefont
  {P.}~\bibnamefont {Dahle}}, \bibinfo {author} {\bibfnamefont {E.~K.}\
  \bibnamefont {Dalskov}}, \bibinfo {author} {\bibfnamefont {U.}~\bibnamefont
  {Ekstr{\"{o}}m}}, \bibinfo {author} {\bibfnamefont {T.}~\bibnamefont
  {Enevoldsen}}, \bibinfo {author} {\bibfnamefont {J.~J.}\ \bibnamefont
  {Eriksen}}, \bibinfo {author} {\bibfnamefont {P.}~\bibnamefont {Ettenhuber}},
  \bibinfo {author} {\bibfnamefont {B.}~\bibnamefont {Fern{\'{a}}ndez}},
  \bibinfo {author} {\bibfnamefont {L.}~\bibnamefont {Ferrighi}}, \bibinfo
  {author} {\bibfnamefont {H.}~\bibnamefont {Fliegl}}, \bibinfo {author}
  {\bibfnamefont {L.}~\bibnamefont {Frediani}}, \bibinfo {author}
  {\bibfnamefont {K.}~\bibnamefont {Hald}}, \bibinfo {author} {\bibfnamefont
  {A.}~\bibnamefont {Halkier}}, \bibinfo {author} {\bibfnamefont
  {C.}~\bibnamefont {H{\"{a}}ttig}}, \bibinfo {author} {\bibfnamefont
  {H.}~\bibnamefont {Heiberg}}, \bibinfo {author} {\bibfnamefont
  {T.}~\bibnamefont {Helgaker}}, \bibinfo {author} {\bibfnamefont {A.~C.}\
  \bibnamefont {Hennum}}, \bibinfo {author} {\bibfnamefont {H.}~\bibnamefont
  {Hettema}}, \bibinfo {author} {\bibfnamefont {E.}~\bibnamefont
  {Hjerten{\ae}s}}, \bibinfo {author} {\bibfnamefont {S.}~\bibnamefont
  {H{\o}st}}, \bibinfo {author} {\bibfnamefont {I.~M.}\ \bibnamefont
  {H{\o}yvik}}, \bibinfo {author} {\bibfnamefont {M.~F.}\ \bibnamefont
  {Iozzi}}, \bibinfo {author} {\bibfnamefont {B.}~\bibnamefont {Jans{\'{i}}k}},
  \bibinfo {author} {\bibfnamefont {H.~J.~A.}\ \bibnamefont {Jensen}}, \bibinfo
  {author} {\bibfnamefont {D.}~\bibnamefont {Jonsson}}, \bibinfo {author}
  {\bibfnamefont {P.}~\bibnamefont {J{\o}rgensen}}, \bibinfo {author}
  {\bibfnamefont {J.}~\bibnamefont {Kauczor}}, \bibinfo {author} {\bibfnamefont
  {S.}~\bibnamefont {Kirpekar}}, \bibinfo {author} {\bibfnamefont
  {T.}~\bibnamefont {Kj{\ae}rgaard}}, \bibinfo {author} {\bibfnamefont
  {W.}~\bibnamefont {Klopper}}, \bibinfo {author} {\bibfnamefont
  {S.}~\bibnamefont {Knecht}}, \bibinfo {author} {\bibfnamefont
  {R.}~\bibnamefont {Kobayashi}}, \bibinfo {author} {\bibfnamefont
  {H.}~\bibnamefont {Koch}}, \bibinfo {author} {\bibfnamefont {J.}~\bibnamefont
  {Kongsted}}, \bibinfo {author} {\bibfnamefont {A.}~\bibnamefont {Krapp}},
  \bibinfo {author} {\bibfnamefont {K.}~\bibnamefont {Kristensen}}, \bibinfo
  {author} {\bibfnamefont {A.}~\bibnamefont {Ligabue}}, \bibinfo {author}
  {\bibfnamefont {O.~B.}\ \bibnamefont {Lutn{\ae}s}}, \bibinfo {author}
  {\bibfnamefont {J.~I.}\ \bibnamefont {Melo}}, \bibinfo {author}
  {\bibfnamefont {K.~V.}\ \bibnamefont {Mikkelsen}}, \bibinfo {author}
  {\bibfnamefont {R.~H.}\ \bibnamefont {Myhre}}, \bibinfo {author}
  {\bibfnamefont {C.}~\bibnamefont {Neiss}}, \bibinfo {author} {\bibfnamefont
  {C.~B.}\ \bibnamefont {Nielsen}}, \bibinfo {author} {\bibfnamefont
  {P.}~\bibnamefont {Norman}}, \bibinfo {author} {\bibfnamefont
  {J.}~\bibnamefont {Olsen}}, \bibinfo {author} {\bibfnamefont {J.~M.~H.}\
  \bibnamefont {Olsen}}, \bibinfo {author} {\bibfnamefont {A.}~\bibnamefont
  {Osted}}, \bibinfo {author} {\bibfnamefont {M.~J.}\ \bibnamefont {Packer}},
  \bibinfo {author} {\bibfnamefont {F.}~\bibnamefont {Pawlowski}}, \bibinfo
  {author} {\bibfnamefont {T.~B.}\ \bibnamefont {Pedersen}}, \bibinfo {author}
  {\bibfnamefont {P.~F.}\ \bibnamefont {Provasi}}, \bibinfo {author}
  {\bibfnamefont {S.}~\bibnamefont {Reine}}, \bibinfo {author} {\bibfnamefont
  {Z.}~\bibnamefont {Rinkevicius}}, \bibinfo {author} {\bibfnamefont {T.~A.}\
  \bibnamefont {Ruden}}, \bibinfo {author} {\bibfnamefont {K.}~\bibnamefont
  {Ruud}}, \bibinfo {author} {\bibfnamefont {V.~V.}\ \bibnamefont {Rybkin}},
  \bibinfo {author} {\bibfnamefont {P.}~\bibnamefont {Sa{\l}ek}}, \bibinfo
  {author} {\bibfnamefont {C.~C.}\ \bibnamefont {Samson}}, \bibinfo {author}
  {\bibfnamefont {A.~S.}\ \bibnamefont {de~Mer{\'{a}}s}}, \bibinfo {author}
  {\bibfnamefont {T.}~\bibnamefont {Saue}}, \bibinfo {author} {\bibfnamefont
  {S.~P.}\ \bibnamefont {Sauer}}, \bibinfo {author} {\bibfnamefont
  {B.}~\bibnamefont {Schimmelpfennig}}, \bibinfo {author} {\bibfnamefont
  {K.}~\bibnamefont {Sneskov}}, \bibinfo {author} {\bibfnamefont {A.~H.}\
  \bibnamefont {Steindal}}, \bibinfo {author} {\bibfnamefont {K.~O.}\
  \bibnamefont {Sylvester-Hvid}}, \bibinfo {author} {\bibfnamefont {P.~R.}\
  \bibnamefont {Taylor}}, \bibinfo {author} {\bibfnamefont {A.~M.}\
  \bibnamefont {Teale}}, \bibinfo {author} {\bibfnamefont {E.~I.}\ \bibnamefont
  {Tellgren}}, \bibinfo {author} {\bibfnamefont {D.~P.}\ \bibnamefont {Tew}},
  \bibinfo {author} {\bibfnamefont {A.~J.}\ \bibnamefont {Thorvaldsen}},
  \bibinfo {author} {\bibfnamefont {L.}~\bibnamefont {Th{\o}gersen}}, \bibinfo
  {author} {\bibfnamefont {O.}~\bibnamefont {Vahtras}}, \bibinfo {author}
  {\bibfnamefont {M.~A.}\ \bibnamefont {Watson}}, \bibinfo {author}
  {\bibfnamefont {D.~J.}\ \bibnamefont {Wilson}}, \bibinfo {author}
  {\bibfnamefont {M.}~\bibnamefont {Ziolkowski}},\ and\ \bibinfo {author}
  {\bibfnamefont {H.}~\bibnamefont {{\AA}gren}},\ }\href
  {https://doi.org/10.1002/wcms.1172} {\bibfield  {journal} {\bibinfo
  {journal} {Wiley Interdisciplinary Reviews: Computational Molecular Science}\
  }\textbf {\bibinfo {volume} {4}},\ \bibinfo {pages} {269} (\bibinfo {year}
  {2014})}\BibitemShut {NoStop}%
\bibitem [{\citenamefont {J{\o}rgensen}\ \emph {et~al.}(1988)\citenamefont
  {J{\o}rgensen}, \citenamefont {Jensen},\ and\ \citenamefont
  {Olsen}}]{Jorgensen1988}%
  \BibitemOpen
  \bibfield  {author} {\bibinfo {author} {\bibfnamefont {P.}~\bibnamefont
  {J{\o}rgensen}}, \bibinfo {author} {\bibfnamefont {H.~J.~A.}\ \bibnamefont
  {Jensen}},\ and\ \bibinfo {author} {\bibfnamefont {J.}~\bibnamefont
  {Olsen}},\ }\href {https://doi.org/10.1063/1.454885} {\bibfield  {journal}
  {\bibinfo  {journal} {J. Chem. Phys.}\ }\textbf {\bibinfo {volume} {89}},\
  \bibinfo {pages} {3654} (\bibinfo {year} {1988})}\BibitemShut {NoStop}%
\bibitem [{\citenamefont {Olsen}\ \emph {et~al.}(1989)\citenamefont {Olsen},
  \citenamefont {Yeager},\ and\ \citenamefont {J{\o}rgensen}}]{Olsen1989}%
  \BibitemOpen
  \bibfield  {author} {\bibinfo {author} {\bibfnamefont {J.}~\bibnamefont
  {Olsen}}, \bibinfo {author} {\bibfnamefont {D.~L.}\ \bibnamefont {Yeager}},\
  and\ \bibinfo {author} {\bibfnamefont {P.}~\bibnamefont {J{\o}rgensen}},\
  }\href {https://doi.org/10.1063/1.457471} {\bibfield  {journal} {\bibinfo
  {journal} {J. Chem. Phys.}\ }\textbf {\bibinfo {volume} {91}},\ \bibinfo
  {pages} {381} (\bibinfo {year} {1989})}\BibitemShut {NoStop}%
\bibitem [{\citenamefont {Limaye}\ and\ \citenamefont
  {Gadre}(1994)}]{Limaye1994}%
  \BibitemOpen
  \bibfield  {author} {\bibinfo {author} {\bibfnamefont {A.~C.}\ \bibnamefont
  {Limaye}}\ and\ \bibinfo {author} {\bibfnamefont {S.~R.}\ \bibnamefont
  {Gadre}},\ }\href {https://doi.org/10.1063/1.466659} {\bibfield  {journal}
  {\bibinfo  {journal} {J. Chem. Phys.}\ }\textbf {\bibinfo {volume} {100}},\
  \bibinfo {pages} {1303} (\bibinfo {year} {1994})}\BibitemShut {NoStop}%
\bibitem [{\citenamefont {Tang}\ and\ \citenamefont
  {Edmiston}(1970)}]{Tang1970}%
  \BibitemOpen
  \bibfield  {author} {\bibinfo {author} {\bibfnamefont {K.~C.}\ \bibnamefont
  {Tang}}\ and\ \bibinfo {author} {\bibfnamefont {C.}~\bibnamefont
  {Edmiston}},\ }\href {https://doi.org/10.1063/1.1673090} {\bibfield
  {journal} {\bibinfo  {journal} {J. Chem. Phys.}\ }\textbf {\bibinfo {volume}
  {52}},\ \bibinfo {pages} {997} (\bibinfo {year} {1970})}\BibitemShut
  {NoStop}%
\bibitem [{\citenamefont {Bender}(1972)}]{Bender1972}%
  \BibitemOpen
  \bibfield  {author} {\bibinfo {author} {\bibfnamefont {C.}~\bibnamefont
  {Bender}},\ }\href {https://doi.org/10.1016/0021-9991(72)90010-1} {\bibfield
  {journal} {\bibinfo  {journal} {J. Comput. Phys.}\ }\textbf {\bibinfo
  {volume} {9}},\ \bibinfo {pages} {547} (\bibinfo {year} {1972})}\BibitemShut
  {NoStop}%
\bibitem [{\citenamefont {Sherrill}(2010)}]{Sherrill2010}%
  \BibitemOpen
  \bibfield  {author} {\bibinfo {author} {\bibfnamefont {C.~D.}\ \bibnamefont
  {Sherrill}},\ }\href@noop {} {\bibfield  {journal} {\bibinfo  {journal}
  {Notes}\ }\textbf {\bibinfo {volume} {12}},\ \bibinfo {pages} {1} (\bibinfo
  {year} {2010})}\BibitemShut {NoStop}%
\bibitem [{\citenamefont {Casida}(1995)}]{CASIDA1995}%
  \BibitemOpen
  \bibfield  {author} {\bibinfo {author} {\bibfnamefont {M.~E.}\ \bibnamefont
  {Casida}},\ }in\ \href {https://doi.org/10.1142/9789812830586_0005} {\emph
  {\bibinfo {booktitle} {Recent Advances in Density Functional Methods}}},\
  \bibinfo {editor} {edited by\ \bibinfo {editor} {\bibfnamefont {D.~P.}\
  \bibnamefont {Chong}}}\ (\bibinfo  {publisher} {World Scientific},\ \bibinfo
  {year} {1995})\ pp.\ \bibinfo {pages} {155--192}\BibitemShut {NoStop}%
\end{thebibliography}%


\begin{thebibliography}{0}%
\makeatletter
\providecommand \@ifxundefined [1]{%
 \@ifx{#1\undefined}
}%
\providecommand \@ifnum [1]{%
 \ifnum #1\expandafter \@firstoftwo
 \else \expandafter \@secondoftwo
 \fi
}%
\providecommand \@ifx [1]{%
 \ifx #1\expandafter \@firstoftwo
 \else \expandafter \@secondoftwo
 \fi
}%
\providecommand \natexlab [1]{#1}%
\providecommand \enquote  [1]{``#1''}%
\providecommand \bibnamefont  [1]{#1}%
\providecommand \bibfnamefont [1]{#1}%
\providecommand \citenamefont [1]{#1}%
\providecommand \href@noop [0]{\@secondoftwo}%
\providecommand \href [0]{\begingroup \@sanitize@url \@href}%
\providecommand \@href[1]{\@@startlink{#1}\@@href}%
\providecommand \@@href[1]{\endgroup#1\@@endlink}%
\providecommand \@sanitize@url [0]{\catcode `\\12\catcode `\$12\catcode
  `\&12\catcode `\#12\catcode `\^12\catcode `\_12\catcode `\%12\relax}%
\providecommand \@@startlink[1]{}%
\providecommand \@@endlink[0]{}%
\providecommand \url  [0]{\begingroup\@sanitize@url \@url }%
\providecommand \@url [1]{\endgroup\@href {#1}{\urlprefix }}%
\providecommand \urlprefix  [0]{URL }%
\providecommand \Eprint [0]{\href }%
\providecommand \doibase [0]{https://doi.org/}%
\providecommand \selectlanguage [0]{\@gobble}%
\providecommand \bibinfo  [0]{\@secondoftwo}%
\providecommand \bibfield  [0]{\@secondoftwo}%
\providecommand \translation [1]{[#1]}%
\providecommand \BibitemOpen [0]{}%
\providecommand \bibitemStop [0]{}%
\providecommand \bibitemNoStop [0]{.\EOS\space}%
\providecommand \EOS [0]{\spacefactor3000\relax}%
\providecommand \BibitemShut  [1]{\csname bibitem#1\endcsname}%
\let\auto@bib@innerbib\@empty
\end{thebibliography}%

\end{document}


\title{Supplementary Material for ``A new framework for frequency-dependent polarizable force fields''}

\date{\today}

\author{YingXing Cheng}
\author{Toon Verstraelen}\email{toon.verstraelen@ugent.be}
\affiliation{Center for Molecular Modeling (CMM), Ghent University - Technologiepark-Zwijnaarde 46, B-9052 Gent, Belgium}

\maketitle

\section{Additional display items}

\begin{figure}[htbp]
\centering
    \includegraphics[scale=0.9]{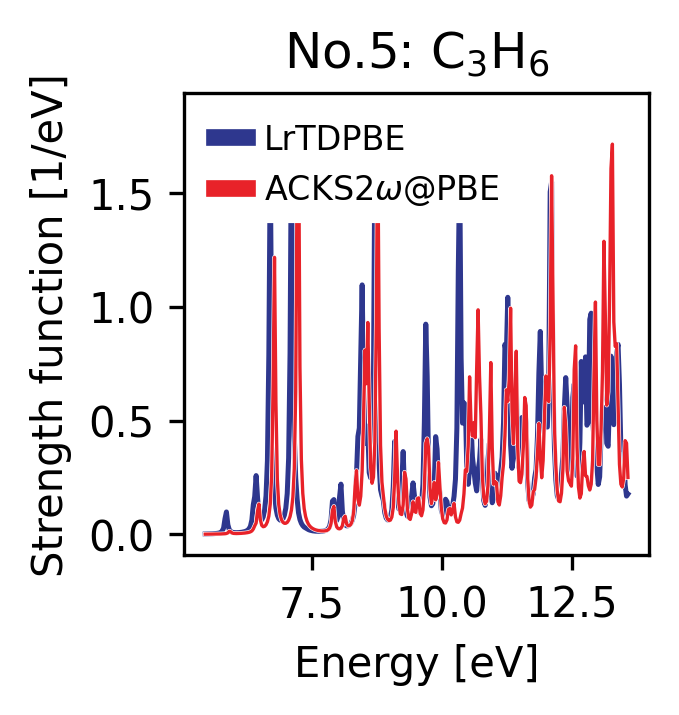}
    \includegraphics[scale=0.9]{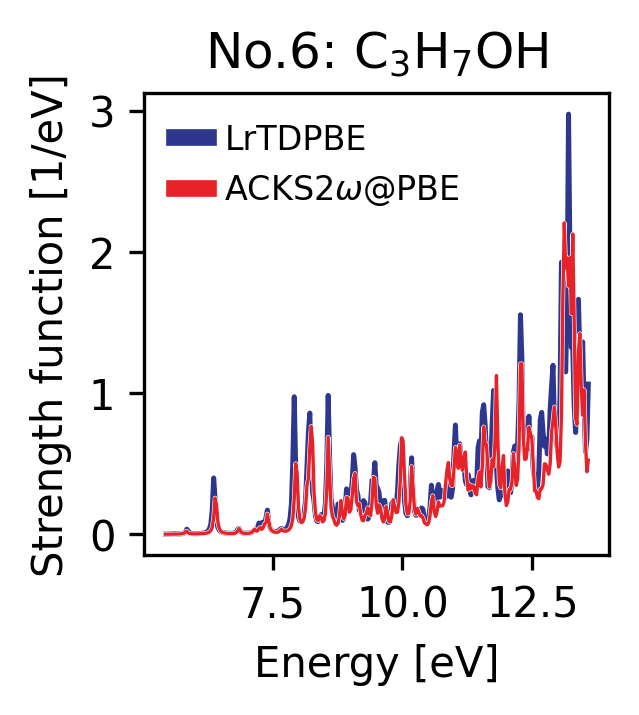}
    \includegraphics[scale=0.9]{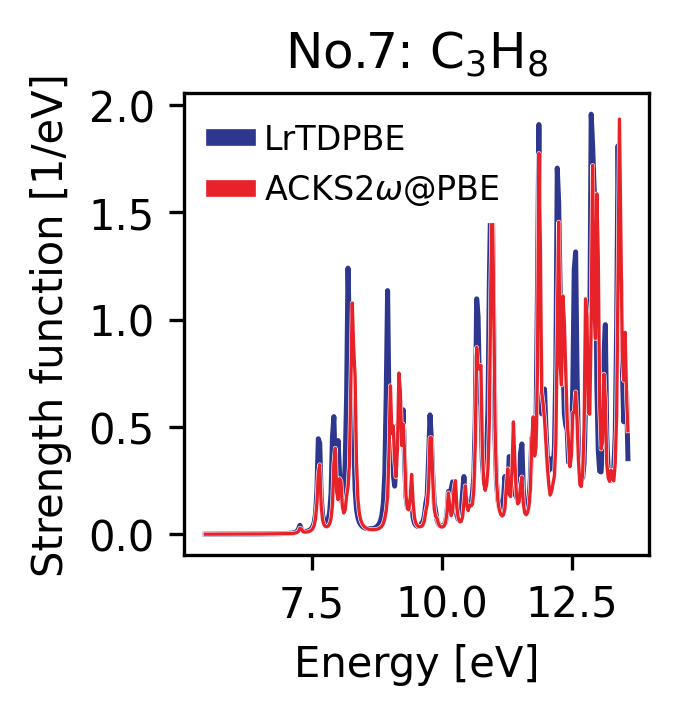}
    \\
    \includegraphics[scale=0.9]{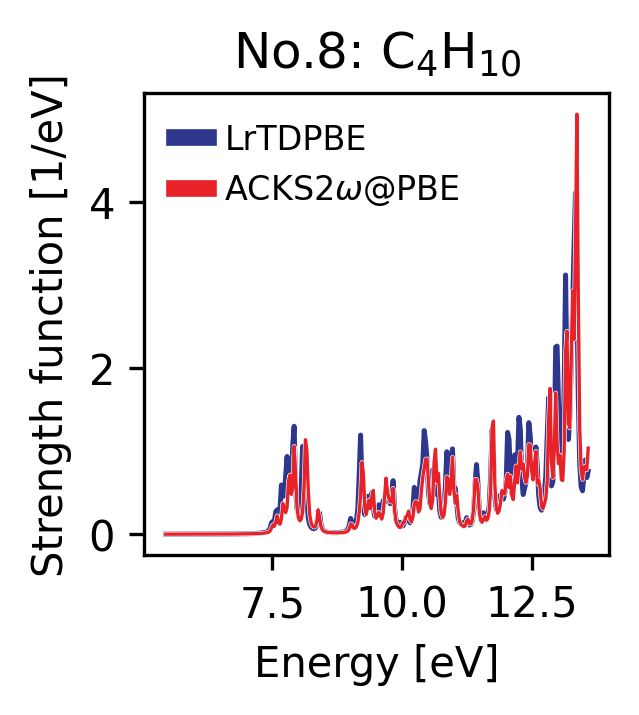}
    \includegraphics[scale=0.9]{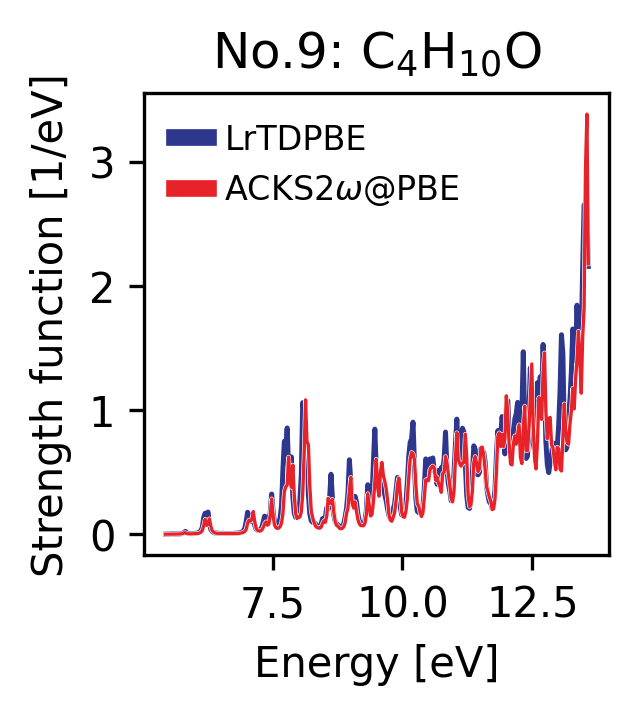}
    \includegraphics[scale=0.9]{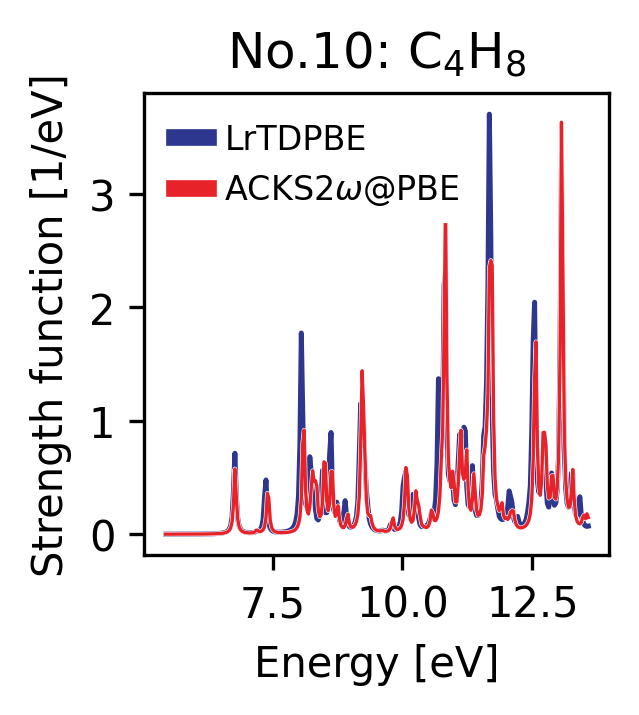}
    \\
    \includegraphics[scale=0.9]{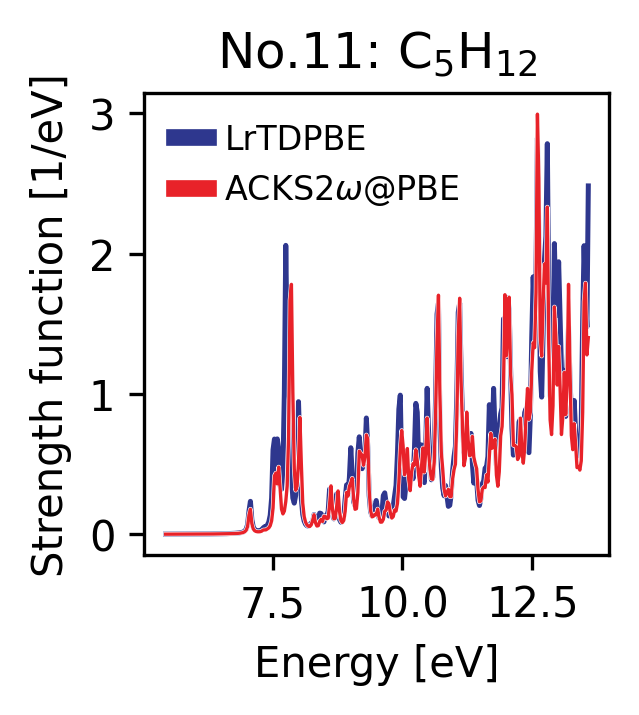}
    \includegraphics[scale=0.9]{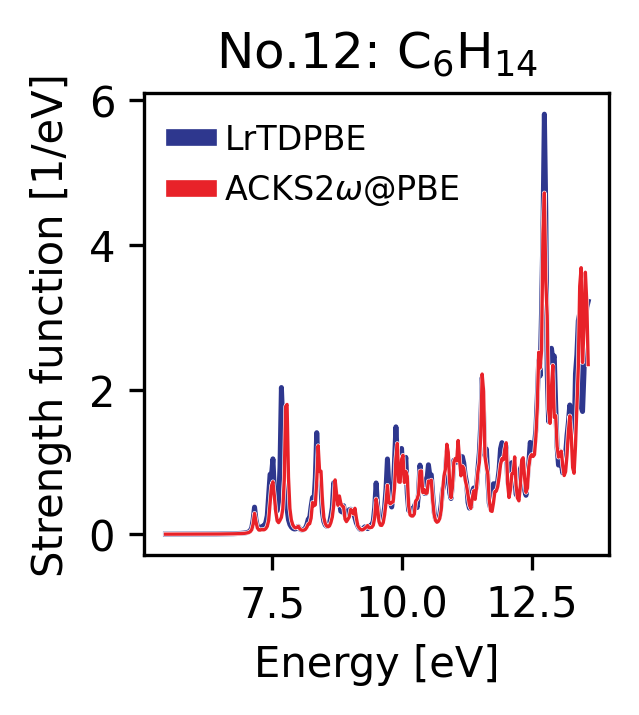}
    \includegraphics[scale=0.9]{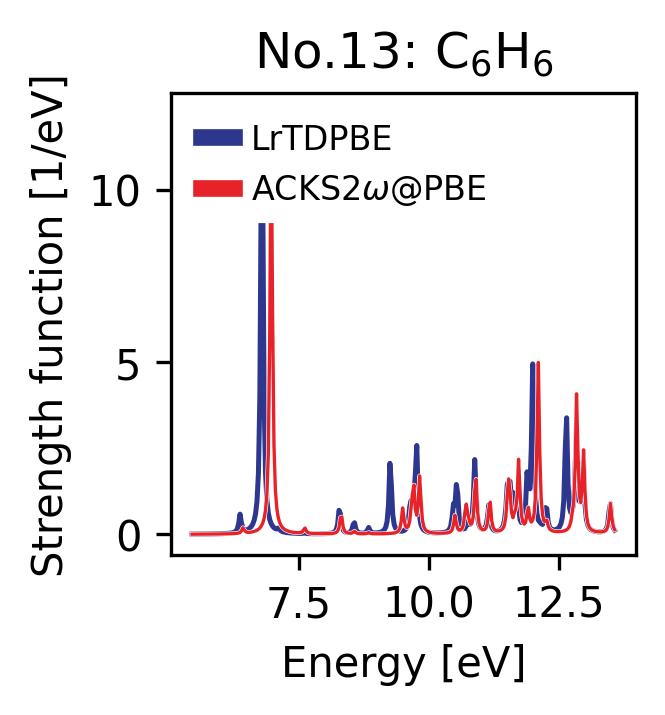}
    \caption{
    Comparison of spectra calculated by ACKS2$\omega$@PBE and LrTDPBE methods for \ce{C3H6}, \ce{C3H7OH}, \ce{C3H8}, \ce{C4H10}, \ce{C4H10O}, \ce{C4H8}, \ce{C5H12}, \ce{C6H14}, and \ce{C6H6}.
    }
\end{figure}

\begin{figure}[htbp]
\centering

    \includegraphics[scale=0.9]{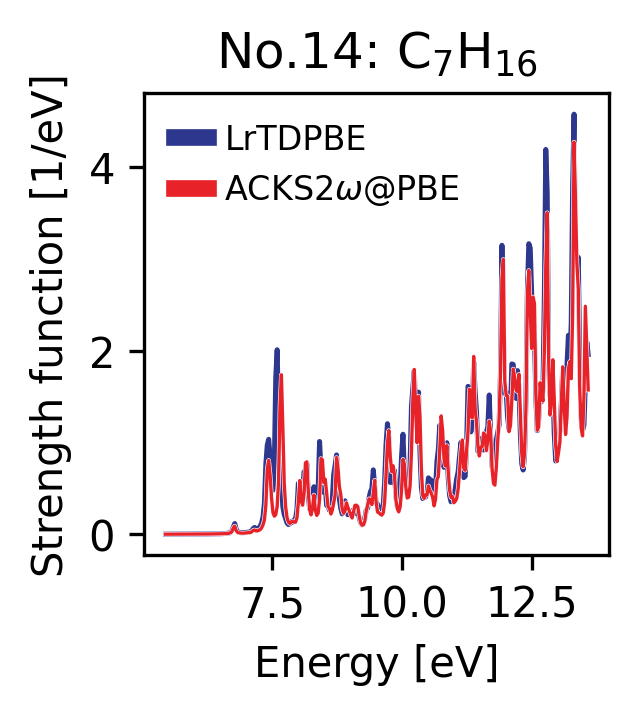}
    \includegraphics[scale=0.9]{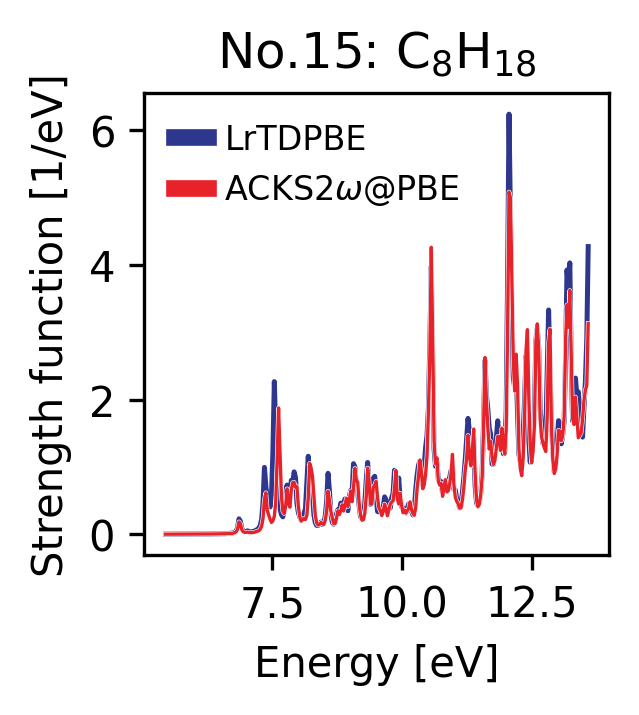}
    \includegraphics[scale=0.9]{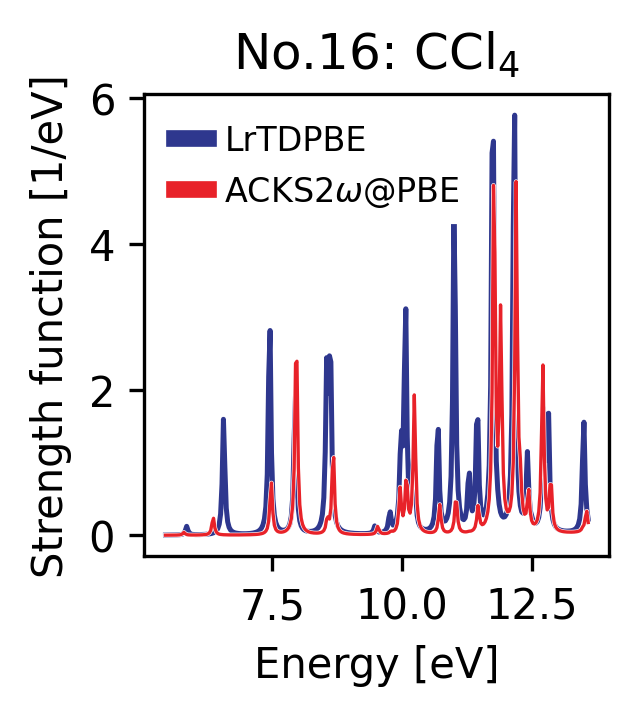}
    \\
    \includegraphics[scale=0.9]{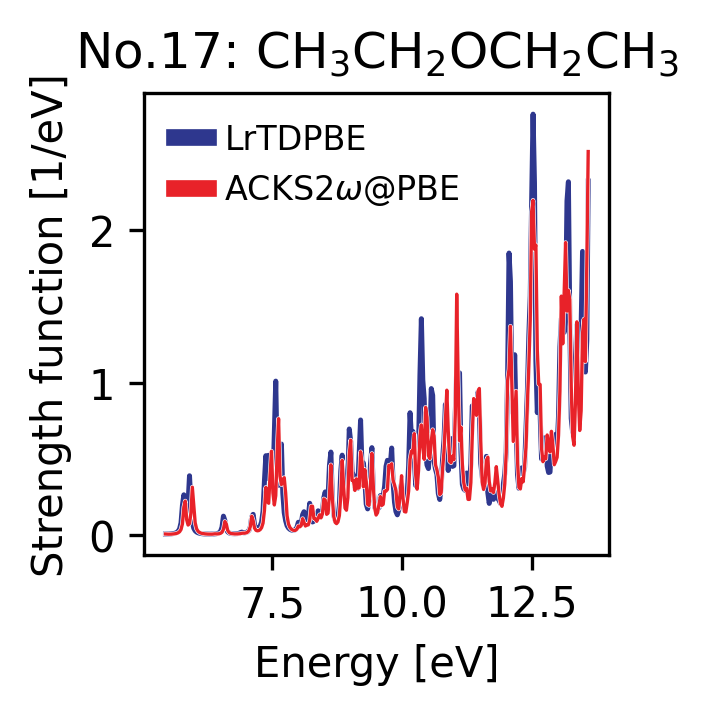}
    \includegraphics[scale=0.9]{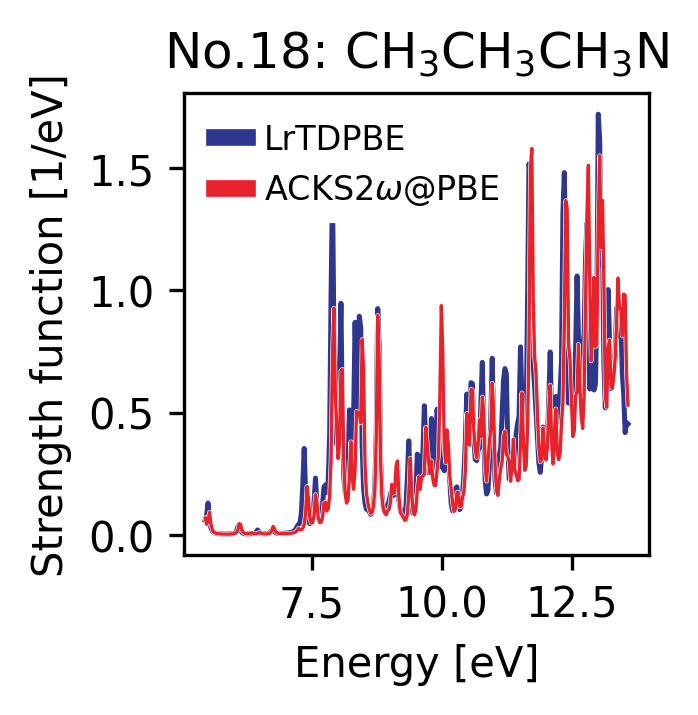}
    \includegraphics[scale=0.9]{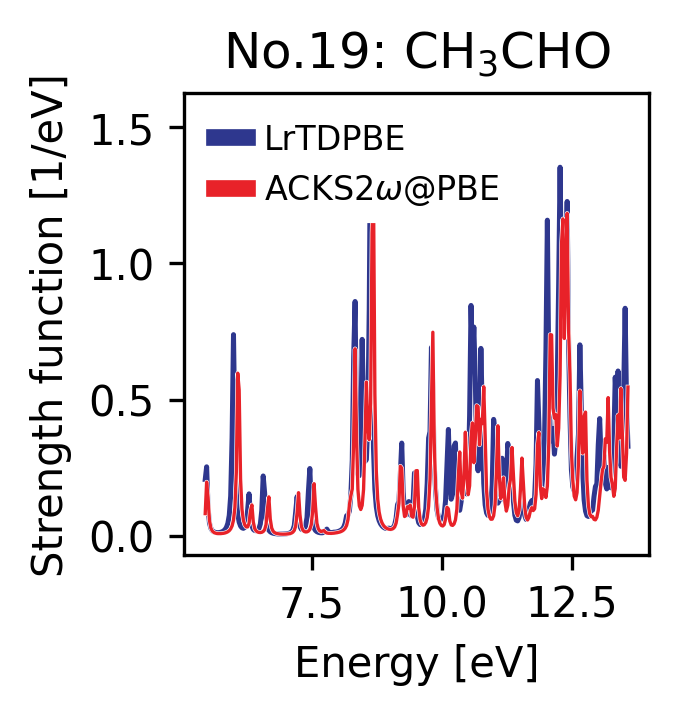}
    \\
    \includegraphics[scale=0.9]{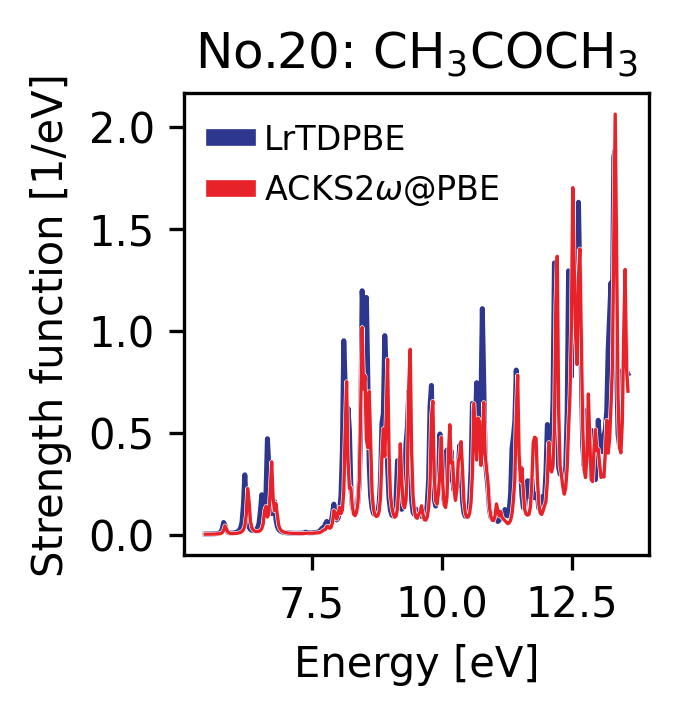}
    \includegraphics[scale=0.9]{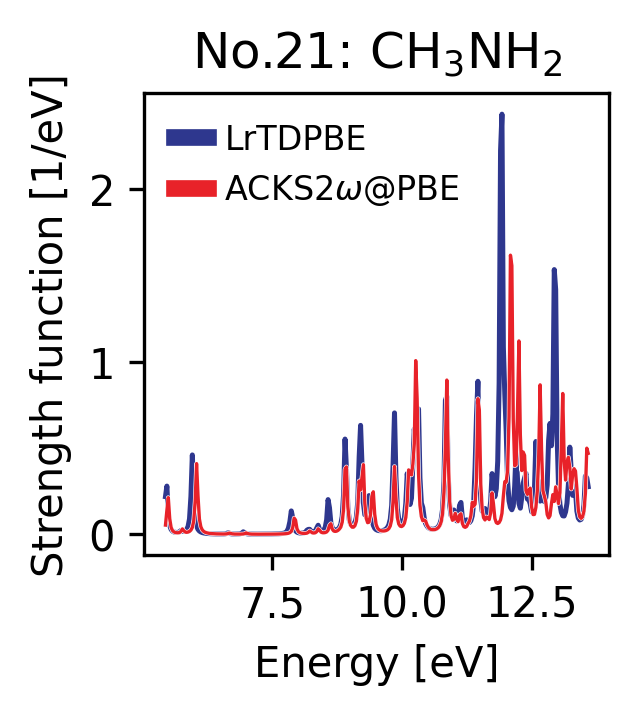}
    \includegraphics[scale=0.9]{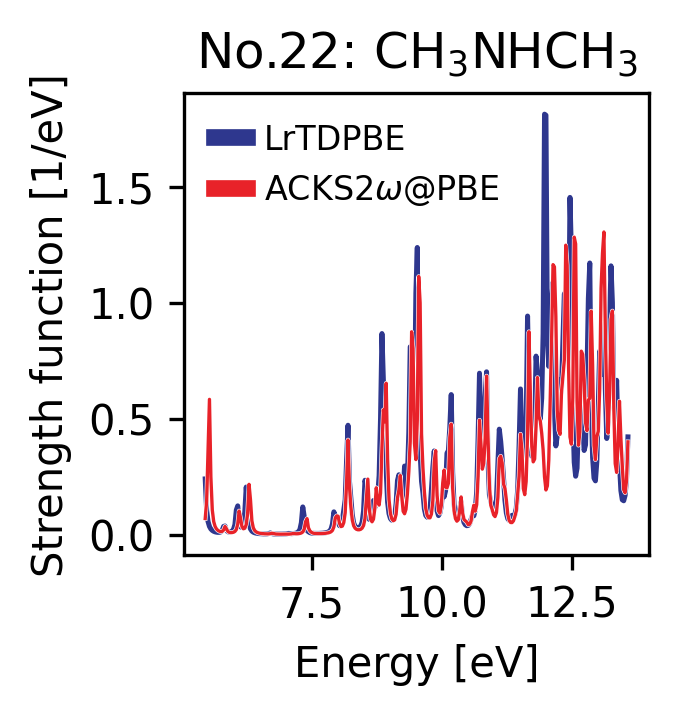}
    \caption{
    Comparison of spectra calculated by ACKS2$\omega$@PBE and LrTDPBE methods for \ce{C7H16}, \ce{C8H18}, \ce{CCl4}, \ce{CH3CH2OCH2CH3}, \ce{CH3CH3CH3N}, \ce{CH3CHO}, \ce{CH3COCH3}, \ce{CH3NH2}, and \ce{CH3NHCH3}.
    }
\end{figure}

\begin{figure}[htbp]
\centering
    \includegraphics[scale=0.9]{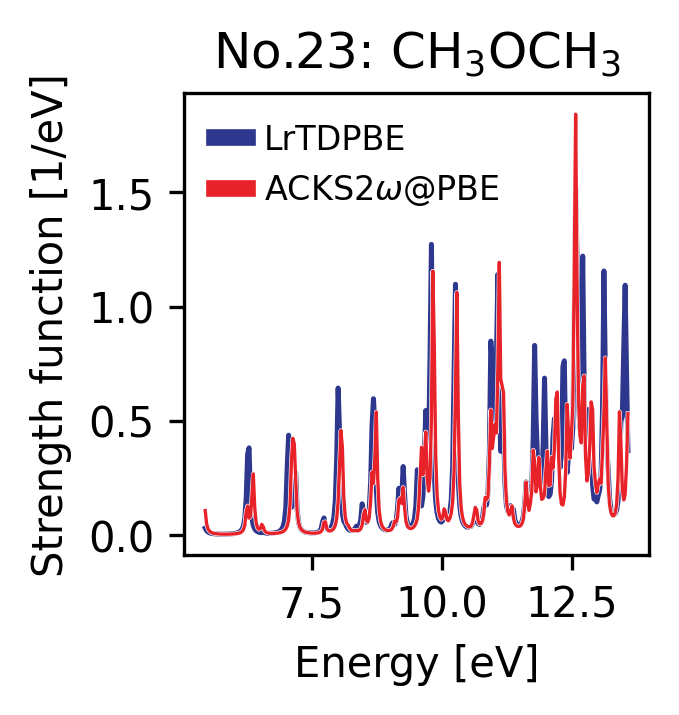}
    \includegraphics[scale=0.9]{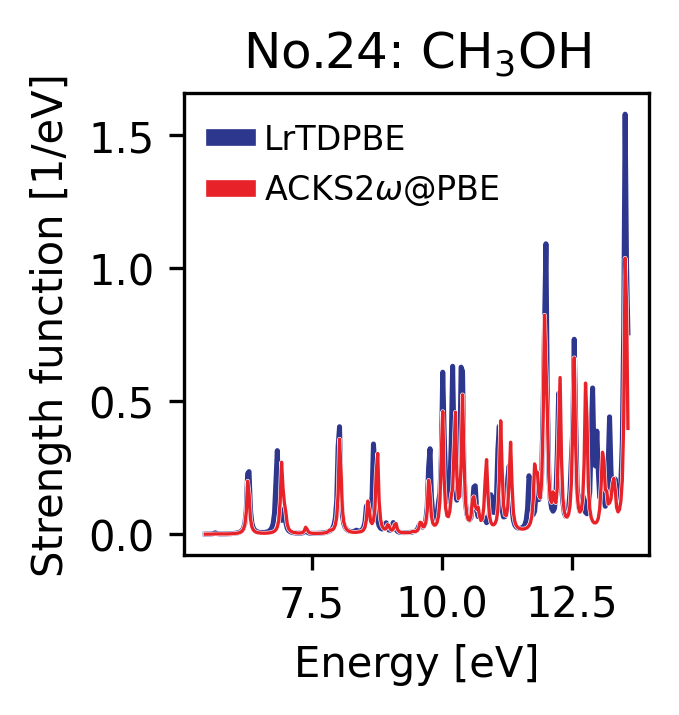}
    \includegraphics[scale=0.9]{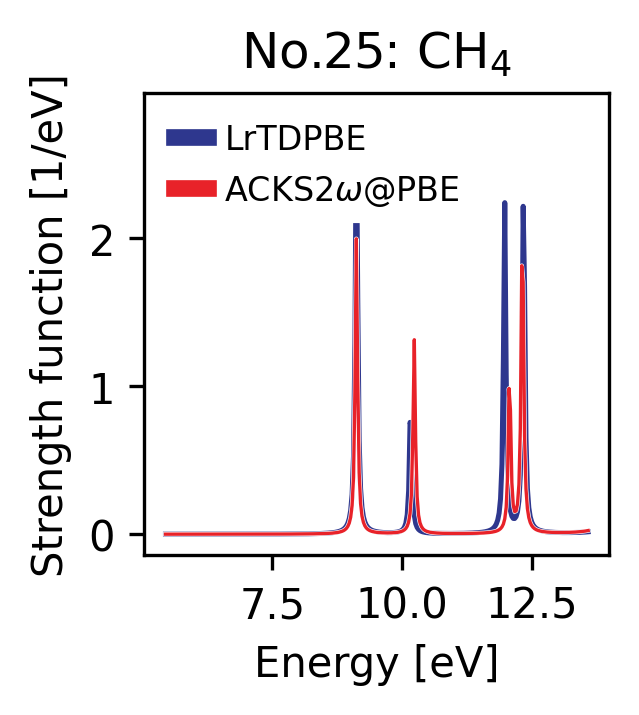}\\
    \includegraphics[scale=0.9]{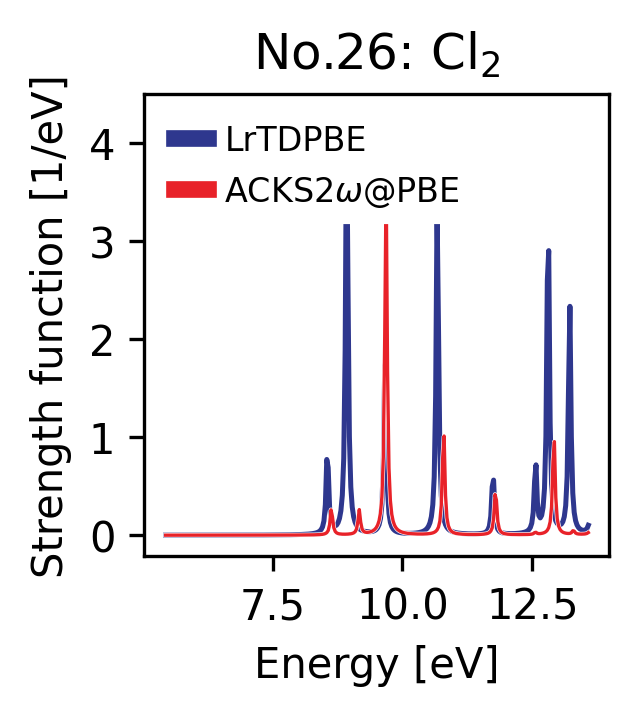}
    \includegraphics[scale=0.9]{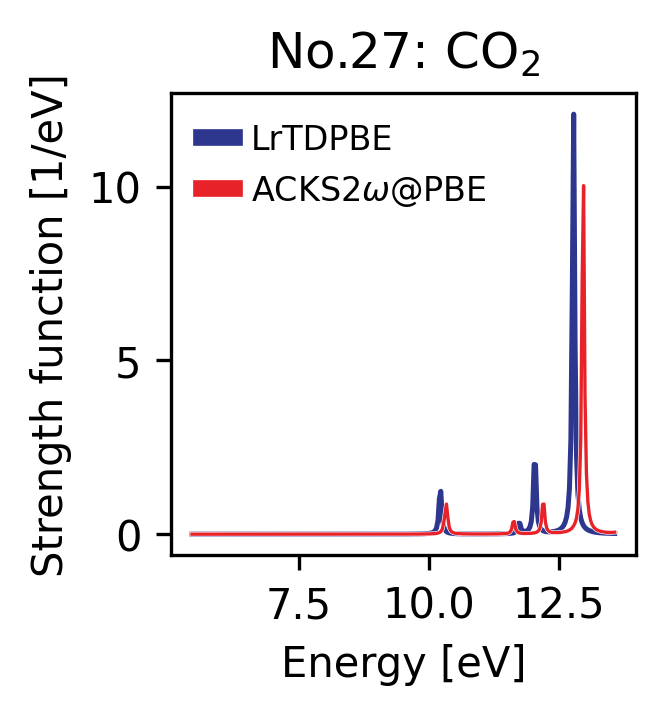}
    \includegraphics[scale=0.9]{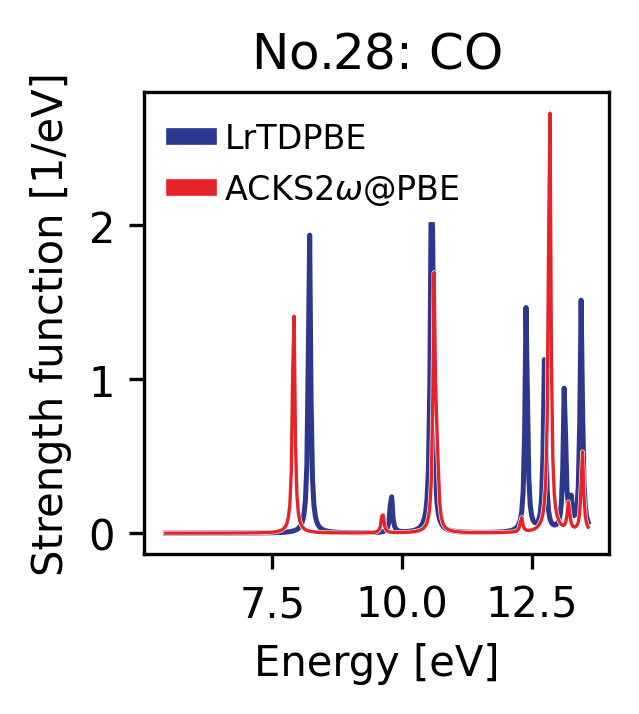}
    \\
    \includegraphics[scale=0.9]{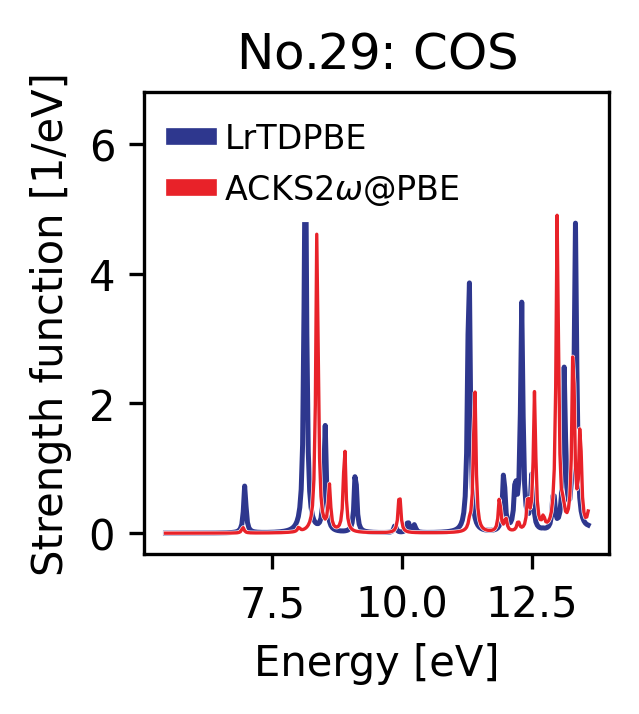}
    \includegraphics[scale=0.9]{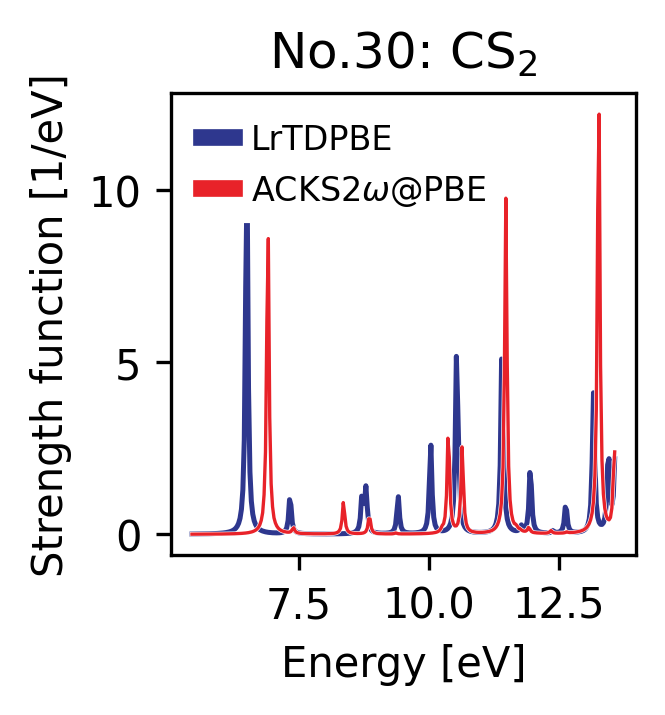}
    \includegraphics[scale=0.9]{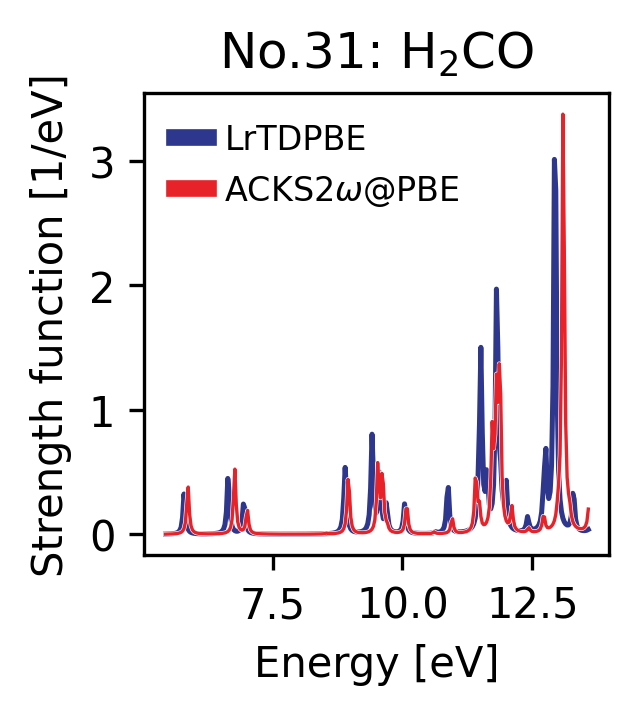}    
    \caption{
    Comparison of spectra calculated by ACKS2$\omega$@PBE and LrTDPBE methods for \ce{CH3OCH3},\ce{CH3OH}, \ce{CH4}, \ce{Cl2}, \ce{CO2}, \ce{CO}, \ce{COS}, \ce{CS2}, and \ce{H2CO}.
    }
\end{figure}

\begin{figure}[htbp]
\centering
    \includegraphics[scale=0.9]{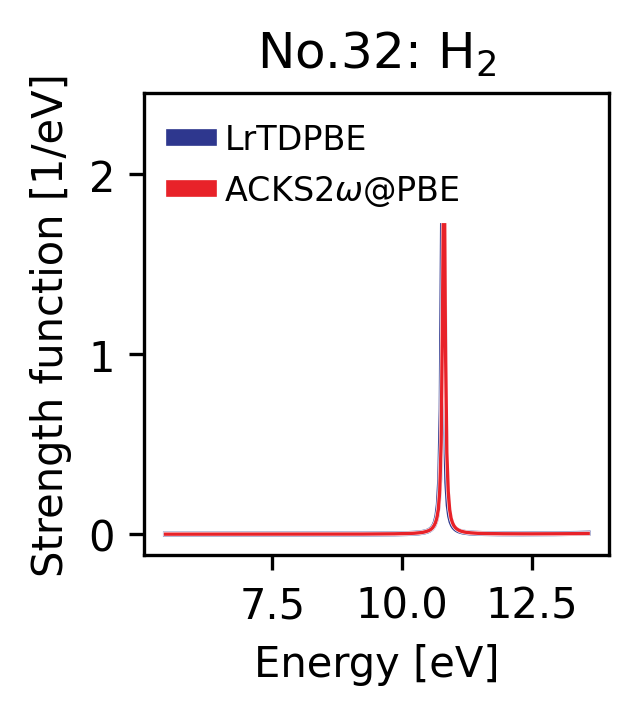}
    \includegraphics[scale=0.9]{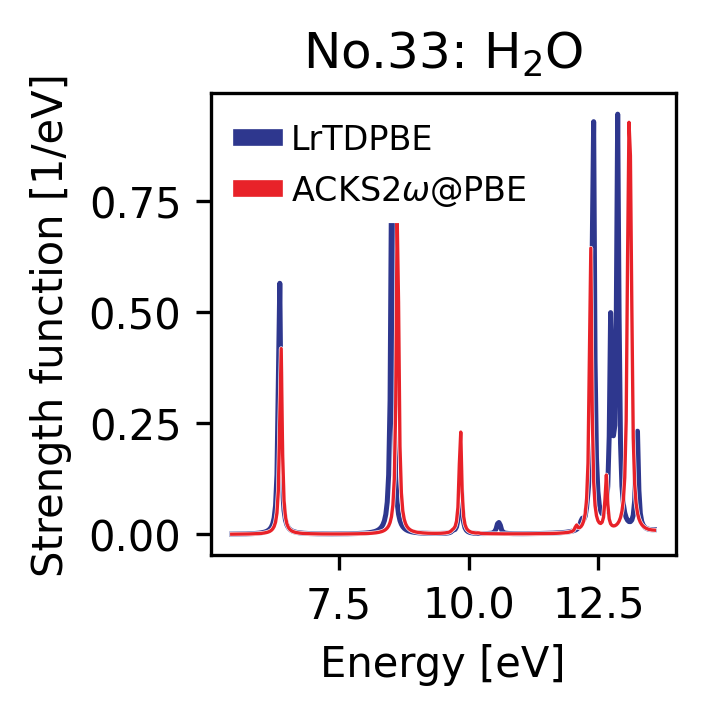}
    \includegraphics[scale=0.9]{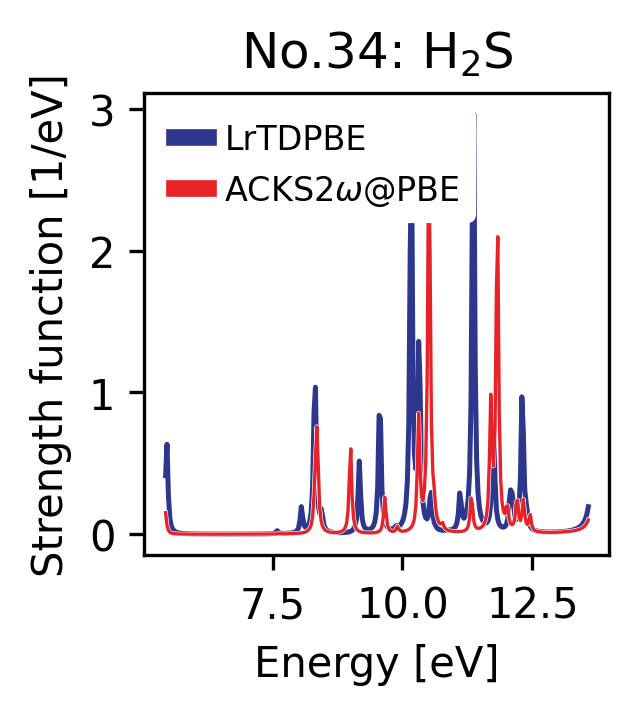}
    \\
    \includegraphics[scale=0.9]{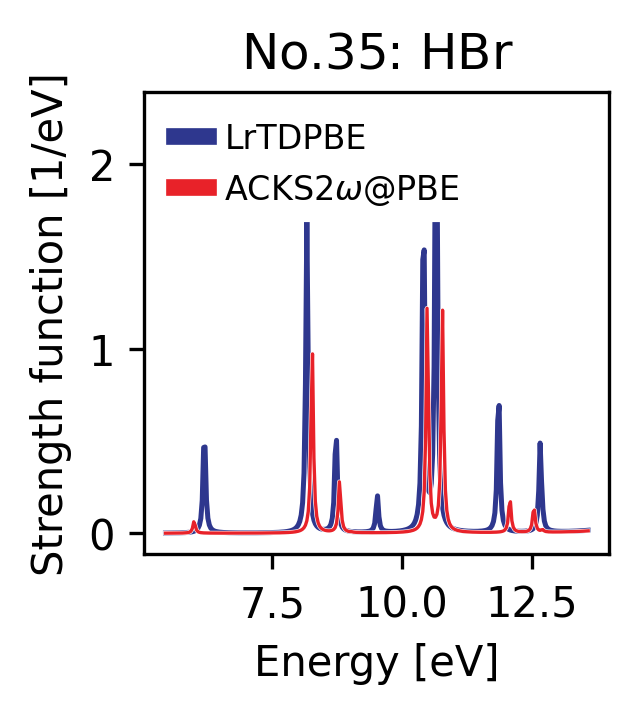}
    \includegraphics[scale=0.9]{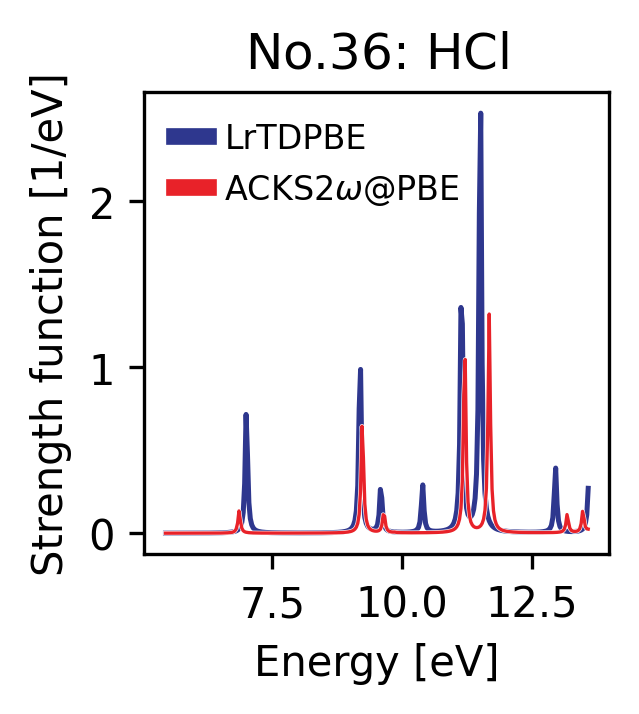}
    \includegraphics[scale=0.9]{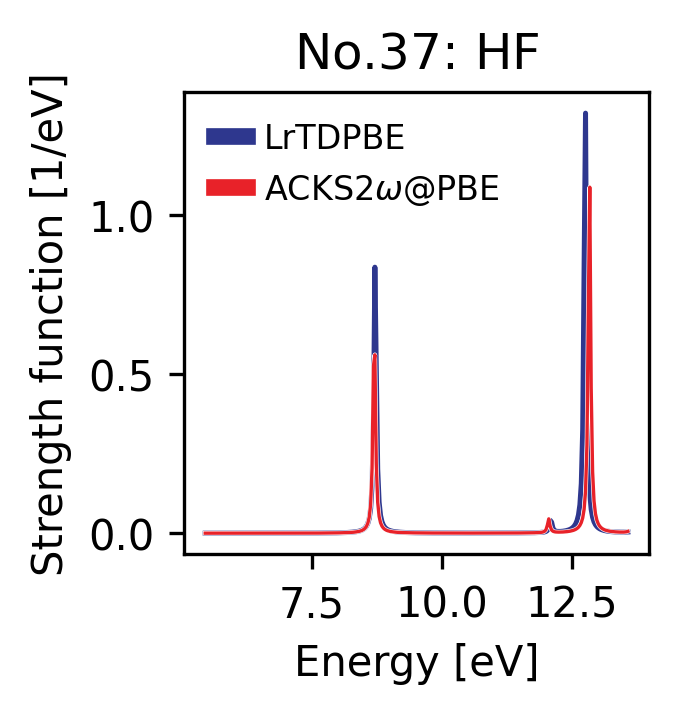}
    \\
    \includegraphics[scale=0.9]{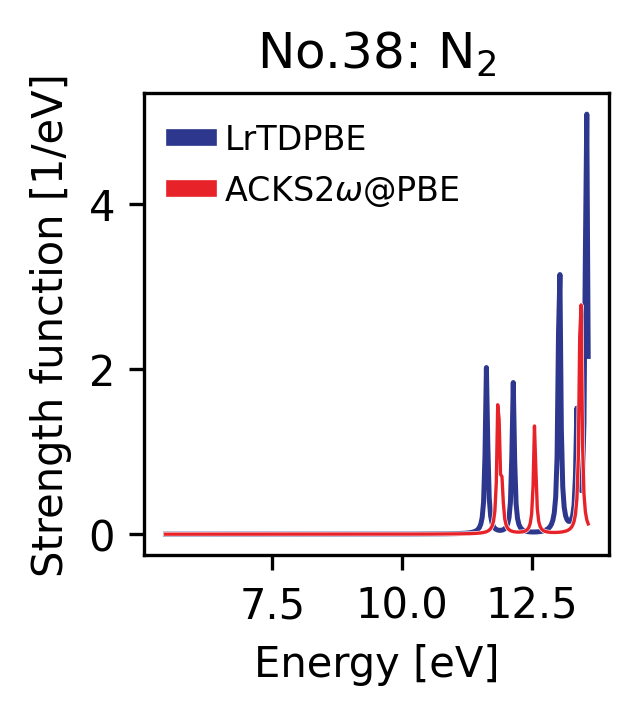}
    \includegraphics[scale=0.9]{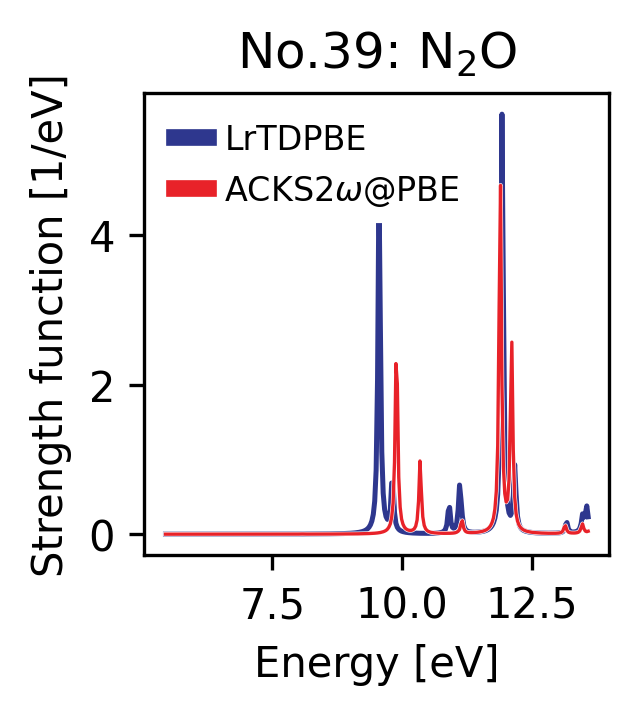}
    \includegraphics[scale=0.9]{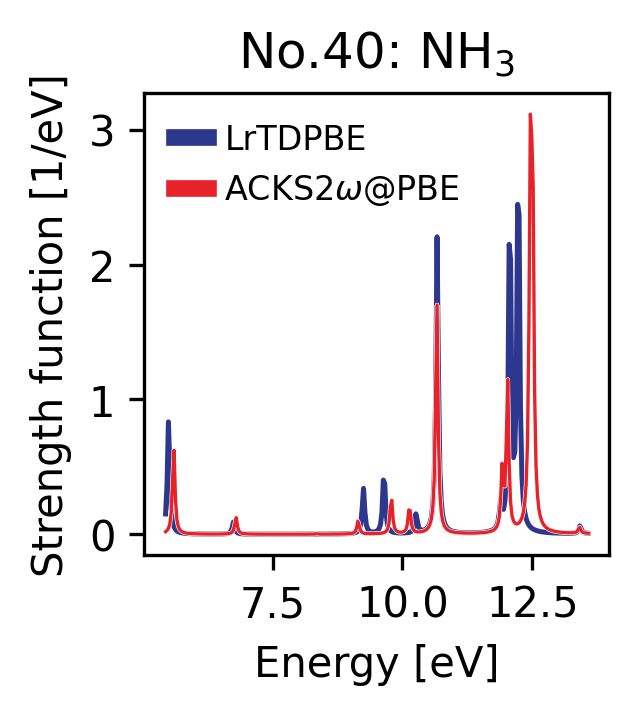}
    \caption{
    Comparison of spectra calculated by ACKS2$\omega$@PBE and LrTDPBE methods for \ce{H2}, \ce{H2O}, \ce{H2S}, \ce{HBr}, \ce{HCl}, \ce{HF}, \ce{N2}, \ce{N2O}, and \ce{NH3}.
    }
\end{figure}

\begin{figure}[htbp]
\centering
    \includegraphics[scale=0.9]{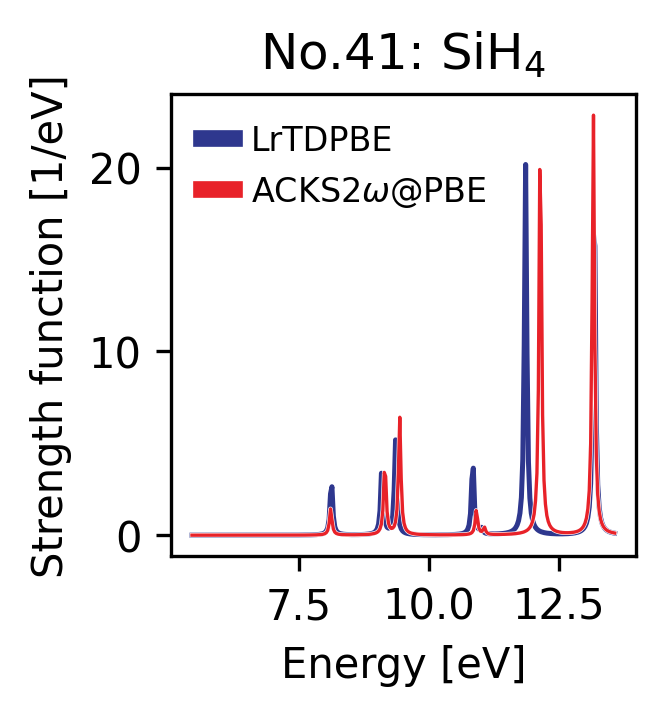}
    \includegraphics[scale=0.9]{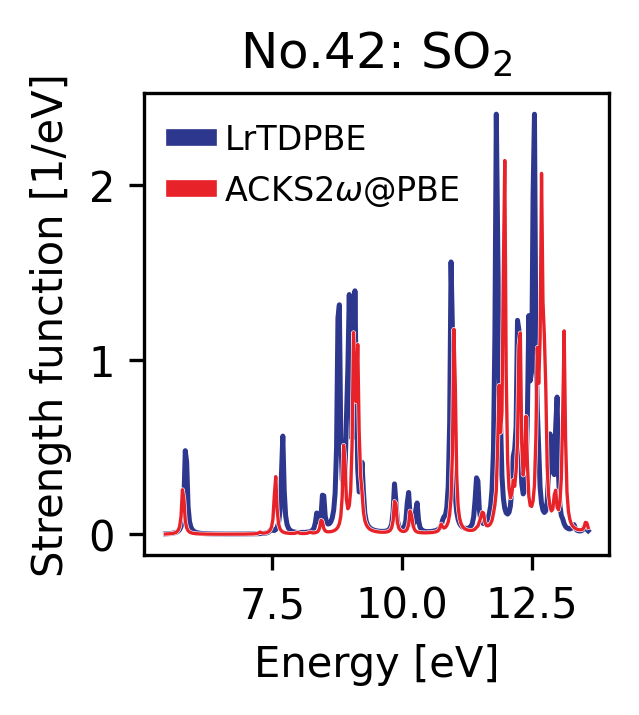}
    \caption{
    Comparison of spectra calculated by ACKS2$\omega$@PBE and LrTDPBE methods for \ce{SiH4} and \ce{SO2}.
    }
\end{figure}

\bibliography{}